\documentclass[nologo,11pt,a4paper,hidelinks]{ETHpaper}
\usepackage[english]{babel}
\usepackage{tabularx}
\usepackage{MnSymbol}
\usepackage{booktabs}
\usepackage{caption}
\usepackage{hhline}
\usepackage[table,dvipsnames]{xcolor}
\usepackage{graphicx}
\usepackage[square,numbers,sort&compress]{natbib}
\usepackage{setspace}
\usepackage{cleveref}

\usepackage{appendix}
\usepackage{amsmath,mathtools}
\usepackage[utf8]{inputenc}
\usepackage{xcolor} \usepackage{soul}
\usepackage{subcaption}
\usepackage{hyperref}
\usepackage{xparse,tikz,calc} \usepackage{pifont}

\renewcommand{\epsilon}{\varepsilon}

\newcommand{\mean}[1]{\left\langle #1 \right\rangle}
\newcommand{\abs}[1]{\left| #1 \right|}
\newcommand{\sign}{\mathrm{sign}}
\newcommand{\inlinequote}[1]{{``#1''}}

\usepackage{tikz}
\usetikzlibrary{arrows}
\usetikzlibrary{arrows.meta}
\usetikzlibrary{backgrounds}
\usetikzlibrary{calc}
\usetikzlibrary{fit}
\usetikzlibrary{positioning}
\usetikzlibrary{trees}

\definecolor{firstcolor}{RGB}{168,50,45}
\definecolor{colorSG}{HTML}{881f24}
\definecolor{customY}{HTML}{bb9d79}
\definecolor{customG}{HTML}{608065}
\definecolor{customB}{HTML}{274f66}
\definecolor{customP}{HTML}{954c35}

\usepackage[framemethod=tikz]{mdframed}
\newenvironment{custombox}[2]{
	\mdfsetup{
		frametitle={\colorbox{white}{\sffamily \space #1\space}},
		innertopmargin=6pt,
		frametitleaboveskip=-.8\ht\strutbox,
		roundcorner=10pt,
		linewidth=2pt,
		linecolor=#2
	}
	\begin{mdframed}\vspace{-3mm}
	}{
		\vspace*{0mm}\end{mdframed}
}

\usepackage{enumerate}
\usepackage{enumitem}
\setlist[itemize]{%
  labelsep=3mm, noitemsep, topsep=0pt, leftmargin=10.5mm}

\setlist[enumerate]{noitemsep, topsep=0pt, labelsep=3mm, leftmargin=10.5mm}

\title{Modeling social resilience: \\ Questions, answers, open problems}
\titlealternative{Modeling social resilience: Questions, answers, open problems} 
\renewcommand*{\thefootnote}{\fnsymbol{footnote}}
\author{Frank Schweitzer$^{1,2,}$\footnote{Corresponding author, \texttt{fschweitzer@ethz.ch}},
  Georges Andres$^{1}$,
  Giona Casiraghi$^{1}$,
  Christoph Gote$^{1,3}$,
  Ramona Roller$^{1}$,
  Ingo Scholtes$^{3,4}$,
  Giacomo Vaccario$^{1}$,
  Christian Zingg$^{1}$}
\authoralternative{F. Schweitzer, G. Andres, G. Casiraghi, C. Gote, \\ R. Roller, I. Scholtes, G. Vaccario, C. Zingg}
\address{$^{1}$Chair of Systems Design, ETH Zurich, Switzerland\\
  $^{2}$Complexity Science Hub, Vienna, Austria \\
  $^{3}$Department of Informatics, University of Zurich, Switzerland \\
  $^{4}$Chair of Computer Science XV,
Julius-Maximilians-Universität Würzburg, Germany}
\www{\url{http://www.sg.ethz.ch}}

\reference{%
 To appear in: \emph{Advances in Complex Systems}, vol. 25, no. 8 (2022)} 
\makeframing

\usepackage{todonotes}

\begin{document}

\maketitle
\renewcommand*{\thefootnote}{\arabic{footnote}}

\begin{abstract}
Resilience denotes the capacity of a system to withstand shocks and its ability to recover from them.
We develop a framework to quantify the resilience of highly volatile, non-equilibrium social organizations, such as collectives or collaborating teams.
It consists of four steps: (i) \emph{delimitation}, i.e., narrowing down the target systems, (ii)
\emph{conceptualization}, i.e., identifying how to approach social organizations, (iii)
formal \emph{representation} using a combination of agent-based and network models,
(iv) \emph{operationalization}, i.e. specifying measures and demonstrating how they enter the calculation of resilience. 
Our framework quantifies two dimensions of resilience, the \emph{robustness} of social organizations and their \emph{adaptivity}, and combines them in a novel resilience measure.
It allows monitoring resilience instantaneously using longitudinal data instead of an ex-post evaluation.

\end{abstract}

\section{Introduction}
\label{sec:Introduction}

Why do some social organizations succeed to persist and thrive in the presence of crises and shocks, while others fail under the same conditions?
They have different levels of \emph{resilience} that can be most generally described as \emph{a system's capacity to withstand shocks and its ability to recover from them.}
Such a description already implies different features:
\begin{enumerate}[noitemsep,label=(\roman*)]
\item resilience is a \emph{systemic} property, as opposed to a property of system elements, 
\item resilience is not restricted to a \emph{specific} system, it rather seems to be a general property of different systems, 
\item resilience is described as a \emph{response} to a \emph{shock}, i.e., it can be only recognized in the presence of shocks, or perturbations, 
\item resilience is not a static property because shocks and recovery imply time dependent \emph{processes}.
\end{enumerate}
Resilience has to consider not only the \emph{magnitude} of shocks, but also different \emph{types} of shocks the system has to absorb.
The same system can be robust to, e.g., the impact of an earthquake, but not to the spreading of a disease.
More importantly, to be resilient a system also needs to have the \emph{ability to recover}.
A system may be robust even in the presence of various perturbations, but once it is impacted by a critical shock it will not recover, so it cannot be seen as resilient. 

Both the response to a shock and ability to recover are system specific and therefore difficult to generalize.
This also hampers 
a more precise or formal definition of resilience.
We should not expect that there is a universal concept of resilience, applicable to various types of systems \citep{Carpenter2001, Meerow2016, Walker2012}.
In fact, resilience concepts diverge across and sometimes even within scientific disciplines \citep{Baggio2015,annarelli2014a,2011-viabil-resil}.

We therefore do not provide a review of existing resilience concepts and refer to the literature already available \citep{Hosseini2016,Fraccascia2018a,Baggio2015}.
Instead, with our paper 
we want to broadly \emph{inspire} researchers from different scientific disciplines who already study social organizations by providing new modeling perspectives. 
The analysis of collaborative teams and collectives is a core topic of social psychology \citep{beyerlein2003beyond} and organizational theory \citep{guimera2005}.
Hence, case studies inform about, e.g., collective decisions, coordination and conflict resolution in teams.

But the models used are most often \emph{descriptive} models, not \emph{generative} models.  
Descriptive models include \emph{statistical} models, e.g., regression models, or \emph{database} models, e.g., conceptual entity-relationship models that indeed resemble our knowledge graphs (see Figure~\ref{fig:knowledge}).
In organizational psychology there are \emph{mental} models of teams and team members to describe perceived relationships or the collective representation of knowledge.
Descriptive models try to include as much detail as \emph{possible}.
But generative models try to include as much detail as \emph{necessary} to generate a macro-social behavior from the micro dynamics of the constitutive individuals.
This methodological approach is advocated in \emph{analytical sociology}~\citep{HedstromBearman2009}.

Agent-based and network models belong to the class of generative models.
Stochastic actor-oriented models (SAOM) \cite{snijders-2017-stoch-actor} and exponential random graph models (ERGM) \cite{leifeld-cranmer-2019,Robins2007,koskinen-daraganova-2012-expon-random}
aim at combining agent-based and network approaches \cite{flache2017models}. 
They also belong to the class of \emph{data-driven models}, using methods similar to logistic regression.
With their focus on link prediction to detect reciprocity, transitivity or homophily they are less suited to study systemic properties of social systems, such as resilience.
Computational issues, in particular scalability,
assumptions about utility maximization of actors and problems in model specification prevent a broader range of applications \citep{leifeld-cranmer-2022}.

But there are better solutions. 
In this overview paper we want to sketch a framework how to utilize them for the study of social organizations.
This framework provides interfaces for mining larger and more fine grained data about interactions between individuals (Section \ref{sec:what-data}), an analytically tractable network ensemble to avoid computational issues (Section \ref{sec:prob-appr}), statistical methods to infer signed relations and significant interactions from observed data (Section \ref{sec:prob-appr}) and formal ways to estimate the social impact of agents involved in these relations (Section \ref{sec:quant-agent-prop}).

But most of all, this framework allows to calculate robustness and adaptivity  \emph{instantaneously}, to estimate  the resilience of the organization.
The attention is shifted from the micro configurations, the dyads and triads, towards the macro-properties of social networks.
These are no longer reduced to topological features, but involve a dynamic component to describe the response to shocks, namely 
the potential for change in an organization.
With this the formal modeling of social organizations can be moved to a new level.
It will also impact research on resilience which is dominated by two paradigmatic views, engineering and ecological resilience (Sections \ref{sec:infr-syst}, \ref{sec:ecological-systems}).

So far, theoretical research on resilience and empirical research on resilience indicators have been largely segregated~\cite{Scheffer2012}.
Many studies restrict their focus on specific systems, in particular engineered systems, ecological, or urban systems \citep{Henry2012,Gunderson2000,Meerow2016,shamsuddin-2020-resil,Dinh2012}.
There are also studies about the resilience of socio-economic systems and organizations \citep{Hosseini2016,lampel2014a}.
As we detail below, 
they are of little help for the problems discussed in this paper, for two reasons:
(i) When referring to 
social systems, most often our modern human society is addressed \citep{lazega2022introduction}.
This bears a complexity way too large to be captured in a formal modeling approach and restricts the discussion to a discourse level.
(ii) Our research interest are social organizations at smaller scale. Organizational resilience studies have pointed out ``factors'' for improving the resilience of such systems, e.g., integration or redundancy \citep{mamounilimnios2014a,Sheffi2005} or social capital \cite{Shenk_2019,kerr-2018-social-capit,petzold-2017-social-capit}. 
But they do not instruct us \emph{what to do} if such factors shall be modeled and quantified.

A major aim of this paper is to provide a framework to overcome this research gap.
To develop a broader foundation, 
we will specifically address the problems of conceptualizing social organizations.
One main issue is their \emph{volatility}, i.e., the continuous change of their structure and dynamics that makes it difficult to define stability, to measure the impact of shocks, or to distinguish recovery from change.
A second and probably more ambitious aim is to  revise the premature
anxiety-laden understanding of ``shocks'' and ``breakdowns'' that dominate the discussion of societal resilience.  
To cope with social organizations, 
we need to shift the focus from the \emph{fear to breakdown} towards the \emph{faith to recover}.

\section{What do we know about resilience?}
\label{sec:what-do-we}

In order to investigate the resilience of social organizations, we should first take a look at two of the most prominent resilience concepts in engineering and in ecology.
They may provide already good starting points to formalize robustness and adaptivity.
If so, we have to test weather such formalization could be utilized to model social organizations.
But even if that is not the case we learn from the shortcomings about requirements for social resilience concepts. 
This helps to further clarify the underlying assumptions. 

\parbox{\textwidth}{

  \begin{custombox}{Questions}{black!50}

    \begin{itemize}
    \item Is there a difference between robustness, stability, and resilience?
    \item What is the relation between adaptivity and recovery? 
    \item How is robustness related to existing systemic risk measures? 
    \item Should a resilient system always return to equilibrium?  
    \end{itemize}
  \end{custombox}
}

\subsection{Robustness and adaptivity}
\label{sec:robustn-adapt}

\paragraph{Constituting dimensions. }

\begin{figure}[htbp]
    \centering
%

\begin{tikzpicture}[edge from parent fork down,
                    level 1/.style={sibling distance=6cm},
                    level 2/.style={sibling distance=0.8cm}]
    \tikzstyle{Resilience}=[rectangle, very thick, draw=customB, fill=customB!20, rounded corners, minimum height=0.7cm]
    \tikzstyle{RobAda}=[rectangle, very thick, draw=customG, fill=customG!20, rounded corners, minimum height=0.7cm, align=center]
    \tikzstyle{Dimension}=[anchor=east, rotate=45]
    \node [Resilience] (Resilience) {\textbf{Resilience}} {
        child {
            node[RobAda] (Robustness) {\textbf{Robustness:} \textbf{Structural}}
                child[child anchor=east] {
                    node[Dimension] {\small Absorptive \citep{Francis2014,Vugrin2011}}
                        edge from parent[thick]
                }
                child[child anchor=east] {
                    node[Dimension] {\small Buffer \citep{IfejikaSperanza2014}}
                        edge from parent[thick]
                }
                child[child anchor=east] {
                    node[Dimension] {\small Coping \citep{Keck2013}}
                        edge from parent[thick]
                }
                child[child anchor=east] {
                    node[Dimension] {\small Redundant \citep{Bruneau2003a}}
                        edge from parent[thick]
                }
                child[child anchor=east] {
                    node[Dimension] {\small Robust \citep{Bruneau2003a}}
                        edge from parent[thick]
                }
            edge from parent[very thick]
        }
        child {
            node[RobAda] (Adaptivity) {\textbf{Adaptivity:} \textbf{Dynamic}}
                child[child anchor=east] {
                    node[Dimension] {\small Adaptive \citep{Keck2013,Francis2014}}
                        edge from parent[thick]
                }
                child[child anchor=east] {
                    node[Dimension] {\small Learning \citep{IfejikaSperanza2014}}
                        edge from parent[thick]
                }
                child[child anchor=east] {
                    node[Dimension] {\small Recoverability \citep{Francis2014}}
                        edge from parent[thick]
                }
                child[child anchor=east] {
                    node[Dimension] {\small Self-organization \citep{IfejikaSperanza2014}}
                        edge from parent[thick]
                }
                child[child anchor=east] {
                    node[Dimension] {\small Transformative \citep{Keck2013}}
                        edge from parent[thick]
                }
            edge from parent[very thick]
        }
    };
\end{tikzpicture}
    \caption{
        Examples of resilience factors used in the literature, which we assigned to the structural and the dynamic dimension of resilience without claiming consistency across  different systems and shocks.
    }\label{fig:intro-robustness-adaptivity-by-papers}
\end{figure}
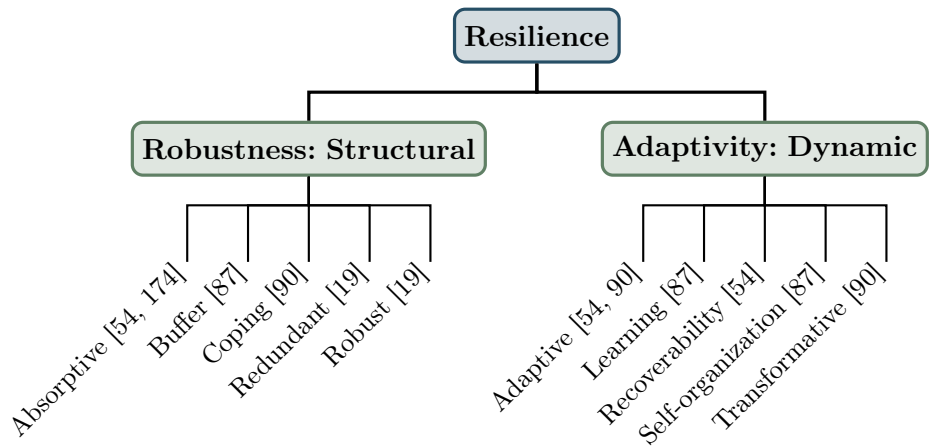

While the concrete meaning of resilience may vary across scientific disciplines,
there are also conceptual commonalities.
As pointed out in Figure~\ref{fig:intro-robustness-adaptivity-by-papers},  resilience bears relations to  concepts of \emph{robustness} and \emph{adaptivity}.
The former relates to the property of a system to withstand shocks, the latter to its ability to overcome their impact.
Robustness  represents the \emph{structural} and adaptivity  the \emph{dynamic} dimension of resilience.
However, we need to understand how they constitute resilience as a function defined on these two dimensions. 

\paragraph{Topology. }

We argue that both dimensions need to be combined to explain resilience, but most often they  have been studied as \emph{stand-alone concepts}.
For instance, the robustness of 
interconnected systems is defined with respect to the failure of nodes or links. 
Robustness measures then estimate the size of failure cascades
after a shock \citep{albert2004structural,Cohen2000,Peixoto2012,Bascompte2007}. 
This approach uses the \emph{complex networks} perspective that we also utilize in our framework.
But it assumes that simple topological features, like connectedness, are sufficient to describe the robustness of a system.
This leaves out the dependence on individual properties which becomes a problems when applying the approach to social systems.

\paragraph{Equilibrium. }

In addition to topological robustness, other robustness measures build on dynamical stability.
It analyzes how small perturbations of an assumed equilibrium state affect a system.
If the system is able to regain its equilibrium state, it is robust at least against this type of perturbations.
This approach has several limitations when applying it to social organizations.
It requires to know the dynamics of the system, to estimate its response to shocks.
Further, the underlying assumption of different types of equilibrium states can hardly be justified for very volatile social systems.

\paragraph{Response. }

Adaptivity is often simply understood as \emph{dynamics}, which neglects the quality of  change.
Adaptive systems do not simply react to a shock.
Their response aims at preserving the system's functionality, to ensure its persistence.
Recovery after a shock therefore implies a certain directedness which depends on the type of shock.
Most important, the system needs to have several options to adapt even to unexpected challenges.
Therefore, adaptivity should estimate how many options exist in a given situation.
This points towards the problems of quantification and measurement discussed later. %

\paragraph{Creative destruction. }

Additionally, simply adopting existing concepts of robustness and adaptivity may lead to misconceptions about  social systems. 
Quite often, the understanding of these concepts builds on a negative perception of shocks.
This ignores the role of ``creative destruction'' that, according to the economist Joseph Schumpeter, is instrumental for renewing and further developing the economy.
Stable systems do not evolve.
Therefore, challenging their stability is one of the driving forces of evolution. 
This also regards social organizations.
The leave of established members is not only a threat, it is also an opportunity for newcomers.
Creative organizations often respond to shocks with innovations.
Therefore, the discussion about robustness and adaptivity should not just focus on maintaining the \emph{status quo}.
Resilience means to cope with change in a sustainable manner.

\subsection{Infrastructure systems}
\label{sec:infr-syst}

\paragraph{Critical functionality. }

With respect to critical technical infrastructures such as power grids or communication networks, the design of resilient systems has been studied extensively \citep{%
Hollnagel2007, Tamvakis2013, Smith2011a, Sterbenz2010, albert2004structural, buldyrev2010catastrophic}.
Here resilience, sometimes also called ``resiliency'', refers to a system's capacity to \inlinequote{maintain an acceptable level of service in the presence of [\dots] challenges} \citep{Smith2011a}.
Such challenges include software and hardware faults, human mistakes, malicious attacks or large-scale natural disasters.

\begin{figure}[htbp]
  \centering
  \includegraphics[width=0.41\textwidth]{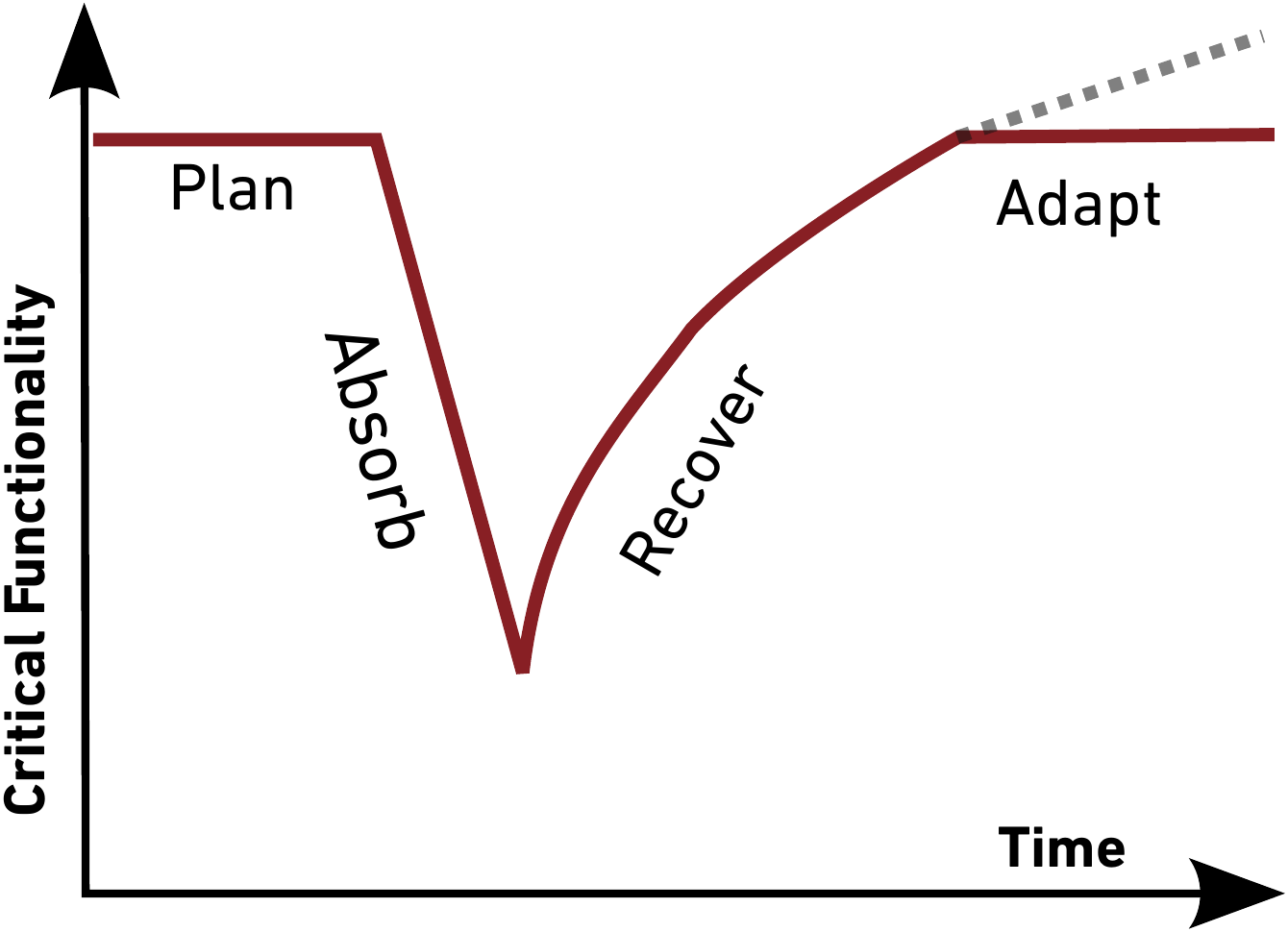}
  \caption{The common perception of resilience for engineered systems.
  }
  \label{fig:resilience1}
\end{figure}

This concept of resilience is goal-oriented, preserving a system's \emph{functionality} after the shock as illustrated in 
Figure~\ref{fig:resilience1}. 
A power system has a defined functionality which remains as long as no critical shocks are caused by \emph{internal} malfunction, e.g., lack of maintenance, or by \emph{external} disruptions, e.g., an earthquake.
If the system has to ``absorb'' a shock, this functionality is partially or entirely destroyed because the system lost its robustness, to some degree.

\paragraph{Adaptivity. }

We note that functionality is assumed here as a function of an underlying network structure, in this case the power grid.
Consequently, the focus is on the \emph{robustness} of this underlying structure.
\emph{Adaptivity} refers to possible changes in the \emph{processes running on this structure}, and not to the structure itself.
This can be illustrated by the following example:
During the 9/11 attacks in 2001, the Internet infrastructure in downtown Manhattan was largely destroyed.
This posed a severe shock to the global Internet because of (i) the termination of transatlantic cables in the basement of the World Trade Center (WTC) and (ii) the failure of the major Internet exchange point NYIIX next to the WTC, which was responsible for 70 \% of transatlantic traffic.
Despite these combined failures, the attack caused only minor disruptions and the global routing infrastructure continued to operate normally within a few hours~\citep{Eisenberg2003,Sterbenz2010}.

This resilience of the Internet was not obtained by the robustness of the hardware, which was heavily affected in the above example, but by the adaptive capacity of the dynamics \emph{on} the network. 
However, \emph{redundancies} built into the underlying \emph{network}, i.e., a level of \emph{robustness},
were essential to allowing for the quick rerouting of the communication flow, i.e., \emph{adaptivity}.
Hence, a minimum level of robustness can be a precondition for the adaptive  capacity.

\paragraph{Recovery. }

What distinguishes resilience from robustness is the ability to recover.
Infrastructure systems do not recover by themselves \cite{Gonzalez2017}.
They have to be rebuilt by human activity, reestablishing the functionality and often even improving it.
Thus, attributing the ability to recover to the infrastructure would reduce the concept of resilience to absurdity.
Every bridge that was rebuilt after a collapse would then be ``resilient''.
In engineering, hence resilience
is often used as a synonym for robustness. 
For example,  \citet{Dinh2012} define a ``resilient reactor'' as one achieved by \inlinequote{[using] a tank designed to withstand high pressures and temperatures}.
However, once a tank has exploded, it has no ability to recover a functional state.

\paragraph{Systemic risk. }

Quantifying the impact of shocks  requires to model the system's response.
Only such models allow to estimate whether large parts of the system will be destroyed, which is commonly denoted as \emph{systemic risk}.
Based on the theory of extreme events \citep{de2007extreme}, one can calculate probabilities for rare, but severe \emph{external} shocks, e.g., floods or earthquakes.
Using engineering knowledge about materials \citep{nash1998} and constructions, then allows to propose critical values for system properties, e.g., the minimum width of walls, or the maximum height of buildings, etc.
Systemic risk, from this perspective, is reduced to the risk that an external event with a critical magnitude may occur.
Consequently, robustness is defined only with respect to such rare events,
for instance, 
\inlinequote{the ability of power systems to withstand low-probability high-impact incidents in an efficient manner}
\citep{Khodaei2014}. 

\paragraph{Loss estimation. }

In infrastructure systems, such as transportation networks, it is common to measure robustness by employing loss estimation models.
These models  evaluate the ability of the system elements to withstand a given level of stress or demand without suffering degradation or loss of function \citep{Bruneau2003a}.
Quantification allows to \emph{control}  the robustness of such systems to some degree, e.g., counterbalancing shocks by a central control station. 
Engineered systems  are \emph{designed} systems, top to bottom. Their functions are clearly defined, and therefore the impact of a shock can be estimated.

\paragraph{Failure cascades. }

The focus on external extreme events neglects another major cause for systemic collapse, namely the \emph{amplification} of a few small failures inside the system into a failure cascade.
If such cascades reach a critical size, they can also destroy the system from inside.
Robustness in this case has to be defined with respect to (i) the probability that a few system elements fail, and (ii) the mechanisms that can amplify such failures, e.g., by redistributing load.
Both may depend, in addition to internal conditions, also on external feedback processes.  
Large scale blackouts in the U.S. have been explained by such failure cascades in the power grid \citep{albert2004structural,wang2015}.

To study failure cascades we need an appropriate representation of the engineered system.
For power grids, communication networks or transportation infrastructure, the network approach is used most often \citep{Wuellner2010}.
System elements, e.g., transformers  or responders, are represented by nodes, and their physical connections by links in the network.
Many models of failure cascades test the robustness of networks by removing nodes or links either in a random or a targeted manner \citep{buldyrev2010catastrophic,Valdez2020} and measure whether the network breaks down into disconnected components after such attacks. 
A modular network is robust if shocks trigger small failure cascades \citep{Ash2007, brummitt2012suppressing}.
Targeted attacks, however, usually result in large failures as the network becomes disconnected \citep{Shai2014, bagrow2015robustness}.

\paragraph{Amplifying mechanisms. }

Removal tests reduce the problem of systemic risk to mere topological properties.
To \emph{understand} amplifying mechanisms inside a systems, we need models for the dynamics inside the nodes, but also models for the exchange of quantities between nodes~\citep{Lorenz:2009aa,burkholz2016damage,Burkholz2016,Burkholz2018}.
That means we need a framework that couples the dynamics \emph{of} the network, i.e., the failure  of nodes or links, with the dynamics \emph{on} the network, i.e., the rerouting of communication or the redistribution of load. 

Without such a framework, we cannot understand the robustness of the system, even less its adaptivity, and hence, its resilience.
Extreme value theory \cite[]{chavez-demoulin-embrechts-hofert-2015-extrem-value} can at best estimate the probability that large-scale failure cascades happen, whereas network models can provide an explanation \emph{why} they happen.
Resilience, as a systemic property, cannot be reduced to the robustness against extreme shocks, it has to be derived from a broader perspective that helps to understand why and when small events can be turned into big disasters.

\subsection{Ecological systems}
\label{sec:ecological-systems}

\paragraph{The classic view. }

Historically, the notion of resilience first appeared in \emph{ecology}.
Ecosystems constantly change under \emph{dynamic} processes of renewal and reorganization.
Therefore, resilience concepts do not primarily focus on robustness as a static property of the system, but rather on the adaptivity as the \emph{dynamic} component of resilience.
For \citet{Holling1973}, resilience is based on   the ``ability of a system to return to an equilibrium state after a temporary disturbance''.
Because shocks and perturbations are unavoidable, emphasis is on the  \emph{survival}, or the \emph{persistence}, of the ecological system, regardless of the impact of a shock \citep{walker2004resilience}.

\paragraph{Dynamical systems. }

Such a notion of resilience essentially builds on the theory of dynamical systems
and its concept of ``stability''.
In different scientific areas, e.g., nonlinear dynamics, control theory, physico-chemical reaction kinetics, or biological pattern formation, %
a system is said to be stable if it returns to the equilibrium state after a shock.
Approaches to assess resilience in biological and engineered systems are based on this idea~\citep{Kitano2004}.

\begin{figure}[htbp]
  \centering

\def\centerarc[#1](#2)(#3:#4:#5)
    { \draw[#1] ($(#2)+({#5*cos(#3)},{#5*sin(#3)})$) arc (#3:#4:#5); }

\tikzstyle{Potential}=[ultra thick]
\tikzstyle{Axis}=[thick, -Stealth]
\tikzstyle{MotionArrow}=[ultra thick, -{Triangle[length=8pt, width=10pt]}, line width=5pt]
\tikzstyle{RollingBall}=[circle, minimum size=15pt]
\tikzstyle{StableBall}=[circle, minimum size=15pt]

\resizebox{0.485\textwidth}{!}{
    \begin{tikzpicture}
        \draw[Potential, customB] (0, 0.25)
            .. controls (0.95, 0) and (1, 0.85) .. (2, 1)
            .. controls (3.85, 1) and (4.75, 0) .. (5, 3.25);
        \fill[shading=axis, left color=customB!25, right color=customB, shading angle=180] (0, 0.25)
            .. controls (0.95, 0) and (1, 0.85) .. (2, 1)
            .. controls (3.85, 1) and (4.75, 0) .. (5, 3.25)
            -- (5, 0)
            -- (0, 0)
            -- cycle;

        \node[StableBall, fill=customB] (stableball) at (3.8, 1.14) {};
        \node[RollingBall, fill=customB!40] (rollingball) at (2.27, 1.28) {};

        \centerarc[](stableball)(10:40:0.325)
        \centerarc[](stableball)(10:40:0.37)
        \centerarc[](stableball)(10:40:0.415)

        \centerarc[](stableball)(140:170:0.325)
        \centerarc[](stableball)(140:170:0.37)
        \centerarc[](stableball)(140:170:0.415)

        \node[below, yshift=-7pt] at (stableball) {\emph{shock}};
        \node[above, anchor=south, yshift=7pt] at (rollingball) {\emph{low recovery rate}};

        \draw[MotionArrow, draw=customB!20]  ($(stableball)!0.3!(rollingball)$) -- ($(stableball)!0.75!(rollingball)$);

        \node at (2.5, -0.4) {State};
        \node[anchor=south, rotate=90] at (-0.1, 1.625) {Potential};

        \draw[Axis] (0, 0) -- (5.3, 0);
        \draw (1, 0.05) -- (1, -0.05);
        \draw (2, 0.05) -- (2, -0.05);
        \draw (3, 0.05) -- (3, -0.05);
        \draw (4, 0.05) -- (4, -0.05);
        \draw (5, 0.05) -- (5, -0.05);

        \draw[Axis] (0, 0) -- (0, 3.75);
        \draw (0.05, 1) -- (-0.05, 1);
        \draw (0.05, 2) -- (-0.05, 2);
        \draw (0.05, 3) -- (-0.05, 3);
    \end{tikzpicture}
}
  \caption{
    Mechanical analogy for a metastable equilibrium.
    This figure is adapted from Figure~1 in \citep{Lenton2012}.
  }
  \label{fig:eq}
\end{figure}
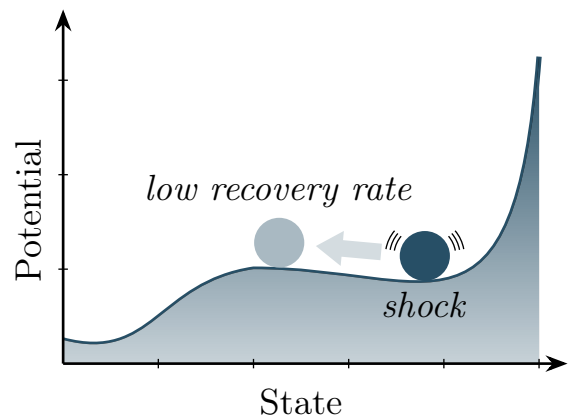

Most systems are in fact \emph{metastable}, i.e., they are stable in the presence of small perturbations, but become unstable if these perturbations exceed a certain critical level.
Then, instead of returning to the previous equilibrium state, the system moves away from it, possibly to another equilibrium state, as Figure~\ref{fig:eq} indicates.
This new equilibrium state corresponds to a ``different'' system, i.e., to a system that has evolved and adapted to the new situation.
This aspect has been studied as \emph{robust adaption}, combining the notions of robustness and adaptivity.
There are analogies to concepts of \emph{phase transitions} in physics and chemistry and \emph{regime shifts} in social and biological systems. 

\paragraph{The meaning of recovery. }

In the resilience concept of ecological systems \emph{recovery} may have different connotations dependent on
whether a system returns to a \emph{previously} attained stable state or to a \emph{different} one.
Even a previous state would be gradually different because evolving systems would never reach identical states.

In more general terms, resilience can be implicitly defined by    \inlinequote{the amount of disturbance that an ecosystem could withstand without changing self-organized processes and structures (defined as alternative stable states)} \citep{Gunderson2000}.
To capture differences in the quality of a resilient ecosystem, the additional factors of ecological  \emph{``vulnerability''} and \emph{``sensitivity''} have been proposed \cite[]{dalziell2004, saavedra2011b, medeiros-allesina-2022-rankin}.
They consider that species are more sensitive to perturbations in certain time periods or under special conditions.
This  complements the notion of ``stability'' of ecosystems or ``robustness'' of engineered systems.

\paragraph{Time scales. }

What matters for ecological resilience is the time scale of recovery.
If it would take forever to return to a previous state, the system is not resilient. 
Therefore, \inlinequote{the inverse of the length of time required for an ecosystem to return to near-normal} \citep{Grodzinski1990} is often used as a resilience measure.
But resilience itself can also be \emph{timescale specific} \citep{Carpenter2001, Dakos2015a}.
On a daily timescale, the system may return to a near-normal state.
But on a time scale of decades, the system may lose this ability.

Hence, resilience cannot be reduced to some concept of ``stability'', 
the dimension of \emph{adaptivity} is essential.  
Resilient ecosystems maintain a critical balance between stability and instability, the former providing a level of robustness against small perturbations, the latter allowing them to adapt to changing environmental conditions~\citep{walker2004resilience}.

\paragraph{Connectedness and potential. }

To further grasp the generic mechanisms underlying ecological resilience, two system properties have been proposed: \emph{potential} and \emph{connectedness}.
``Potential sets limits to what is possible - it determines the number of alternative options for the future'' \citep{HollingGunderson2002}. 
Hence, potential resembles our notion of adaptivity. 
Connectedness, on the other hand, is related to robustness.
``Connectedness is assumed to increase over time, leading to high internal control and limited potential to cope with disturbances''  \citep{HollingGunderson2002}.
This is an important remark as it points out to the fact that robustness above a certain critical value may have a negative impact on resilience.
``When connectedness is low, resilience is high because the system can vary over a wide range of states and respond to disturbances in many different ways.
When connectedness, however, is high, ecosystem resilience is low because the system is more tightly organized and has fewer options for responding to disturbances'' \citep{grimm2011resilience}. 
We will return to this argument when discussing our own concept of social resilience.

\paragraph{Adaptive cycles. }

It was argued that resilience changes in a cyclic manner because an increase in connectedness may lead to a breakdown of the system:
``Naturally such over-connected systems crash into a release period, where they have the potential to reorganize, thereby coping with disturbances.
This development is believed to be cyclic'' \citep{grimm2011resilience}.  
We note that the ability to reorganize is given only if during ``release phases''  of low connectivity the potential, i.e., the adaptivity, is high.
This is in line with our arguments for the recovery of social organizations discussed below.

So far, adaptive cycles have not been found empirically:
``Because of its very general nature, the concept of the adaptive cycle should be considered a metaphor \citep{Carpenter2001} or thinking tool rather than a testable scientific theory'' \citep{grimm2011resilience}.
We may add here that with our approach we move the adaptive cycle from a metaphor to a testable concept, which is also accessible to formal modeling.

Couplings between adaptive cycles on different temporal and spatial scales may lead to a nested hierarchy, called \emph{panarchy} \citep{HollingGunderson2002}.

\paragraph{Losing resilience. }

Instead of specifying properties for resilient states, we could address the complementary question, namely about the properties that indicate a loss of resilience or about states that need to be avoided. 
This reveals critical conditions for stabilizing feedback cycles, critical magnitudes for perturbations or critical levels of diversity. 
To identify factors that lead to the loss of resilience, non-parametric regression models or machine learning tools, e.g., symbolic regression,  can be used to analyze data.

Early warning signals for losing resilience can be obtained from time series.
A slower recovery rate from perturbations, known as \emph{critical slowing down} \cite[]{Lenton2012}, right before a tipping point is a possible indicator. 
Also an increase in the auto-correlation of systemic variables, e.g., order parameters, indicates the vicinity of a regime shift.

\paragraph{Coupling to social systems. }

Insights into ecological resilience are limited if the dynamics of ecological systems is predominantly driven by the coupling to the human sphere \cite[]{barlow-franca-2018}. 
On the global scale social and ecological systems are, in fact, coupled inseparably, mainly because of human interventions. 
In recent years, this has triggered integrative research studying the resilience of \emph{social-ecological} systems \cite{folke2006,walker2006,walker2004resilience}. 
It relates ecological resilience with the resilience of societies responding to environmental challenges that originate from the ecological systems into which they are embedded.

\parbox{\textwidth}{

  \begin{custombox}{Conclusions}{black!50}
Different scientific disciplines have their own understanding of resilience.
We should not expect to find a universal resilience measure.
The key question is \cite[]{Carpenter2001}: Resilience of what to what?
Answering this question requires to have a model of the respective system.
Resilience concepts for social organizations may benefit from ecological concepts, because issues of time scales, different equilibria and adaptive cycles are already addressed. 
\end{custombox}

}

\section{Why are social systems different?}
\label{sec:why-are-social}

Existing concepts of resilience from engineering or ecology do not seem to provide the best basis for social resilience.
Although formal approaches exist, we cannot simply reuse them. 
Before developing suitable alternatives, we may first clarify what makes social systems different from other types of systems.
This leads us to more fundamental questions about defining systems and models.
What seems to be a detour at this point will later allow us to better ground our notion of social resilience from a methodological perspective.

\parbox{\textwidth}{

  \begin{custombox}{Questions}{black!50}

    \begin{itemize}
    \item What types of social systems do we want to investigate? 
      
    \item What characterizes social organizations and collectives? 
    \item What is the focus of organizational resilience? 
    \item What modeling consequences entails the complex systems approach?
    \end{itemize}
  \end{custombox}
}

\subsection{Social organizations}
\label{sec:social-organizations}

\paragraph{No model of society. }
The need to understand social resilience is often motivated by the many crises that our \emph{societies} face, today \citep{teekens2021shaping}.
Their vulnerabilities have been widely recognized, ranging from pandemics to political polarization, from climate change to budget crises, from infrastructure breakdown to poverty migration.
Consequently, the resilience of societies cannot be decoupled from the resilience of ecosystems, political systems, infrastructure systems, financial and economic systems, etc.
While these connections cannot be denied, they raise a methodological question that, unfortunately, is not addressed with the same emphasis:
How should we \emph{model} all of these interdependencies?

\begin{figure}[h!]
  \centering
  \includegraphics[width=0.9\textwidth]{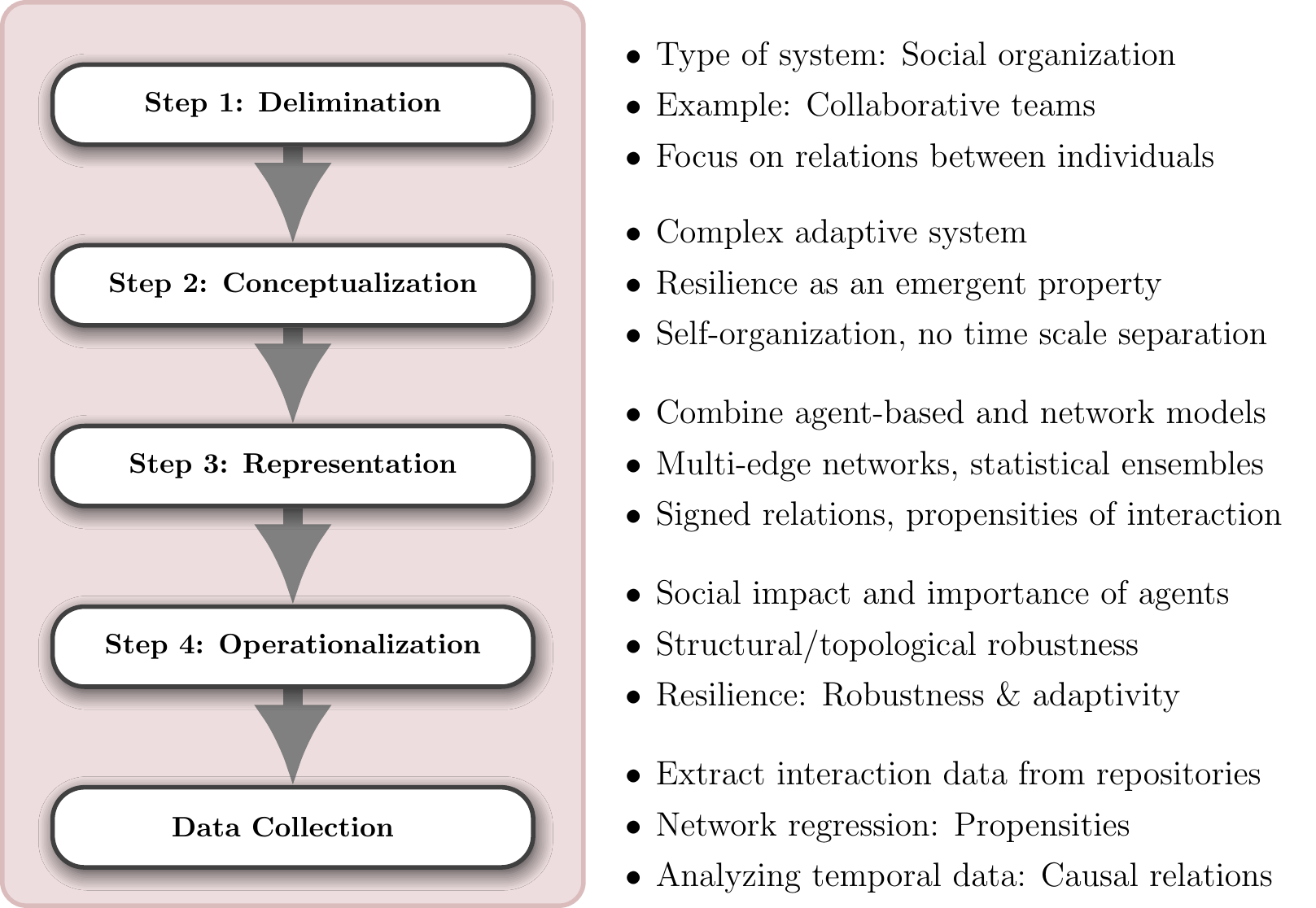}
  \caption{Our framework to quantify  social resilience}
  \label{fig:frame}
\end{figure}

\paragraph{Four steps towards a model. }

We aim at a \emph{quantitative} understanding of social systems.
Therefore, we specify the resilience problem in a tractable manner in \emph{four steps} that are summarized in Figure~\ref{fig:frame}. 
The first step is \emph{delimitation}: Which types of systems should be specifically investigated, and which ones \emph{not}?
This question is discussed in the remainder of Section~\ref{sec:social-organizations}.
In Section~\ref{sec:what-point-organ}, we further distinguish our problem from existing concepts of organizational resilience.

The second step is \emph{conceptualization}: Which approaches should we use to describe social systems? Which of the many possible features will we focus on?
This is discussed in Section~\ref{sec:conceptual-framework}. 
Only this clarification will enable the third step, \emph{representation}: To build a model means to represent the system and its elements in a formal manner.
In Sections~\ref{sec:netw-repr}, \ref{sec:prob-appr} we introduce different network concepts as \emph{candidates} to represent properties of social systems. 
In Section~\ref{sec:workbench}, eventually, we address the fourth step, 
\emph{operationalization}: How do we specify measures such that they can be calculated?
Again, we introduce different solutions to choose from.
Finally, in Section~\ref{sec:how-do-we}, we comment on the data required for the fourth step.

\paragraph{Systems of systems. }

For methodological reasons we distinguish \emph{social resilience} from the resilience of \emph{societies}. 
The latter is not the target of our formal modeling.
Society is a \emph{system of systems}, as described above.
To understand the resilience of \emph{societies} requires modeling the interaction of these systems \cite[]{reza-alavizadeh-tarokj-2012-from-system}. 

Models for systems of systems exist only in rudimentary form \citep{reza-alavizadeh-tarokj-2012-from-system}.
Particular emphasis was given to the coupling between socio-economic systems and ecological systems.
More recently, also the coupling between climate systems and socio-economic systems and/or ecological systems is captured by formal models.
Most of the systemic relations that constitute a ``society'' are not formalized at all 
\cite[]{helbing-balietti-2011-from}. 

\paragraph{Collectives. }

We have to restrict our investigation of social resilience to
clearly defined social entities rather than ``social systems'' in general.
Our focus are social organizations, or collectives. 
With the term \emph{social organization} we refer to \emph{formal or informal groups of interrelated individuals who pursue a collective goal and who are embedded into an environment}~\citep{ostrom2009understanding,hoegl2001teamwork}. 

For illustrative purposes, our running examples are project teams, in particular teams of software developers \cite{Scholtes2016b,hoegl2001teamwork,Xuan2015}.
These teams face numerous shocks during their development.
Competing products, technical evolution, organizational problems, lack of motivation or resources put up challenges and let them fail quite often. 
Their common goal, namely to develop a software for a certain scope,  is important to distinguish this type of social system from a  collection of a hundred persons who, for example, use the subway without being interrelated or contributing to a common good.
Collective goals in most cases go beyond the mere survival \citep{Elster, ostrom2009understanding}, which would characterize ecological systems.

\paragraph{Beyond individual resilience. }

The term ``social'' specifically refers to human individuals in the following, although many animal societies also have a remarkable degree of social organization.
A broad range of psychological studies focuses on the resilience of an individual \citep{egeland1993resilience,kitano2005resilience,arnetz2009trauma,villavicencioayub201545}.
Given that we want to understand resilience as a systemic property, psychological
resilience would require us to model the individual as a system ~\citep{luthar2003resilience}.
That could be possible, but is not aligned with our aims.

Instead, we are interested in \emph{collectives} of many individuals, for which we use the term \emph{social organization} synonymously.
We assume collectives of the order of $10^{2}$ individuals, small enough that the impact of a few individuals can still matter, but large enough to distinguish individual and collective in a meaningful manner.
This implies that resilience, as a systemic property, is neither identical to, nor the mere combination of the psychological resilience of its members.

\subsection{Organizational resilience}
\label{sec:what-point-organ}

Different from engineering, in \emph{psychology} and \emph{organizational science} the concept of resilience focuses on the \emph{dynamic} component, i.e., adaptivity as the system's ability to cope with shocks. 
But in social organizations resilience does not require to return to a previous state.
Hence, resilience is generally seen as
\inlinequote{the ability of groups or communities to cope with external stresses and disturbances as a result of social, political and environmental change} \citep{Adger2000}. 
We note that in such a definition shocks primarily result from \emph{other systems} an organization is embedded in, rather than from \emph{internal processes}, which is the main focus of our concept of social resilience.  

\paragraph{Community resilience. }

Examples for this outside orientation are studies of \emph{citizen communities} in urban or rural areas \citep{maguire2007disasters,Adger2000}.
Their response to natural hazards or disasters \citep{Bruneau2003a, coles2004} or to climate change \citep{marshall2010understanding}
is of particular interest.
\emph{Community resilience} generally refers to the ability of communities to cope and adjust to stresses
and to engage community resources to overcome adversity \citep{norris2008resilience,Shenk_2019}.
An organization should not only persist after a disturbance, but also manage to strengthen its capability for future adjustments \citep{Sutcliffe2003}. 
The ability to transform challenges into advantages is known as \emph{transformational} resilience.

Whether a new state is resilient may depend on specific positive outcomes that need to be achieved \citep{Obrist2010}.
Hence, resilience comprises more than just persistence: 
\inlinequote{The capacity of actors to access capitals in order to -- not only cope with and adjust to adverse conditions (that is, reactive capacity) -- but also search for and create options (that is, proactive capacity), and thus develop increased competence (that is, positive outcomes) in dealing with a threat} \citep{Obrist2010}.

\paragraph{Resilience factors. }

It is an open question why and how \emph{some} organizations manage to thrive and enhance core capabilities when faced with a crisis, while \emph{others} fail \citep{Vogus2007}.
Recent research highlights three factors that influence the resilience of a social system: (i) its \emph{vulnerability} (or susceptibility) to  disruptions, (ii) its level of \emph{anticipation}, foresight, or situational awareness for such vulnerabilities, and (iii) its \emph{adaptive capacity}, flexibility or fluidity which allows to mitigate vulnerabilities or respond to disruptions~\citep{Mcmanus2007,Goggins2014}.

To better understand the role of these factors, 
most empirical resilience studies have followed a ``hindsight approach'', focusing on organizations which have recovered from a shock and transformed crises into advantages~\citep{Wildavsky1988,Robb2000,Lengnick2005,freemansf2004}.
For such social organizations, \citet{Sutcliffe2003} define \emph{organizational resilience} as the ability to maintain a ``positive adjustment under challenging conditions''.

\paragraph{Adaptive capacity. }

A system's capacity to adapt in a constantly changing environment is also referred to as \emph{adaptive capacity} \cite{Gallopin2006,Smit2006}. 
From a social science perspective, the adaptive capacity is expressed in a number of different ways, for instance in terms of the ability to learn and store knowledge, the ability to anticipate disruptive events, the level of creativity in problem solving, or the dynamics of organizational structures~\citep{Folke2002,Smit2006}.
Some of these aspects have been assessed by means of survey research designs, such as the learning capability~\citep{Svetlik2007}, situational awareness, creativity~\citep{Mcmanus2007}, or the fluidity of structures~\citep{Goggins2014}.

\paragraph{Missing macro-variables. }

Most notions of resilience proposed above reveal their limitations when it comes to \emph{quantification} \citep{Levine2014,lazega2022introduction}. 
Two problems need to be solved, (i) to \emph{define} a measure for resilience that considers also the dynamics of the system, and (ii) to \emph{measure} the defined variables against available \emph{data}.
Many studies of social resilience, e.g., in disaster management \cite{Ilbeigi2019,sahebjamnia2015a,teo2013a,renschler2010,norris2008resilience,maguire2007disasters,coles2004}, monitor resources for basic needs or survey social well-being.
But we lack macro-variables to describe social organizations, e.g., to measure their adaptive capacity and their elasticity.
Such variables exist in economics, for instance productivity and efficiency measures, but also for ecological systems, e.g., biomass production or recovery rates. 
In engineering functional resilience can be  computed through the integral below the function of performance \citep{Bruneau2007, renschler2010}.

In absence of these variables, tools to derive early-warning signals, e.g., the critical slowing down or the increase of auto-correlations mentioned above, cannot be applied.
Therefore, one of our aims is to provide such macro-variables for robustness and adaptivity and to show how they can be monitored over time.
These variables will help separating the ability to resist a shock from the capacity to recover.

\subsection{Conceptualization}
\label{sec:conceptual-framework}

In Section~\ref{sec:social-organizations} we have already processed the first step towards a model of social resilience, namely \emph{delimitation}.
The second step, \emph{conceptualization}, now requires us to specify how to approach highly dynamic social organizations.
This particularly regards the modeling framework that shall foster our understanding of the \emph{micro-processes} to explain social resilience.

\paragraph{Different conceptual frameworks. }

Ecological systems are often modeled using concepts from \emph{system dynamics} \citep{schweitzer-2022-ACS}, where species are described by densities.
Interactions between different species in a food web are then formally expressed by coupled differential equations.
This approach does not focus on individuals, but mostly this is also not needed.

Models of engineered systems often use concepts from  \emph{control theory}. 
This allows to steer system elements, e.g., transformers, from a central perspective, but requires to have precise models of such elements and their relations to others.
This is often the case because engineered systems are designed systems.

Both of these modeling approaches cannot be applied to social organizations the way we see them.
They are much more volatile, more adaptive in response to shocks and, most importantly, have no defined reference state.
Therefore, in the following we specify what concepts we will use to describe their structure and dynamics.

\paragraph{Complex systems. }

We start from the insight  that
social organizations are \emph{complex adaptive systems}.
They comprise a larger number of interacting system elements, commonly denoted as \emph{agents}.
Taking the complex system perspective implies that systemic properties, such as \emph{resilience}, need to be understood as \emph{emerging} from the interaction of agents.

Hence, we have to develop a bottom-up perspective for social resilience, starting from the \emph{micro}, or agent, level rather than from the \emph{macro}, or systemic, level.
This is in line with the methodological principles of \emph{analytical sociology}~\citep{HedstromBearman2009}, which aims at explaining  macro-social phenomena from the micro-processes that generate them \cite{flache2022computational}.

\paragraph{Agent-based and network models. }

To formalize both the dynamics of agents and their relations, we combine agent-based modeling with temporal multi-layer network models.  
The agents, as the nodes of the network, are characterized by different properties, such as status, roles, knowledge, opinions, which depend on other agents and can change over time.
Furthermore, agents are \emph{heterogeneous}.
They can be of different types and even within one type their properties are not identical.
For instance, agents' function and efficiency in solving tasks vary across agents.
We therefore have to model agents explicitly, to overcome  approaches solely based on topological features to describe the functioning of a social system.

Agents' interactions and their social relations are captured in different network layers, which evolve over time.
This requires us to also model interactions explicitly.
In particular, we have to distinguish random from meaningful interactions and to find ways to infer roles and social relations from interaction data.
This paves the ground for a statistical approach based on network ensembles, which also provides the interface for data-driven modeling, which we discuss in \cref{sec:prob-appr}.

\paragraph{Finite systems. }

Our approach explicitly addresses the finite number of agents.
Focusing on larger systems would have the advantage that we could calculate simple  statistical measures, e.g., averages, to overcome details.
For the type of social organizations discussed here details \emph{matter}   and are therefore explicitly addressed.
We need to consider individuals and discrete events instead of continuous variables characterizing a whole system, such as \emph{densities}.

Work teams, online chat groups, or school classes differ from large social systems not only in their interaction structures or perceived goals, they also differ in \emph{size}.
Emergent phenomena  of social systems, such as coherence or cooperation,  
depend on size.
Large systems necessarily behave differently from smaller ones because regime shifts or phase transitions can occur. 
Therefore our models for social resilience are not expected to describe very large social systems, e.g., political parties or urban populations.

In small systems, such as collectives, stochastic influences can have a larger  \emph{relative} impact on the dynamics. 
Further, \emph{path dependent processes} in the evolution of these systems cannot be ignored.
Local effects, such as neighborhood relations become important.
Therefore, known limit cases of formal modeling, such as the \emph{mean-field approach} in which all agents interact in a similar manner, cannot be readily applied to collectives.  
Instead, we need to build agent-based models that reflect agents' heterogeneity.

\paragraph{Self-organized systems. }

An important difference to, e.g., technical systems, is the level of \emph{adaptivity} in social organizations which cannot be simply reduced to ``dynamics''.
Instead, emerging structures in social systems feed back on the interaction of agents and cause further change, often denoted as second-order emergence \citep{schweitzer-2022-ACS}.
This is related to \emph{co-evolution} and \emph{learning}, which occurs on the individual and on the organizational level. 

The outcome of these dynamics can hardly be predicted. 
Social organizations cannot be completely controlled and agents  cannot be forced to behave in a predictable manner when facing changes.
Instead of central control distributed influences and self-organization play a major role.
In general, agents respond to changes both in intended and in unintended ways \citep{Schweitzer2019-bigger}.
This makes the response of social organizations to internal or external shocks so difficult to model.

\begin{figure}[htbp]
  \centering
 \includegraphics[width=0.41\textwidth]{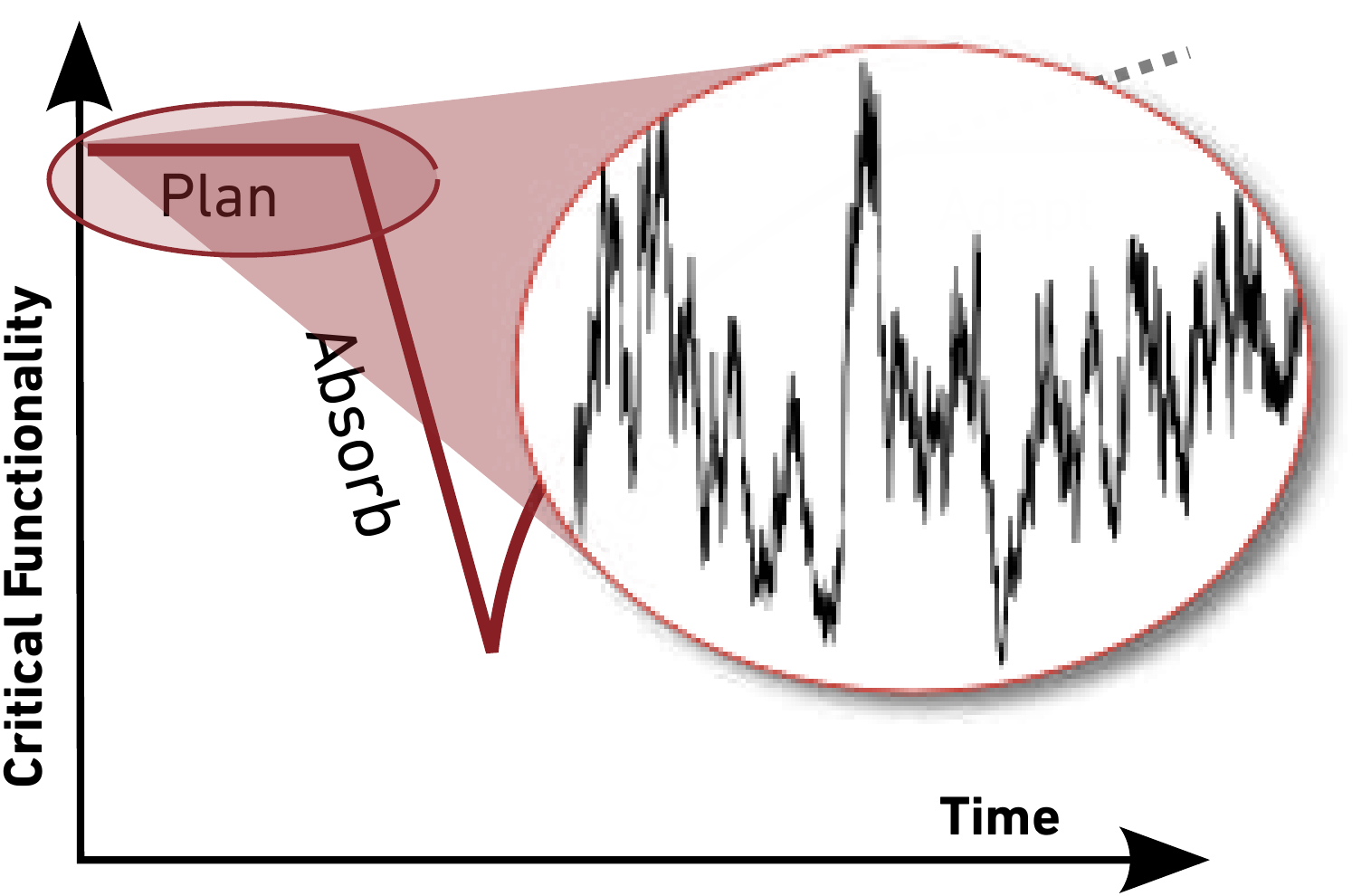}
  \caption{Problems defining a resilient state for volatile organizations.
  }
  \label{fig:resilience2}
\end{figure}

\paragraph{No separation of time scales. }

Social organizations are very volatile systems which makes it almost impossible to define a reference state, as Figure~\ref{fig:resilience2} illustrates.
More importantly, we have to account for the fact that the absorption of shocks and the subsequent recovery cannot be clearly separated as in Figure~\ref{fig:resilience1}.
Instead, changes of robustness and adaptivity follow instantaneously.
This is
a noted difference to ecological systems where the time scale of adaptivity is usually much larger and an out-of-equilibrium state can be clearly separated from the equilibrium and the relaxation time scale is well defined.

\parbox{\textwidth}{

  \begin{custombox}{Conclusions}{black!50}
    Our notion of social resilience focuses on social organizations and teams.
    To develop a formal model we adopt the viewpoint of complex adaptive systems. 
Modeling resilient societies would require a system dynamics approach, instead. 
Existing concepts of organizational resilience mostly take a management perspective.
We aim instead to model resilience bottom up, as an emerging property of organizations. 
Our framework will combine agent-based and network models. 
  \end{custombox}
  }

\section{How shall we model social organizations?}
\label{sec:how-can-we-2}

We continue to go from the general, i.e., delimitation and conceptualization, to the particular, now addressing the problem of system representation.  
Once we agreed upon the complex systems approach with its agent-based and network models, the biggest hurdle is to turn these \emph{concepts} into \emph{formal structures}.
Instead of presenting just one solution, we have to prepare for a broader perspective.
The following descriptions should therefore be seen as \emph{alternatives} for choosing formal approaches.
In Section~\ref{sec:prob-appr}, when we introduce network ensembles, we want to highlight possible \emph{options} of utilizing ensembles. 

\parbox{\textwidth}{

  \begin{custombox}{Questions}{black!50}
    \begin{itemize}
    \item Why is system representation a central problem for modeling? 
    \item What are the differences between the various network representations? 
    \item Why do we need a network ensemble? How shall it be used? 
    \item What types of dynamics do we consider for social organizations? 
    \end{itemize}
  \end{custombox}
  }

  \subsection{Network representation}
\label{sec:netw-repr}

\paragraph{Various options. }

There  is not \emph{the} one way to construct a network,  not only because of  different  topologies.
There are different \emph{types} of networks, as we outline below.
Which network type is most suited to represent the organization depends on the context and the available information, i.e., \emph{data}.

One could argue that we should indeed start our discussion with the latter, to explain what data we \emph{got}. 
This is denoted as the \emph{supply driven} approach in Section~\ref{sec:what-data}.
We instead follow the \emph{demand driven} approach, which requires us to first identify what data we will \emph{need}.
Hence, before collecting data from an empirical system, a suitable formal system representation has to be chosen.
Only then the question about the minimal set of data needed should be addressed.

\paragraph{Link properties. }

\begin{figure}[htbp]
  \centering
  \includegraphics[width=0.5\textwidth]{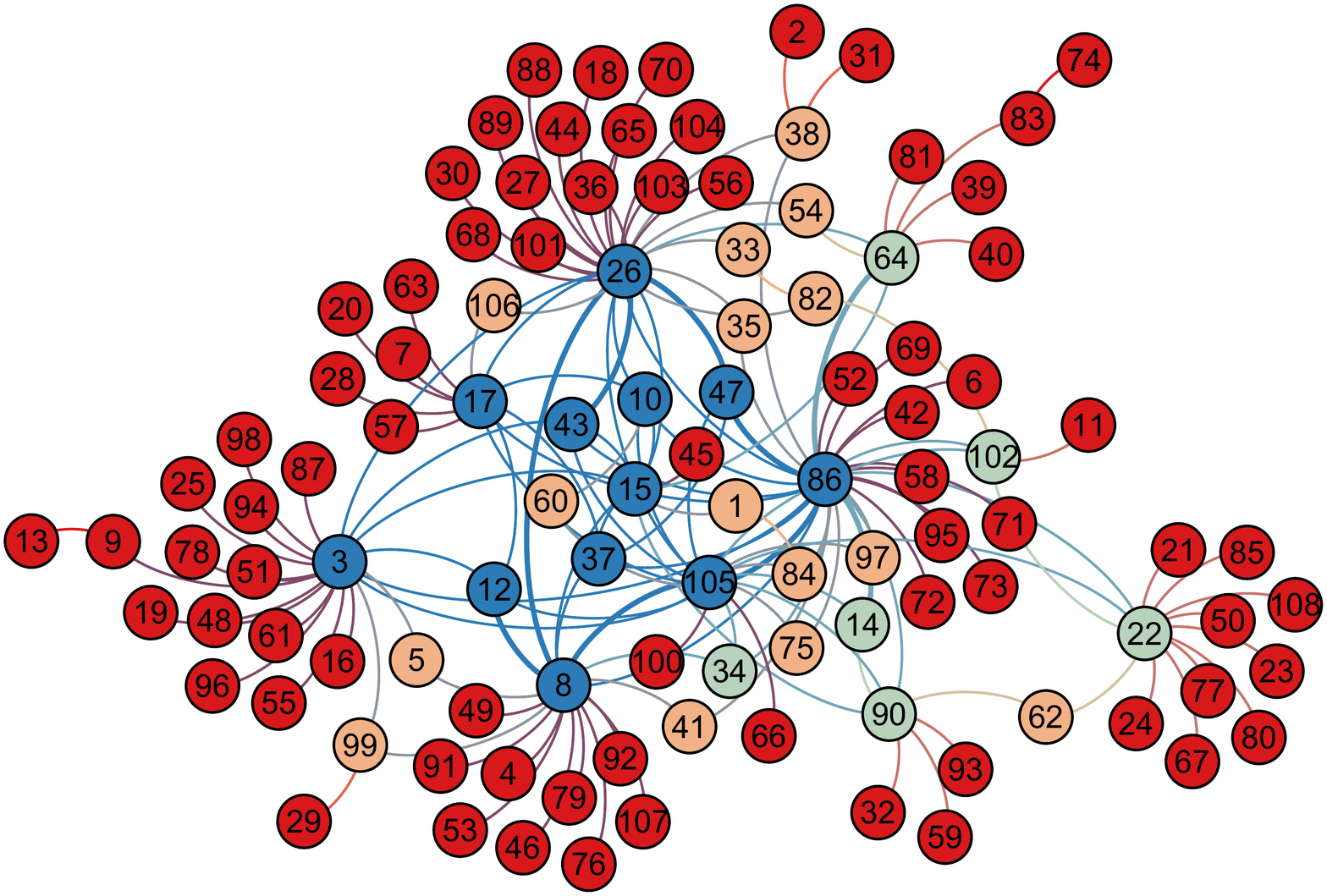}
  \caption{Collaboration network of software developers. Larger link width indicates more interactions. Node color codes individual importance, measured by coreness values as a proxy of network integration. Blue colors correspond to higher coreness.}
  \label{fig:netw}
\end{figure}

Networks are one way of representing complex systems.
The nodes of the network are the agents, and links $a_{ij}$ between nodes $i$ and $j$ represent their relations or interactions.
Figure~\ref{fig:netw} shows one example. 
The network approach focuses on the topological structure, which can be conveniently summarized in an adjacency matrix $\mathbf{A}$ with the entries $a_{ij}$.
Links between agents are usually \emph{directed}, e.g., agent $i$ assigns a task to agent $j$ and $a_{ij}\neq a_{ji}$, \emph{repeated}, e.g., there are multiple links between the same pair of agents, $a_{ij}\geq 1$, and \emph{time bound}, $a_{ij}(t)$, i.e., they have to respect causal ordering or bursts of activities.

\emph{Network inference} builds on the assumption that the topological structure encodes information about agents, i.e., individuals.
Utilizing this information could reduce the model complexity because it allows for operationalizing the structure and dynamics of social organizations. 
But studying the network topology would reveal hidden information about individuals and collectives only to some degree.
Therefore, the network approach has to be extended by explicit models of agents.

\paragraph{Links versus signed relations. }

The network reconstruction is most often based on interaction data to determine links, $a_{ij}$.
Interactions may be frequent and short-lived.
What matters for the resilience of organizations are rather the social \emph{relations} $\omega_{ij}$ between agents.
These are generally signed relations, i.e., they have positive or negative signs.
It takes time to establish social relations and they usually change on a longer time scale. 
Compared to interaction data, data about signed relations is rare.
Therefore we need methods to infer signed relations from interaction data, as described below.

\begin{figure}[htbp]
  \centering
\includegraphics[width=0.5\textwidth]{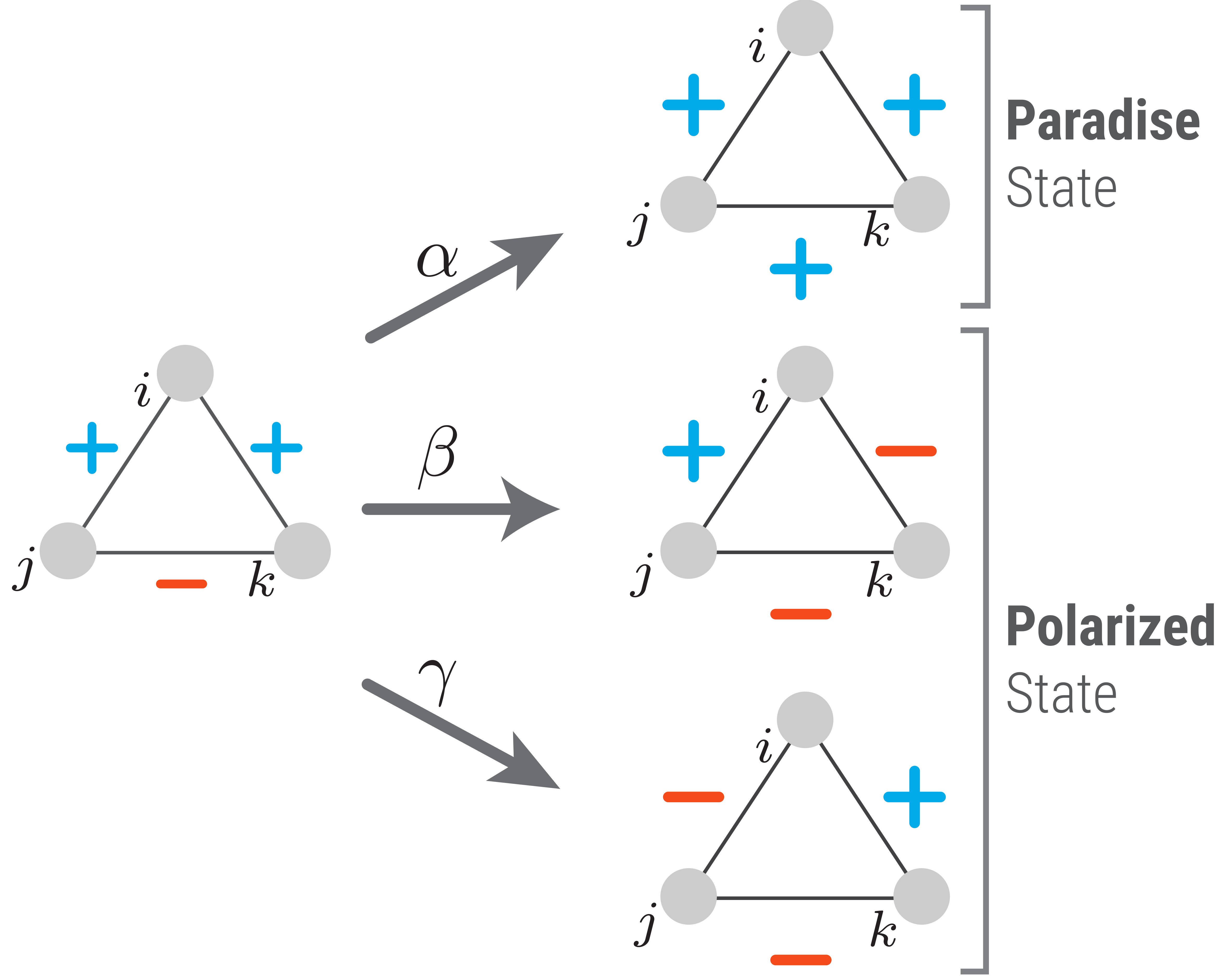}  
\caption{Unbalanced triad and the three ways to obtain a balanced triad.}
  \label{fig:sb}
\end{figure}

Signed relations crucially impact the robustness of a social network.
The theory of structural balance \citep{harary1959measurement,PhysRevLett.125.078302} considers \emph{triads} involving three agents (see Figure~\ref{fig:sb}).
A network is assumed to be robust, i.e., stable, if it contains balanced triads. 
To determine the balanced state, the classical approach only takes the \emph{signs} of the signed relations into account,
$S_{ijk}=
\sign(w_{ij})\ \sign(w_{ik})\ \sign(w_{kj})$.
If $S_{ijk}=1$, triads are balanced, if $S_{ijk}=-1$, they are unbalanced
and have the tendency to change into balanced triads as Figure~\ref{fig:sb} shows.
The line index \citep{harary1959measurement} measures the minimal amount of signs that need to be changed to turn all unbalanced into balanced  triads and can therefore serve as a measure of structural robustness, which is explained later.

\paragraph{Bipartite networks. }

\begin{figure}[htbp]
  \centering
      \includegraphics[height=.2\textheight]{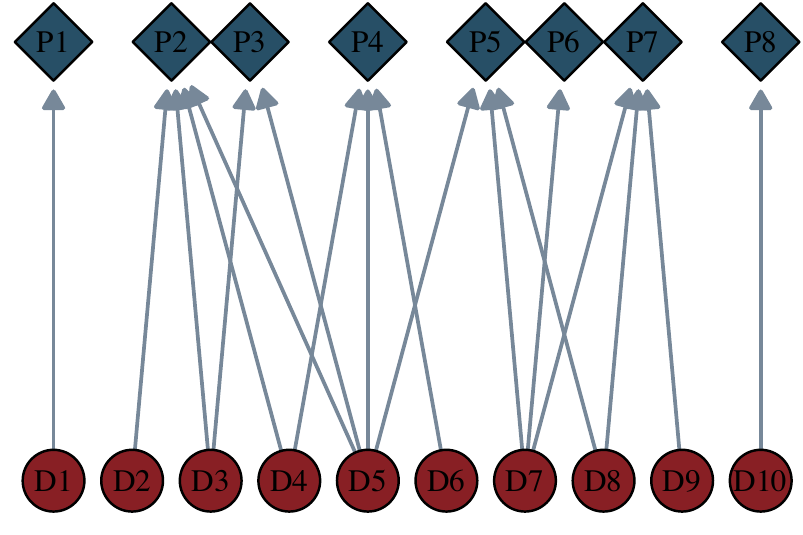}(a) \hfill
      \includegraphics[height=.2\textheight]{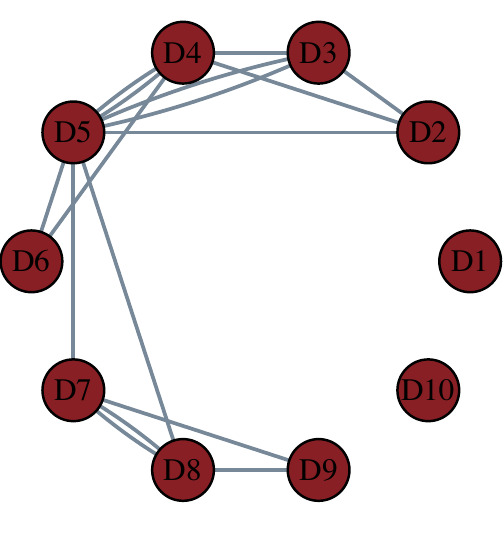}(b)\hfill
      \includegraphics[height=.2\textheight]{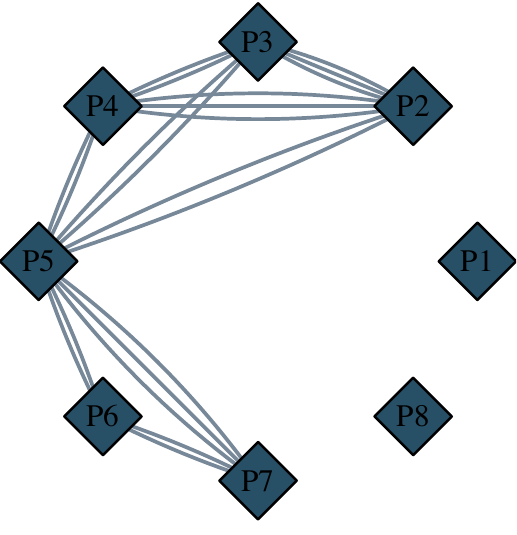}(c)
  \caption{(a) Bipartite network of developers (P) and software code (D). (b) Projected collaboration network of developers. (c) Projected network of code changed.}
  \label{fig:bipartite}
\end{figure}

One of the main challenges in modeling social organizations comes from the vast heterogeneity not only in the agents' properties, but also in their interactions. 
The notion of a link, i.e., a direct interaction, is already an abstraction.
Taking the developer example, collaboration means that two developers work on the same code.
This would be most appropriately represented as a \emph{bipartite} network between different entities, the developers and the pieces of code (see Figure~\ref{fig:bipartite} ).
The collaboration network then is a projection, where developers have a direct link if they have changed the same piece of code.
A second network results from the projection on the code; two pieces of code are connected if they were changed by the same developer. 

\paragraph{Multi-edge networks. }

If two developers collaborate on more than one piece of code, 
nodes in each network projection can be connected by more than one link, i.e., we have a \emph{multi-edge network} (see Figure~\ref{fig:multi-sign}).
From this network we are able to construct an important network ensemble, as we discuss below.

\paragraph{Knowledge graphs. }

\begin{figure}[htbp]
  \centering
  \includegraphics[height=.4\textheight]{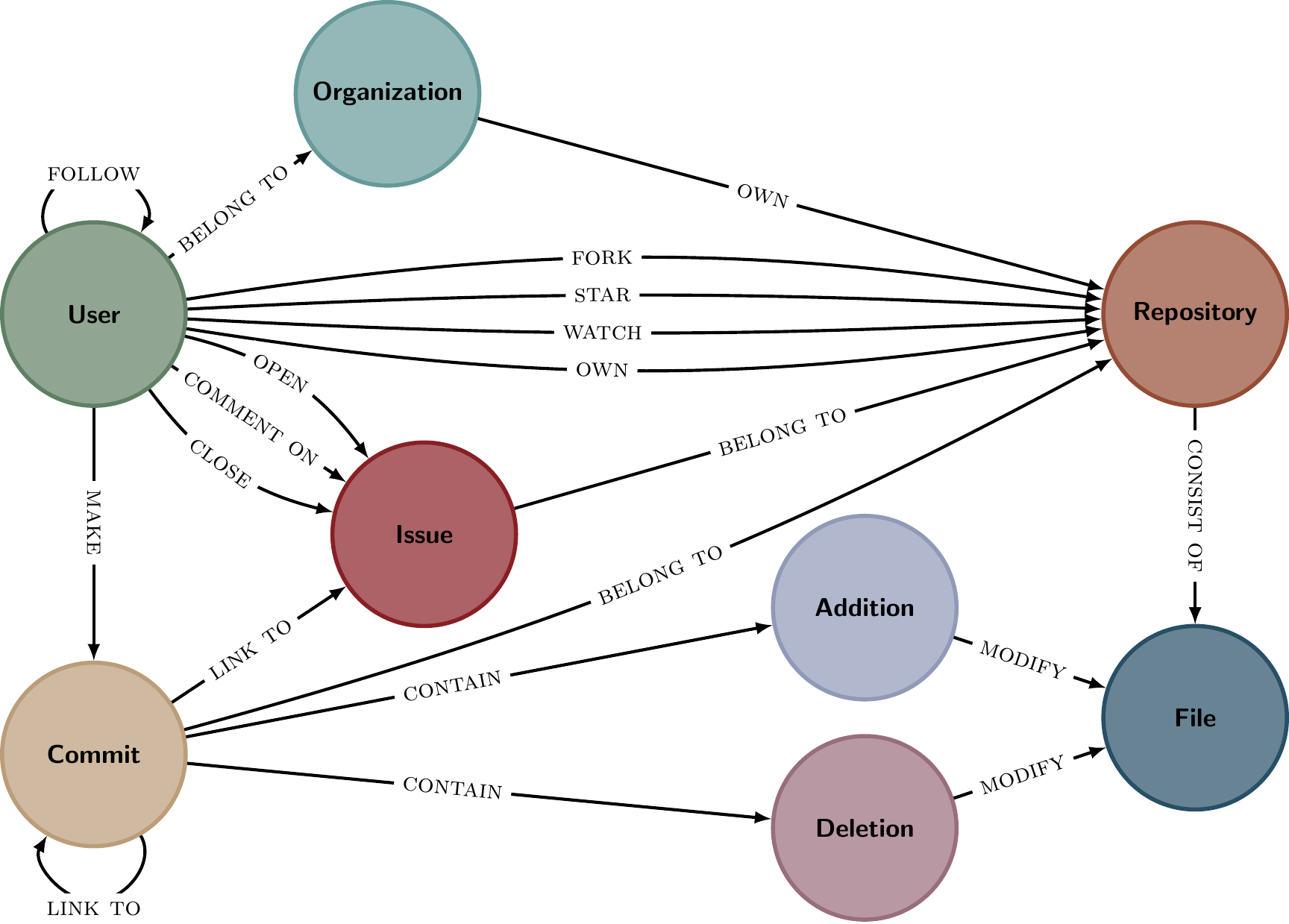}
  \caption{Relational graph of software development activities on GitHub}
  \label{fig:knowledge}
\end{figure}

Following these considerations, 
the starting point for representing organizations by means of networks 
is \emph{not} the social network between agents, which is already a reduction. 
Instead we have to start from a relational graph, also known as \emph{knowledge graph}, that visualizes the various ways of connecting individuals, as shown in Figure~\ref{fig:knowledge}.

\paragraph{Multi-layer networks. }

\begin{figure}[htbp]
  \centering
  \mbox{}\hfill
  \includegraphics[width=0.3\textwidth]{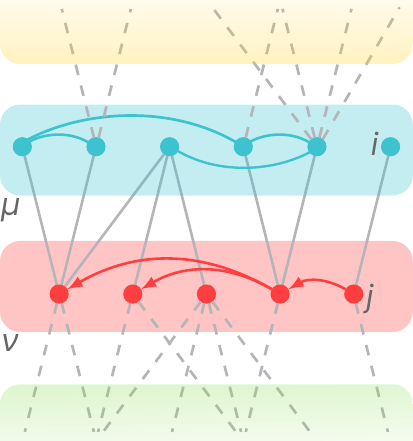}
  \hfill \mbox{}
 \caption{Multilayer network with intra-layer and inter-layer links.}
  \label{fig:multi}
\end{figure}

From a knowledge graph we construct different projections, each of which creates its own network.
These networks are combined in a \emph{multilayer network}, as shown in Figure~\ref{fig:multi}.
In each layer the nodes and their interactions are different.
If the nodes are the same in each layer, but the links represent different types of interactions (e.g., friendship, work relations) this is known as a \emph{multiplex network}.
Hence, we have now \emph{intra-layer links} within each layer and \emph{inter-layer links} between layers \citep{Garas2016}. 

The multilayer network is accessible to mathematical investigations, by representing the topological structure as \emph{tensors}.
This allows to apply methods of spectral analysis \citep{Scholtes2020,de2005spectral}.

\paragraph{Hypergraphs. }

A noted limitation of networks is the decomposition of any type of interactions into bilateral interactions between two agents.
For instance, in a group of five agents this procedure results in ten links.
To overcome this limitation in modeling group interactions, we resort to hypergraphs \citep{papanikolaou2022consensus,battiston2021physics}.
This is a special type of higher-order networks, in which higher-order nodes contain groups of simultaneously interacting agents.
Links between higher-order nodes then capture group interactions.

Similar to multi-layer networks, higher-order networks can have levels of increasing order.
The first order would be then the standard network, the second order level contains groups of two agents, the third order groups of three agents, and so forth.
This way hypergraphs allow to model interactions between groups of different sizes by means of inter-layer links.
For the formation and the dissolution of groups, however, more refined dynamic models are needed \citep{Schweitzer2022,Flamino2021}.

\subsection{Network ensembles}
\label{sec:prob-appr}

\paragraph{Probabilistic approach. }

\begin{figure}[htbp]
  \centering
  \includegraphics[width=0.15\textwidth]{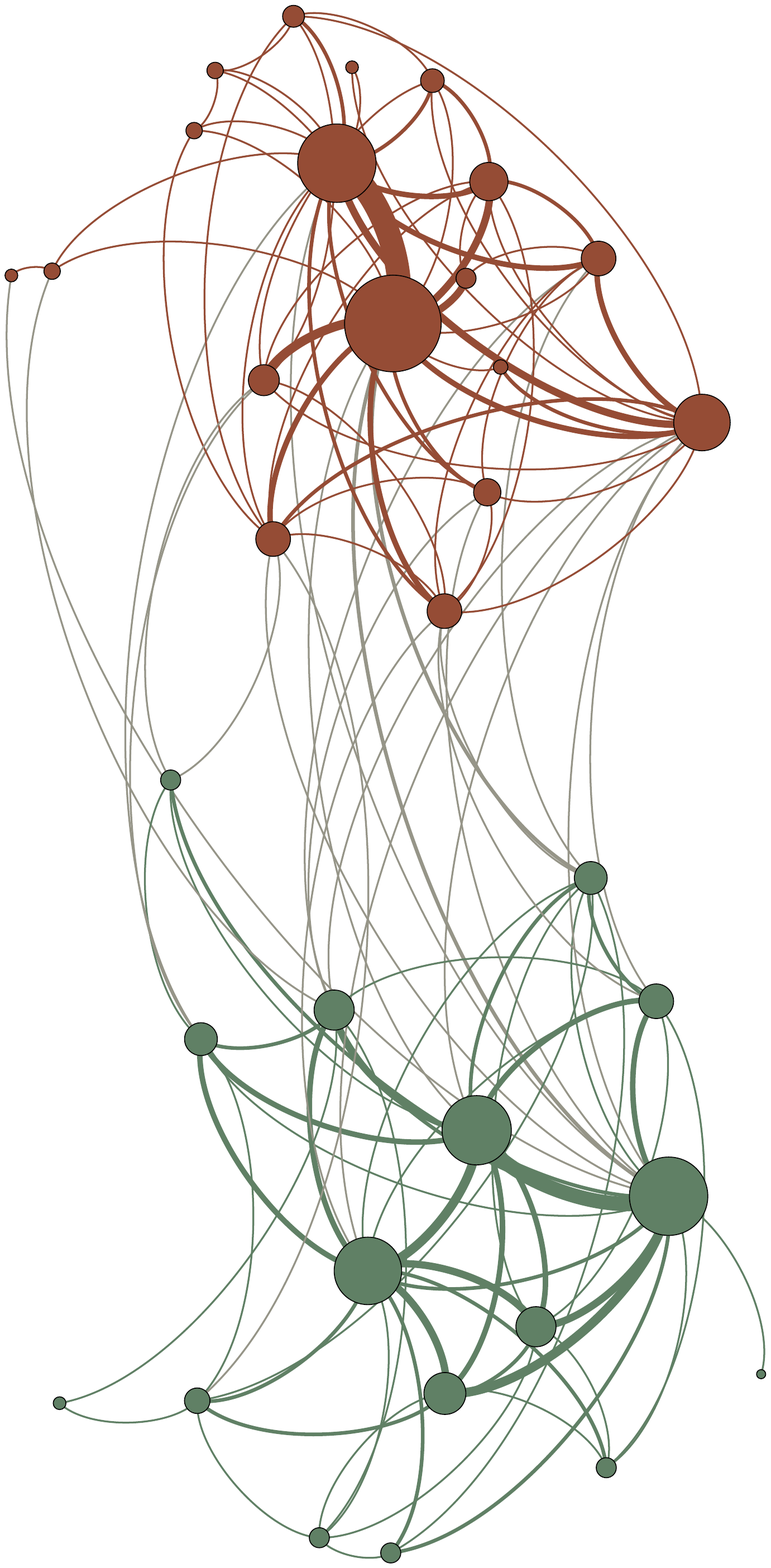}\hfill
  \includegraphics[width=0.15\textwidth]{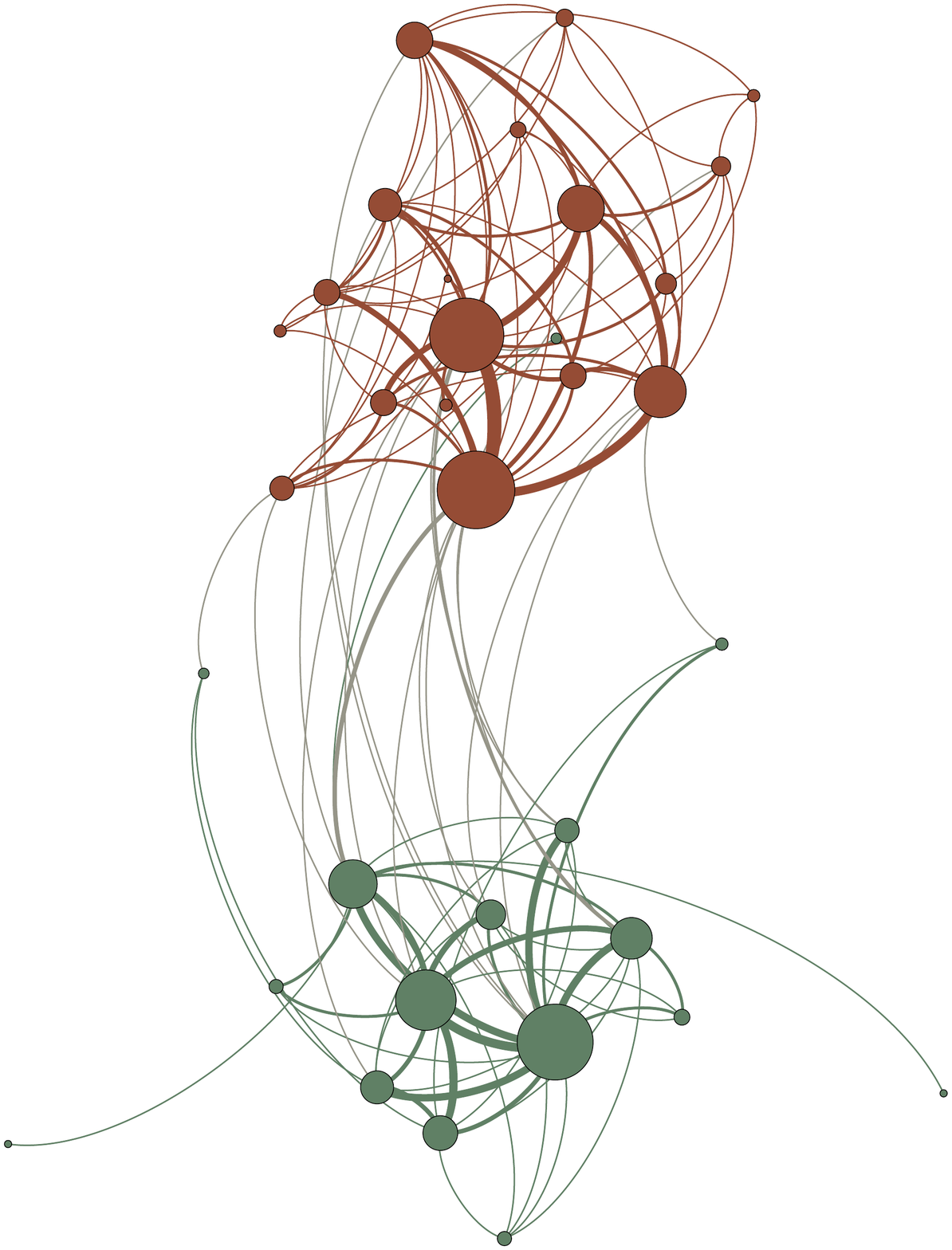}\hfill
  \includegraphics[width=0.15\textwidth]{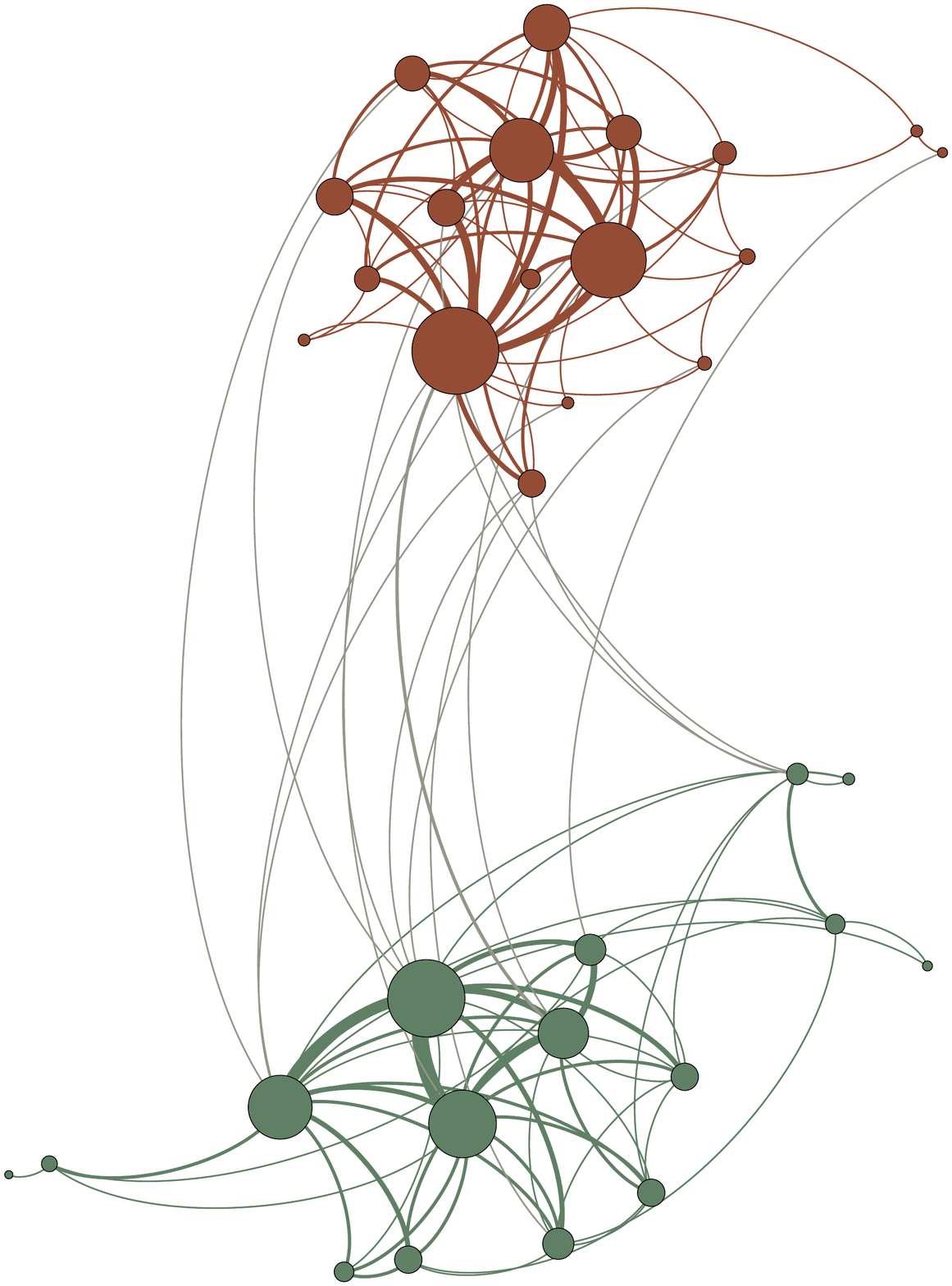}\hfill
  \includegraphics[width=0.15\textwidth]{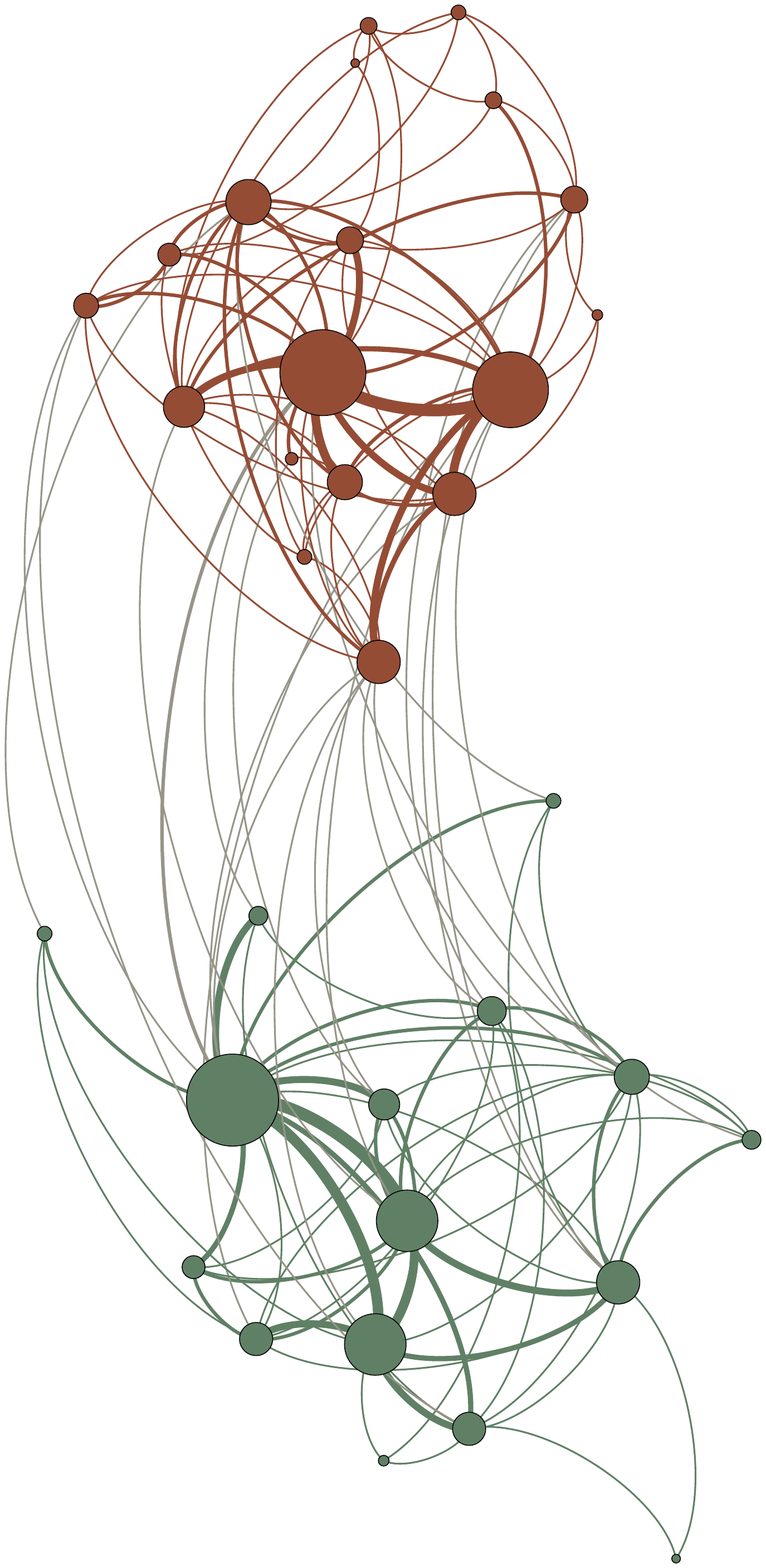}\hfill
  \includegraphics[width=0.15\textwidth]{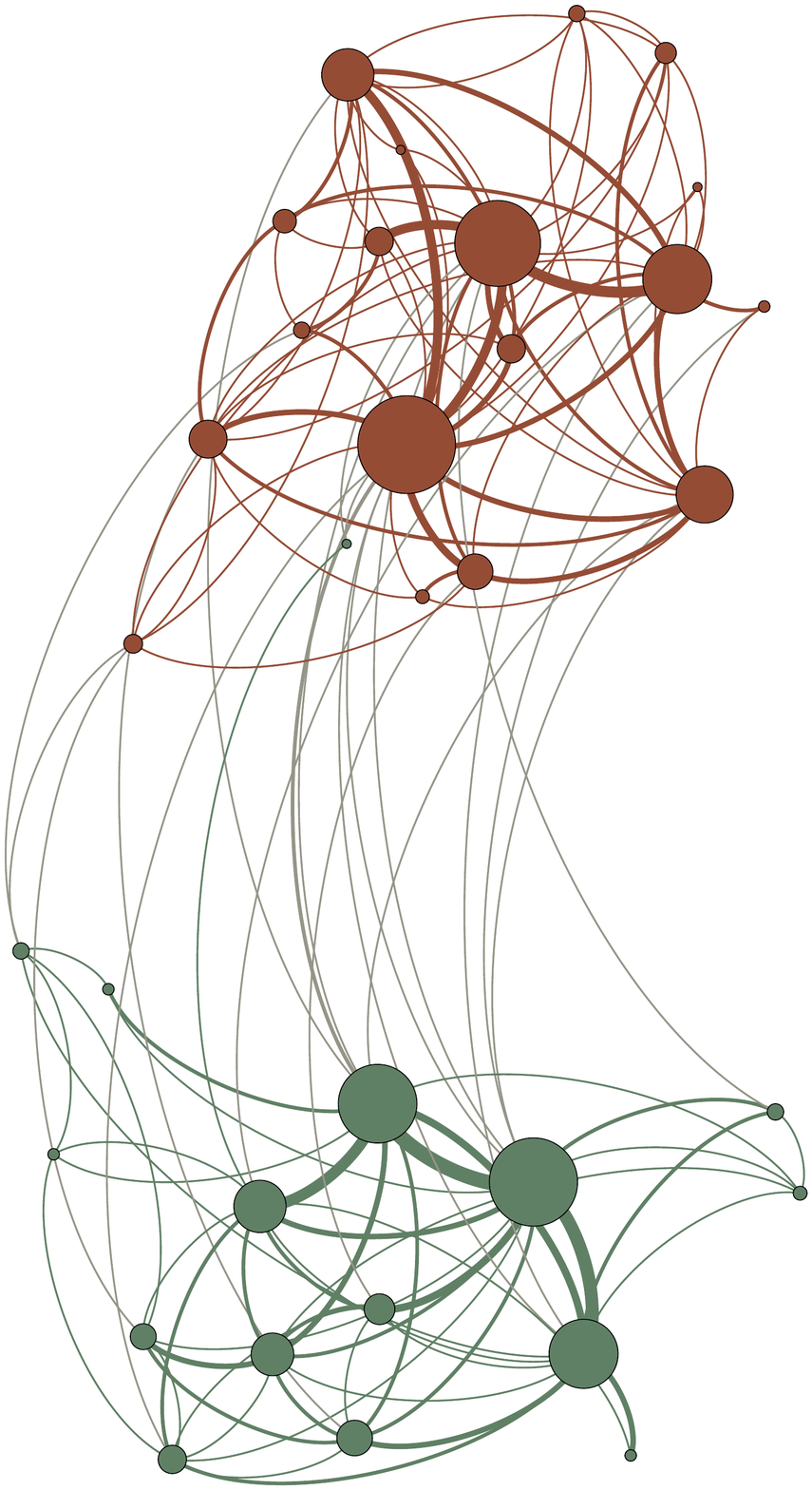}\hfill
  \includegraphics[width=0.15\textwidth]{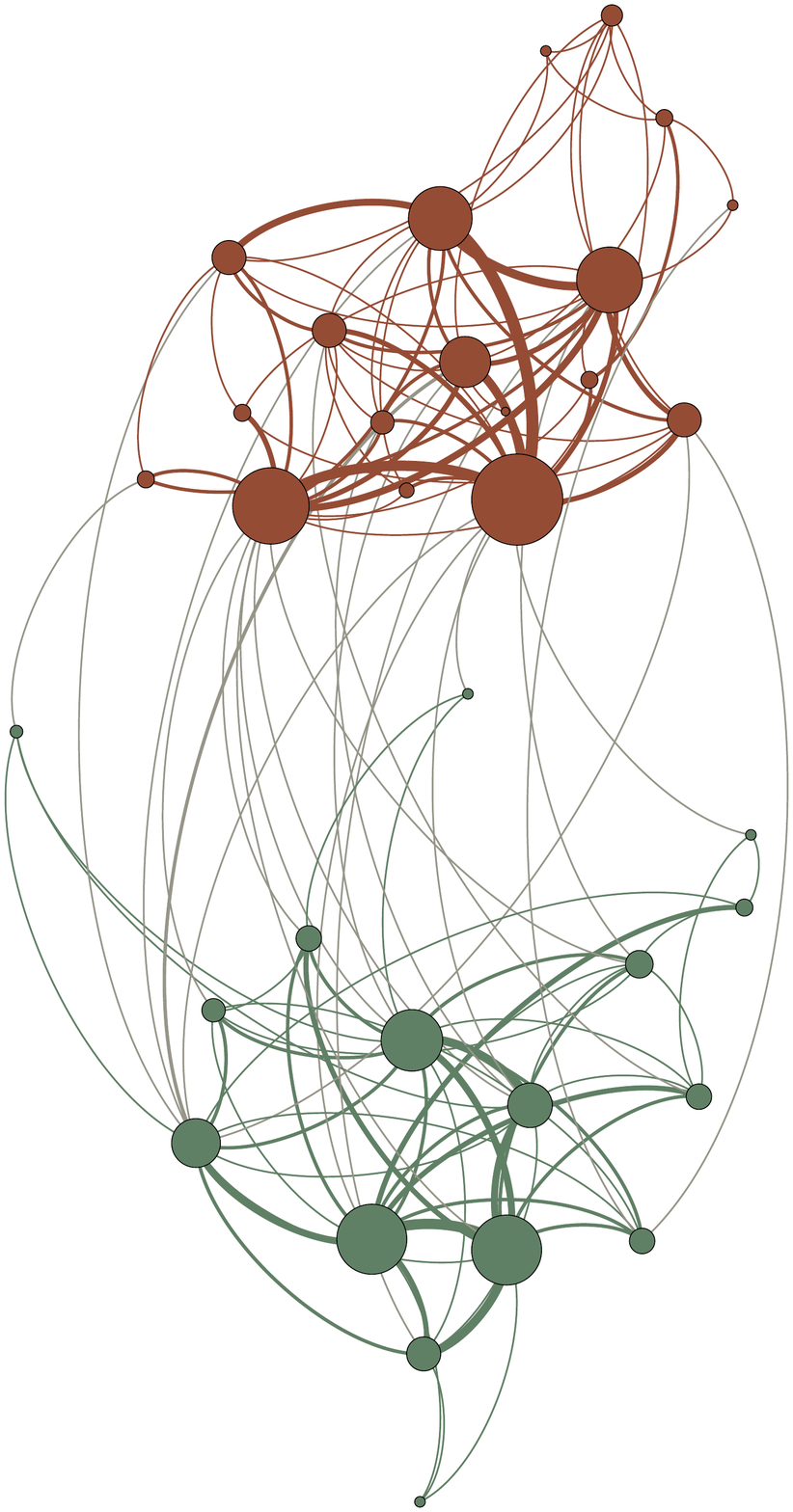}\hfill
  \caption{Six networks sampled from a network ensemble. They look similar, but differ in their details.}
  \label{fig:ensemble}
\end{figure}

A  network representation of the collective constructed from available data will be only \emph{one possible} realization and not necessarily a very \emph{typical} one.  
Ideally, we would need a \emph{probability distribution} that assigns to all possible networks a probability to occur.
Such a \emph{network ensemble} is largely determined by the constraints of agents to form links.
Figure~\ref{fig:ensemble} shows sample networks from such an ensemble.

If no link constraints are taken into account but only the total number of nodes, $n$, and links, $m$, we would arrive at a very large ensemble of random networks that all have the same $n$ and $m$ and the same probability to occur.
The network constructed from data will be part of this ensemble, but it is statistically indistinguishable from the other networks, most of which will look very different.
Hence, we need to incorporate more information to restrict the network ensemble, and to increase the probability for our reconstructed network in comparison to others.

\paragraph{Generalized Hypergeometric Ensemble of Graphs (gHypEG). }

To model multi-edge networks characterized by heterogenous constraints, we have proposed 
gHypEG~\citep{casiraghi2021urn}, a broad class of analytically tractable statistical ensembles of finite, directed, and multi-edge networks.
It introduces dyadic link propensities $\Omega_{ij}$, which capture the preference of nodes to form links.
Precisely, the ratio $\Omega_{ij}/\Omega_{ik}$ is the odds to draw a link $(i,j)$ rather than a link $(i,k)$.
The propensities reflect social mechanisms such as homophily or reciprocity \citep{Brandenberger2019}. 
Furthermore, gHypEG can incorporate formal assignments to classes or communities \citep{Casiraghi2019}.
To do so, it employs propensities $\Omega_{kl}^{B}$ for links between nodes $i,j$ that are in different ``blocks'', i.e., communities, $k,l$.

gHypEG has the benefit of being defined by closed form probability distributions.
Thanks to this, we are able to calculate the weights for all incorporated features by means of efficient \emph{numerical} Maximum Likelihood Estimation (MLE), without the need of expensive Markov Chain Monte Carlo (MCMC) simulations.

\paragraph{Network regression. }

\begin{figure}[htbp]
  \centering
  \includegraphics[width=0.45\textwidth]{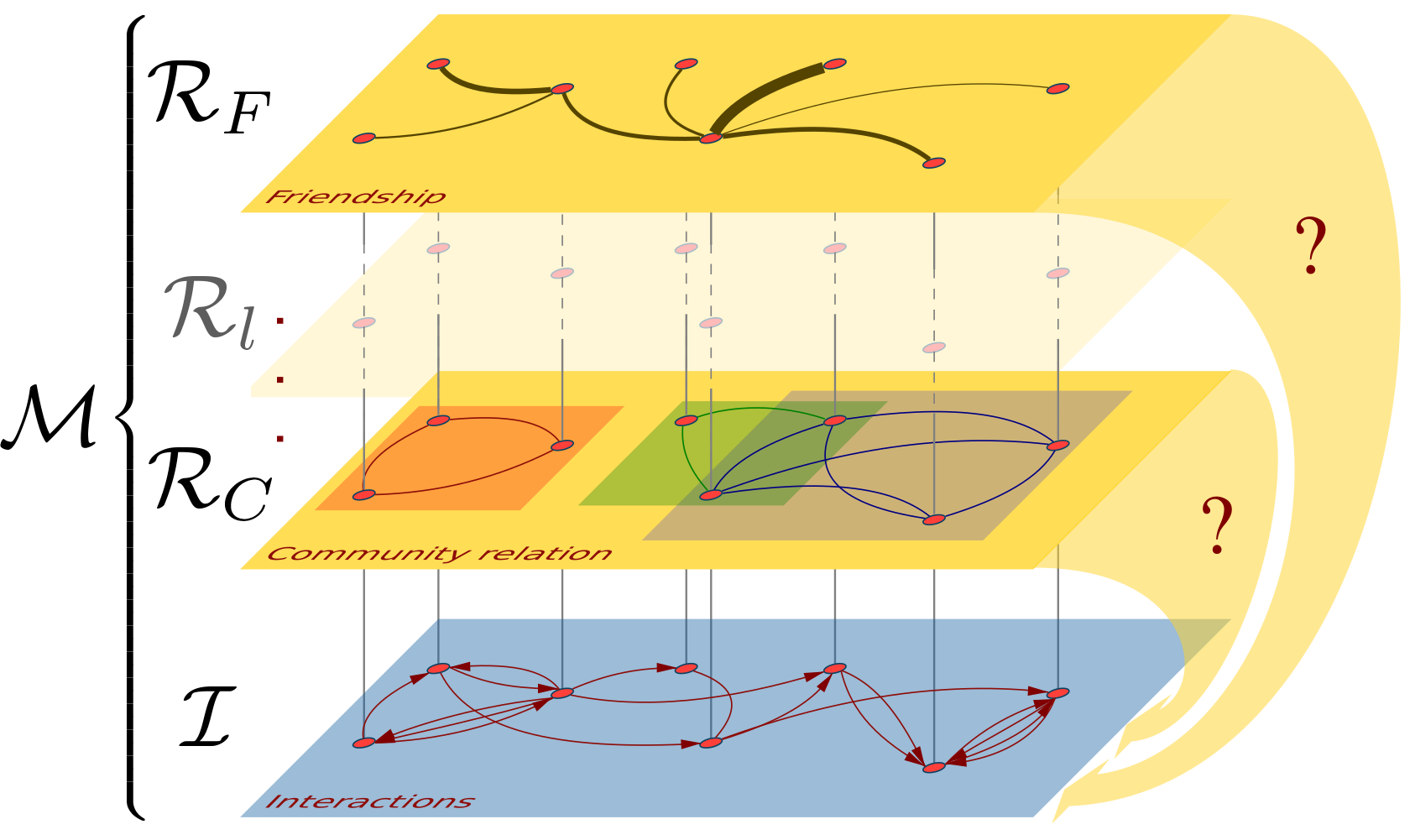} %
  \caption{Illustration of the network regression method \citep{casiraghi2017regression}.}
  \label{fig:regr}
\end{figure}

The challenge to obtain the propensities $\Omega_{ij}$ can be mastered by means of a multiplex network regression~\cite{casiraghi2017regression}.
Each network layer ${l}$ encodes different types of known relations between agents as explanatory variables (see Figure~\ref{fig:regr}). 
The influence of each layer on the interaction counts as the dependent variable is then determined by fitting the $\Omega_{ij}$ such that the observed network has the highest likelihood.
In other words, the optimal propensities are proxies for the constraints that shape the network ensemble. 
As an added benefit of the method, one can test the statistical significance of the explanatory variables for the observed interactions~\cite{casiraghi2021likelihood}.

\paragraph{Potentiality. }

With the calibrated propensities, gHypEG allows to 
calculate how many possible configurations of the observed network exist, given the constraints for links.
This issue becomes of importance if we later want to quantify \emph{adaptivity}, i.e., the ability of an organization to attain different configurations.
Then we need to know not only the number, but also the diversity of possible network configurations.

This information is aggregated in a new measure, \emph{potentiality} which is based on the normalized Shannon entropy~\cite{Zingg2019}.
Importantly, the calculation is feasible without computational problems.
The larger the potentiality, the more alternatives an organization has to respond to shocks.
We discuss below how this will impact the organization's resilience.

\paragraph{Significant relations. }

\begin{figure}[htbp]\centering
  \begin{center}
    \mbox{} \hfill
    \includegraphics[width=0.33\textwidth]{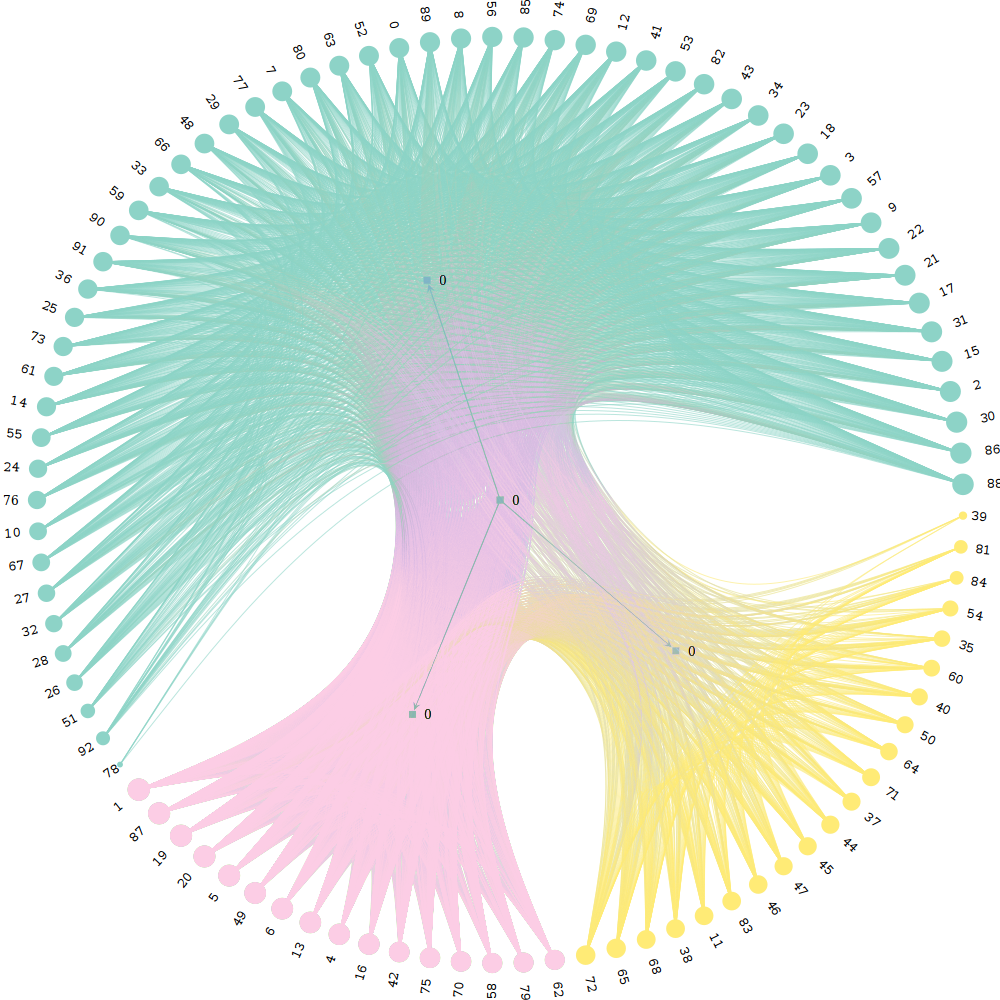}(a) \hfill
    \includegraphics[width=0.33\textwidth]{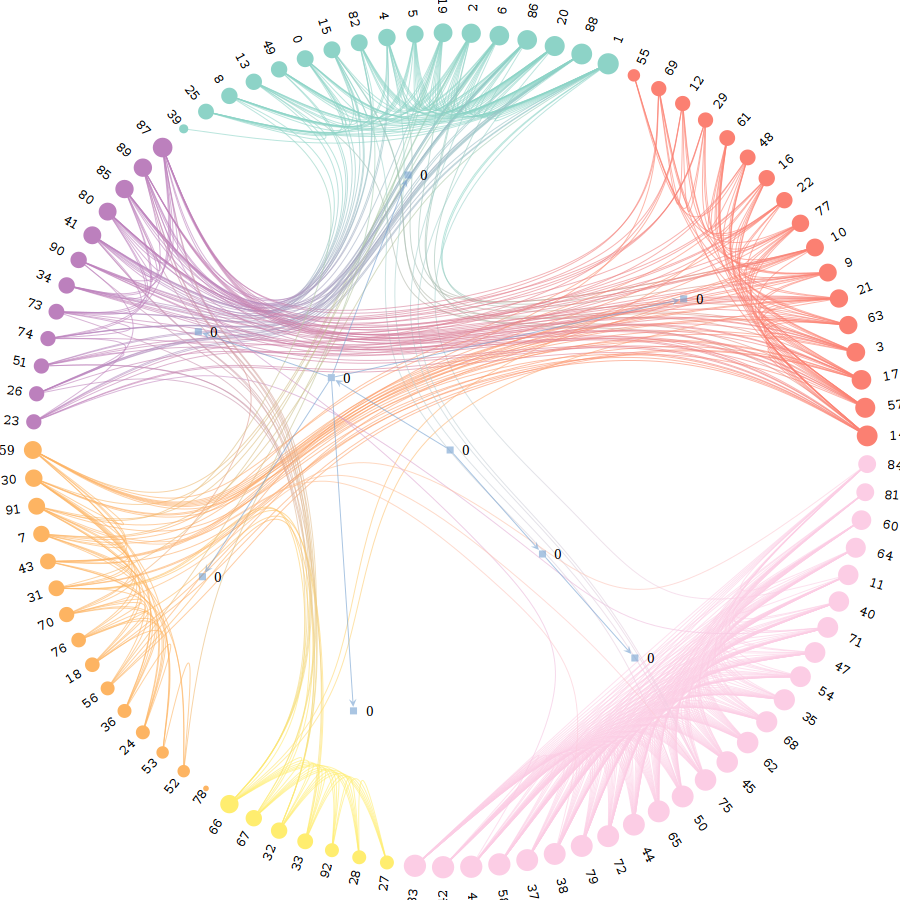}(b)
    \hfill \mbox{}
  \end{center}
  
  \caption{Community detection of a social network considering (a)  all links, (b) only significant links. \citep{Casiraghi2017}}
  \label{fig:significant}
\end{figure}

``Social'' is not  ``random'', 
therefore, social relations should significantly differ from random interactions. 
To test this, we filter the adjacency matrix with the observed number of interactions, $\hat{a_{ij}}$, using a significance threshold $\alpha$ and our probability distribution for the network ensemble.
If $\mathrm{Pr}(A_{ij}\leq \hat{a}_{ij}) > 1- \alpha$, links are significant~\cite{Casiraghi2017}.
Figure~\ref{fig:significant} demonstrates that removing insignificant, i.e., random, links from the network has a considerable impact on determining, for instance, communities.

If the observed network is \emph{not} expected from the network ensemble, we have to  apply an iterative procedure, to refine the probability distribution.
In a first step, we measure the significant deviations. 
This additional information is used in a second step to update the constraints for the network ensemble, i.e., to generate a new ensemble.
The iterative procedure reveals what information is relevant to explain the observed network.

\paragraph{Signed relations. }

Eventually the probability distribution for network ensembles allows to test whether the number of observed  interactions exceeds expectations.
This issue is important if we wish to map \emph{interactions} to social \emph{relations} which have positive or negative signs.
Empirical studies have shown that \emph{more} interactions indicate a positive social relation, e.g., a stronger friendship  \citep{Homans1950,Jones2013,UrenaCarrion2020}, whereas \emph{less} interactions indicate a negative
relation which causes e.g., avoidance behavior \citep{Labianca2006, Harrigan2017}.

\begin{figure}[htbp]
  \centering	\includegraphics[width=0.45\textwidth]{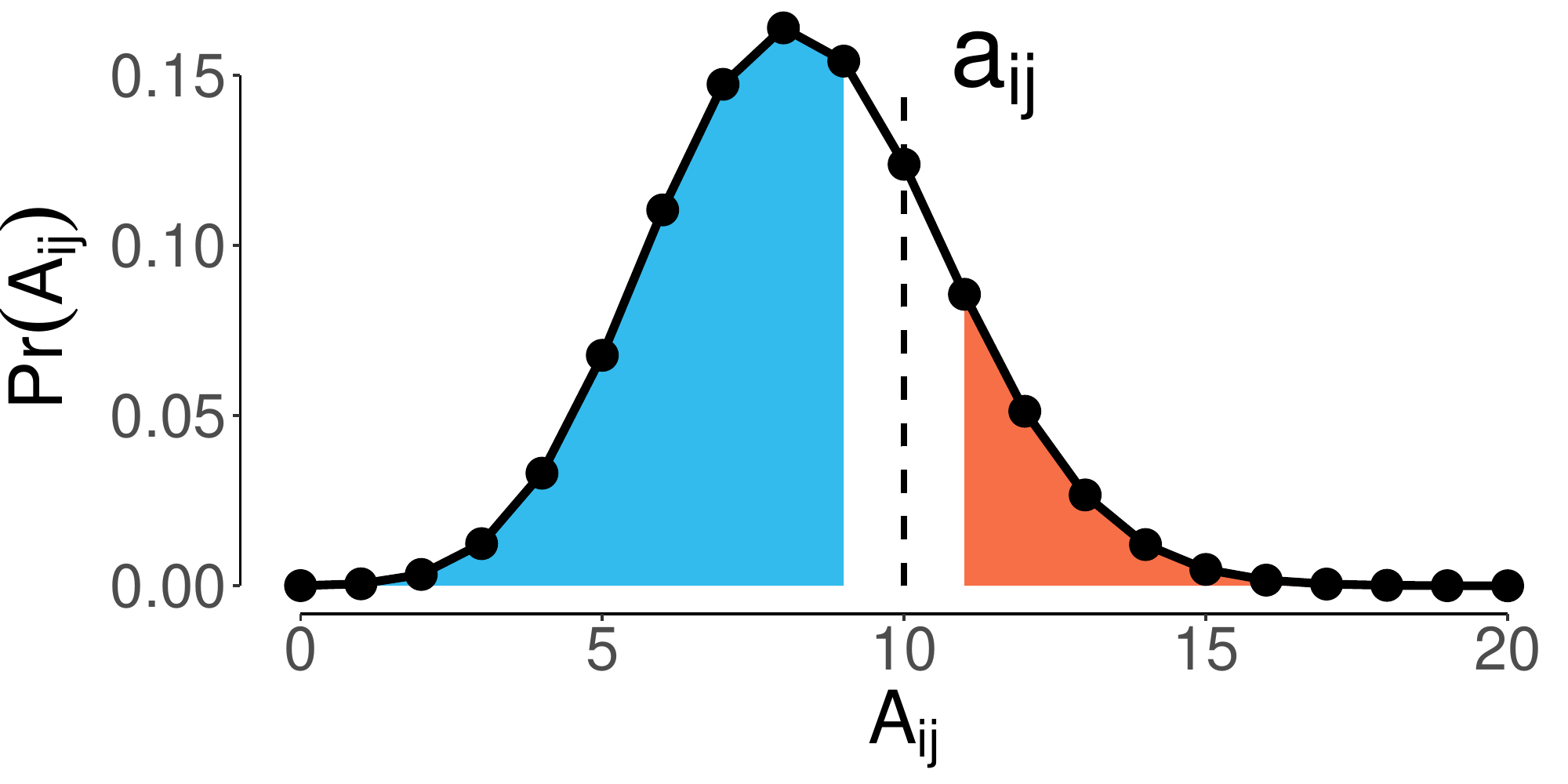}
  \caption{Determining overrepresented interactions \citep{andres2022signed}.}
  \label{fig:phi}
\end{figure}
\begin{figure}[htbp]
  \mbox{}\hfill 
   \includegraphics[width=0.3\textwidth]{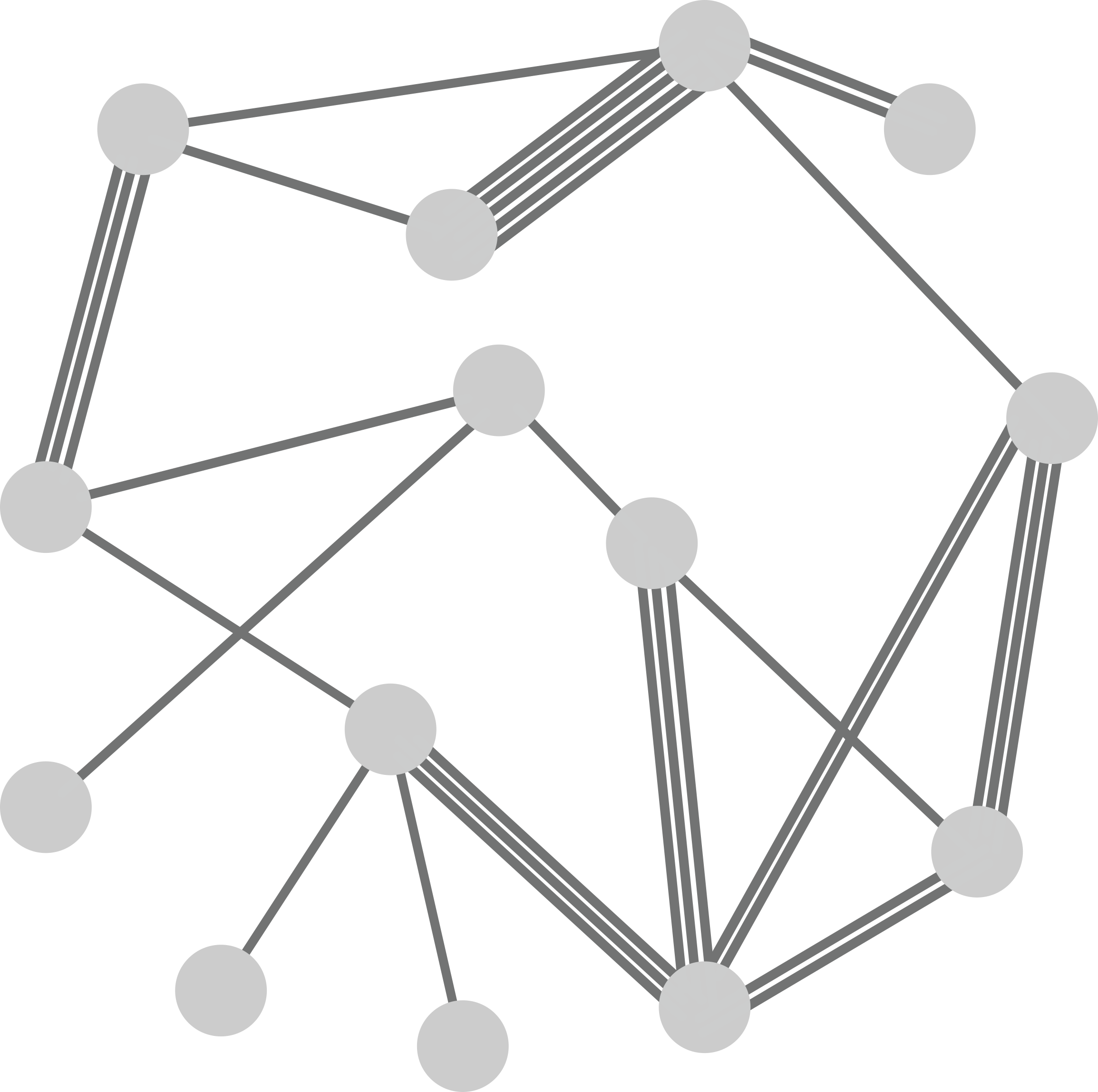}(a)\hfill
\includegraphics[width=0.3\textwidth]{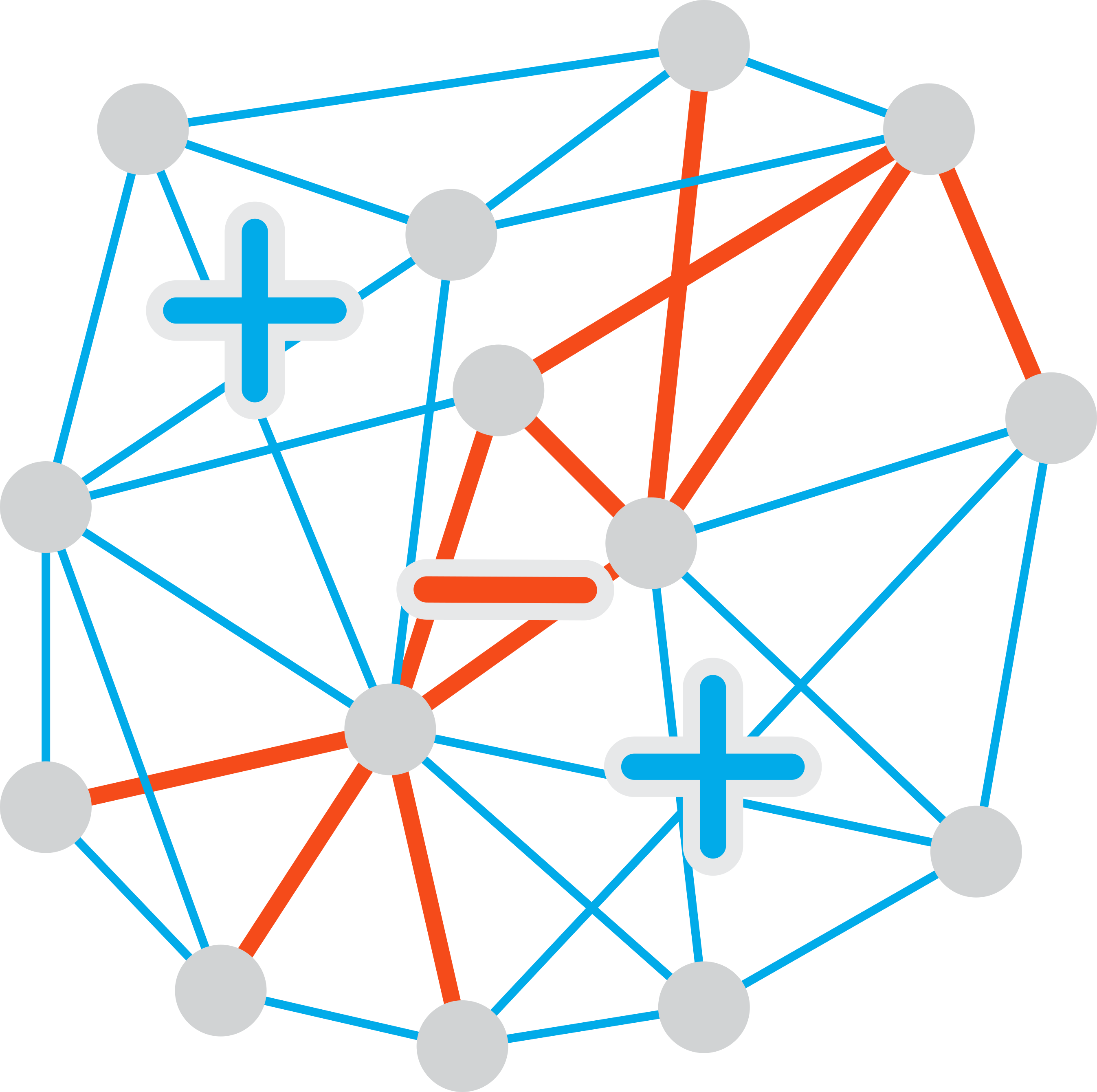}(b)\hfill\mbox{}
\caption{Multi-edge network (a) and the resulting network of signed relations (b) \citep{andres2022signed}.}
\label{fig:multi-sign}
\end{figure}

As illustrated in Figure~\ref{fig:phi}, 
we infer the weight and the sign of the social relation between two agents from \mbox{$\omega_{ij}=\mathrm{Pr}(A_{ij}< \hat{a}_{ij})- \mathrm{Pr}(A_{ij}>\hat{a}_{ij})$}~\citep{andres2022signed}.
This procedure allows us to obtain from a multi-edge network of observed interactions a network with signed relations, as shown in Figure~\ref{fig:multi-sign}.
The weighted signs, on the other hand, will enter the formalism to determine the social impact of agents in a network.

\subsection{Dynamics of social organizations}
\label{sec:temporal}

So far, we have narrowed down our investigations to social organizations of a particular type \emph{(delimitation)}, which are modeled as complex adaptive systems \emph{(conceptualization)}.
For the \emph{representation} we have chosen the network approach, which offers a great variety of network \emph{types}, but also a statistical description using network \emph{ensembles}.

This discussion has focused only on the structure, but not on the dynamics of these networks which will be done in the following. 
Further, we have not addressed yet the agents and their properties which will follow in Section~\ref{sec:quant-agent-prop}.

\paragraph{Concurrent changes. }

Models of complex socio-economic systems often use the concept of separated time scales.
The dynamics on the faster time scale is assumed to reach an equilibrium state, which allows to describe the dynamics on the slower time scale  as a sequence of different equilibrium states.
Similar approaches are used to separate different network dynamics.
For instance changes of the network topology are assumed to be slow, therefore the fast dynamics running on the network can neglect the changing topology.

As already pointed out, we cannot use such assumptions to model the dynamics of collectives.
Instead, the different processes discussed below should be seen as concurrent.
This leads to a number of issues, such as overlapping or sliding time windows, choice of the appropriate time scale for aggregation, etc., which are not discussed here, but should be kept in mind.

\paragraph{Entry and exit dynamics. }

\begin{figure}[htbp]
  \centering
  \includegraphics[width=0.5\textwidth]{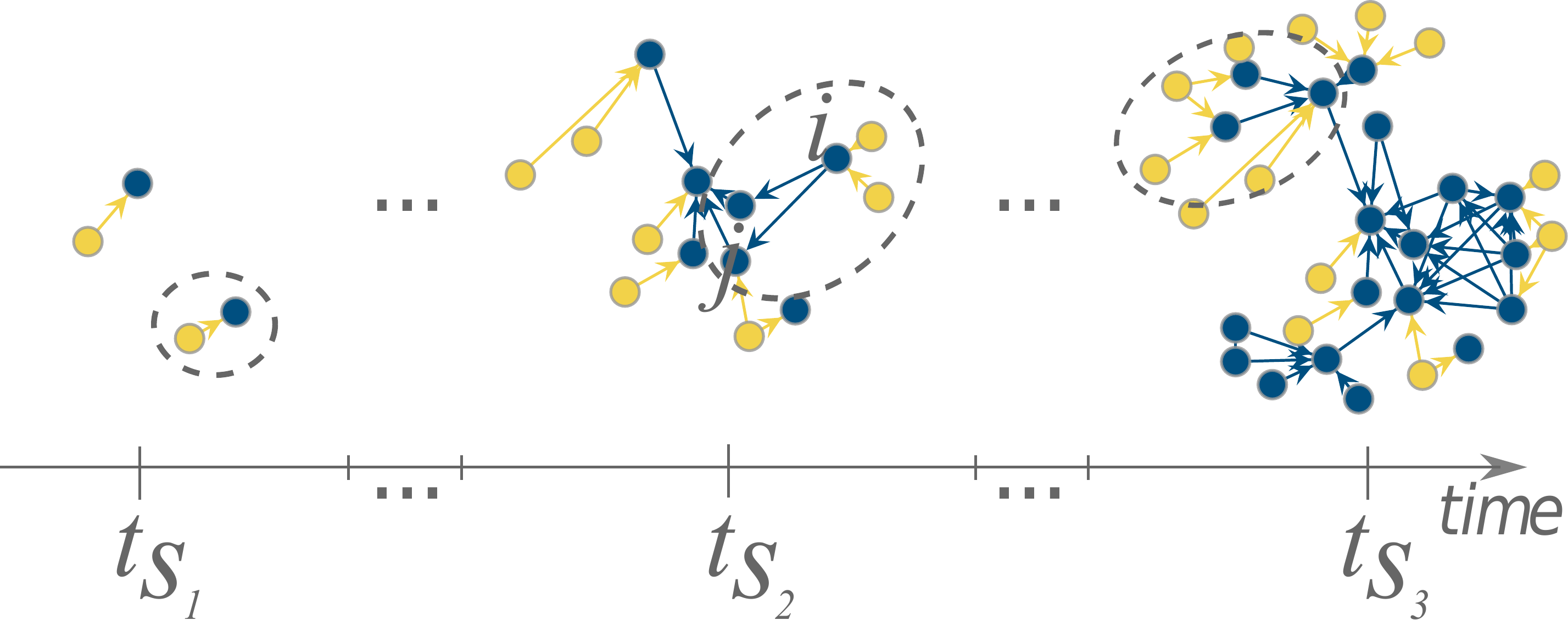}
  \caption{Coupled growth dynamics in a two-layer network. Intra-layer links are between nodes of the same color, inter-layer links between nodes of different color \citep{nanumyan2020multilayer}. }
  \label{fig:temporal}
\end{figure}

The most visible changes regard the network topology.
For social organizations we have to consider an \emph{entry and exit dynamics} of nodes, i.e., newcomers connect to the network \cite{Schweitzer2022,Schweitzer2020-perc}, whereas incumbents may leave.
This implies also the addition and deletion of links, as shown in Figure~\ref{fig:temporal} for the case of a multiplex network. 
Social organizations often exhibit a \emph{life cycle}, i.e., a predominant growth of both nodes and links in early stages is followed by a saturation and a decline caused by many nodes leaving \citep{Garcia2013,Schweitzer2022IPP,schweitzer2021fragile}.

These processes do not occur at random. 
Further, they impact  
the collective as a whole as well as individual agents.
Newcomers may not be able to connect to core nodes initially and thus connect to the periphery.
Their integration into the collective may improve over time, as can be measured by their coreness \citep{Seidman1983,Garas-Havlin-2012,Verginer-ACS}.
If core nodes leave, this may trigger cascades of other nodes leaving as 
empirical and simulation studies have demonstrated \citep{Garcia2013,Casiraghi2021}.

\paragraph{Restructuring. }

\begin{figure}[htbp]
  \centering
  \includegraphics[width=0.28\textwidth,trim=34mm 34mm 34mm 34mm, clip]{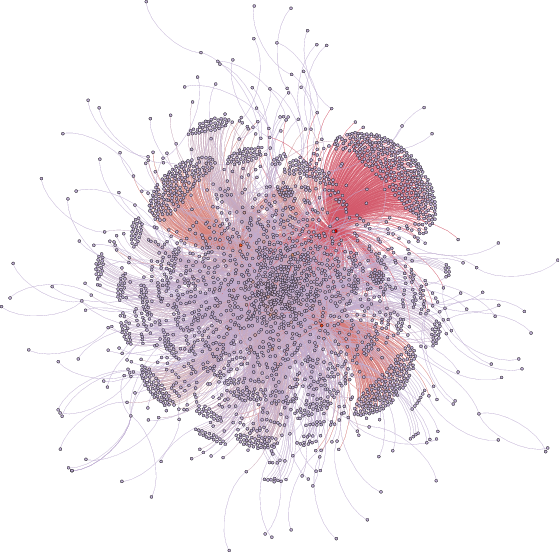}(a)\hfill
  \includegraphics[width=0.28\textwidth,trim=34mm 34mm 34mm 34mm, clip]{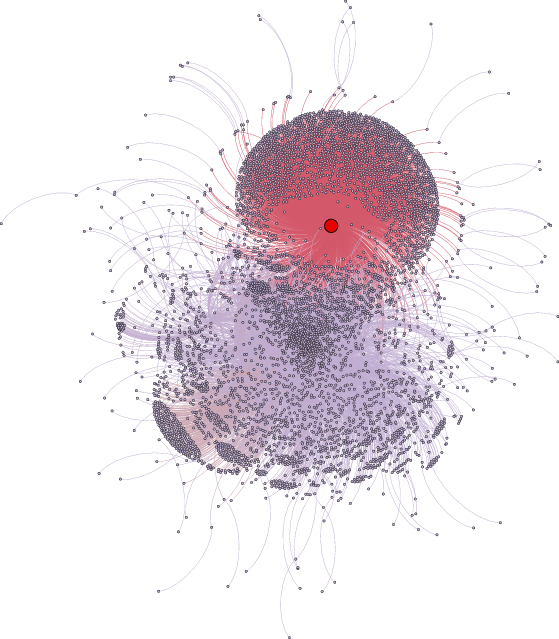}(b)\hfill
  \includegraphics[width=0.28\textwidth, trim=34mm 34mm 34mm 34mm, clip]{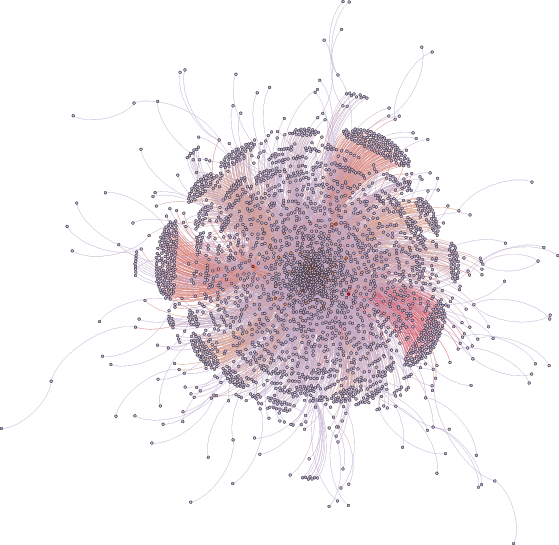}(c)
    \caption{Topological change of a collaboration network of developers. Aggregated interactions (a) before October 2004, (b) between October 2004 and March 2008, (c) after March 2008. \citep{Zanetti2013}.}
  \label{fig:change}
\end{figure}

Next to the addition and deletion of nodes and links, the rewiring of links between nodes plays a major role.
Their impact can be measured by tracking changes in the global and local topological measures discussed below.
Structural changes often reflect changes in the organization, e.g., in responsibilities, hierarchical positions and roles of agents.  
Figure~\ref{fig:change} illustrates such restructuring processes for a developer collective in which a central developer has assumed the main responsibilities for task assignments. 
Already the visual inspection makes clear that this has lead to considerable problems in the robustness of the collective, which eventually lead to a collapse and the establishment of more resilient structures.

\paragraph{Temporal networks. }

Whereas the dynamics \emph{of} networks addressed above changes the topology, the dynamics \emph{on} networks captures \emph{interactions} between agents.
These can be exchange processes, e.g., load redistribution in case of
an agent's failure, but also communication of information.
These processes are strongly \emph{path dependent}, i.e.,  the \emph{sequence} of interaction matters and has to respect causal relations. 
Models of \emph{causal paths} \citep{lambiotte2019networks,Scholtes2014} provide a formal approach.
They build on higher-order networks, where each order captures a causal path of a given length.

\begin{figure}[htbp]
  \centering
  \mbox{}\hfill
  \includegraphics[width=0.35\textwidth]{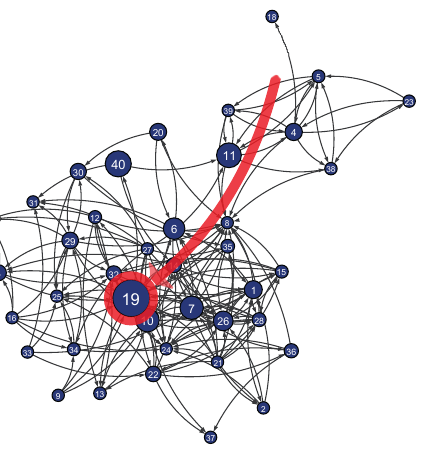}(a)\hfill
    \includegraphics[width=0.35\textwidth]{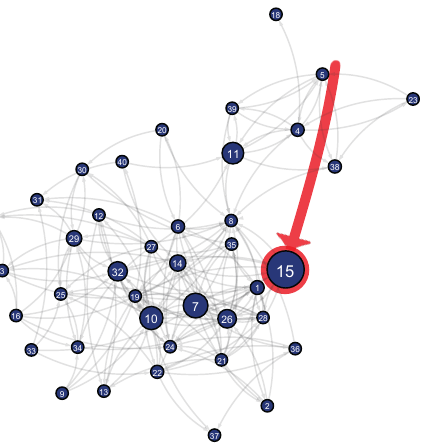}(b)\hfill\mbox{}
  \caption{Identification of important individuals (a) on the aggregated and (b) on the temporal network.}
  \label{fig:centr2}
\end{figure}

In addition to time directedness, 
\emph{temporal networks} also reflect the burstiness of activities \citep{Scholtes2014}, i.e., the fact that not every link in a network is active at all times.
The temporal component significantly impacts the centrality measures of individual agents \citep{Scholtes2016a}, as Figure~\ref{fig:centr2} shows.
\emph{Betweenness preference} \citep{Pfitzner2013} was introduced as an agent-centered measure to quantify its importance in transferring information.

\paragraph{External and internal shocks. }

The different dynamics  described above are continuously perturbed by internal and external shocks of various size and origin.
Internal shocks, for instance, may cause agents to leave, this way triggering cascades of drop-outs and restructuring.
External shocks, e.g., directives during the pandemics,  may change working conditions and collaboration relationships.
Because of the volatile dynamics, we cannot clearly separate shocks from the ``normal'' dynamics, which both occur on the same time scale.

We note that from our modeling perspective we model shocks, but not the \emph{origin} of shocks, e.g., the government that changes the legal regulations. 
But we need to have models for the \emph{impact} of these shocks and for the collective's \emph{response} to different kind of shocks.
In other words, we need to estimate the \emph{robustness}, or the absorptive capacity,  of the collective facing a particular shock, and to estimate the \emph{adaptivity} of the collective to overcome this shock.
Only then we can calculate the social resilience of the collective, as outlined below.

\parbox{\textwidth}{

  \begin{custombox}{Conclusions}{black!50}
    Different types of networks capture different aspects of relations between individuals.
    It depends on the research question and the available data which of these network representations shall be used.
    Building up a network \emph{ensemble} allows us to go beyond the observed network, to include constraints for the social organization.
    In particular, we can distinguish significant from random interactions and infer signed relations from interaction data.
    These link characteristics are important to build an agent-based model, in the next step. 
    Our model has to consider various concurrent dynamics, including growth, entry and exit of individuals, internal restructuring and external shocks.
  \end{custombox}
  
}

\section{What should we do to calculate resilience?}
\label{sec:workbench}

\begin{figure}[htbp]
  \centering
\includegraphics[width=0.99\textwidth]{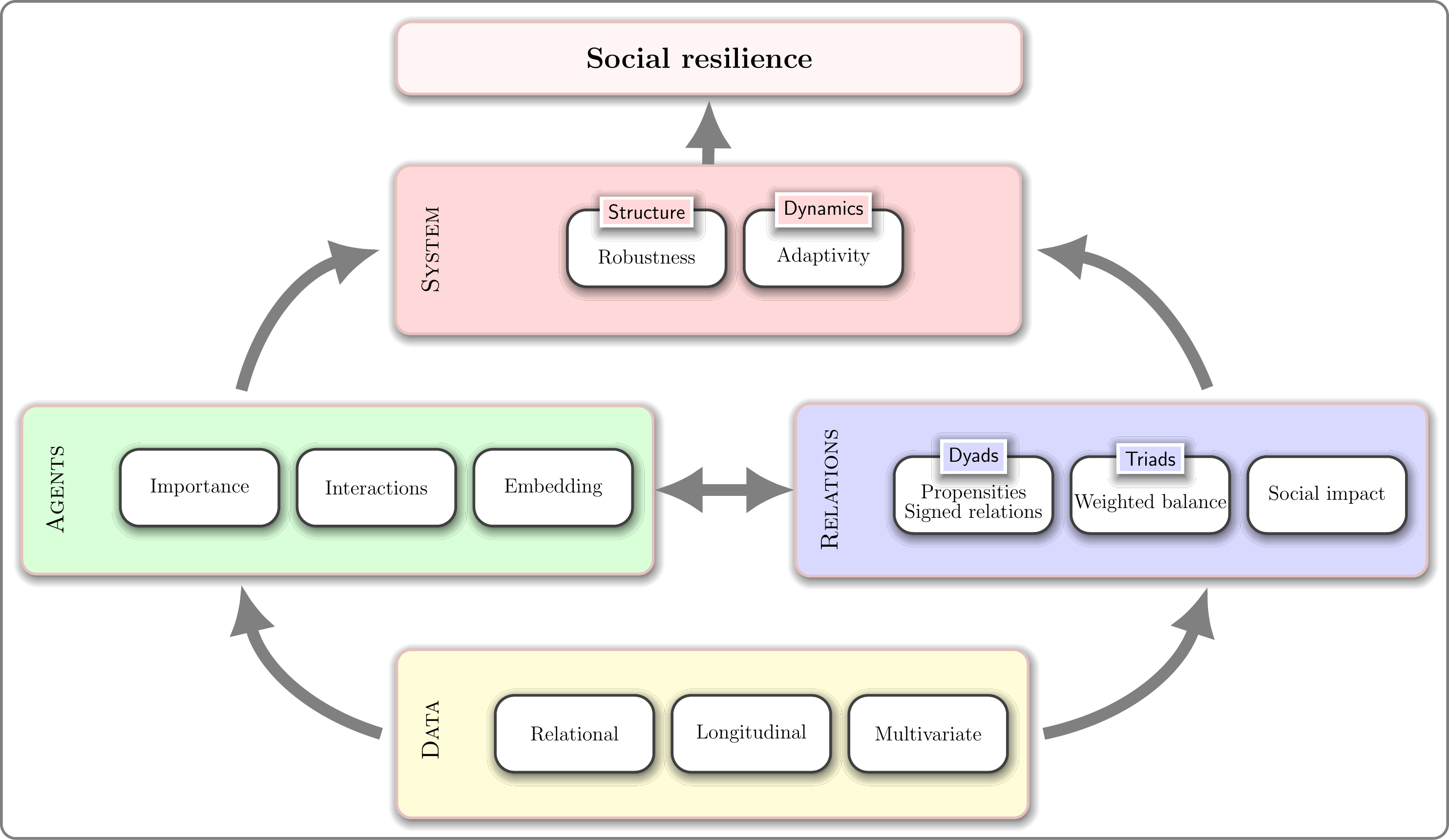}
  \caption{Operationalization to calculate social resilience.}
  \label{fig:outline}
\end{figure}

After completing the steps \emph{delimitation}, \emph{conceptualization} and \emph{representation}, we eventually have to master the last step, \emph{operationalization}, where we merge the network approach with agent-based modeling.
The overview is presented in Figure~\ref{fig:outline}.

\parbox{\textwidth}{

  \begin{custombox}{Questions}{black!50}
    \begin{itemize}
    \item How can we turn concepts into measures for robustness and adaptivity? 
    \item How can we characterize agents, using topological information? 
    \item Why do we consider the social impact of agents? How can we quantify it? 
    \item How is resilience composed of robustness and adaptivity? 
    \end{itemize}
    
  \end{custombox}
  
}

\subsection{Quantifying agent properties} 
\label{sec:quant-agent-prop}

The major goal of our framework is a micro-perspective on \emph{resilience}.
This is an emerging systemic property, 
that means it can neither be reduced to, nor explained by, the dynamics of the agents.
As we demonstrate below, we need to consider the network structure to calculate the robustness and the adaptivity of the social organization.
But the agents' importance, their social impact, will provide  the right weights in calculating these two measures.

\paragraph{Quantifying agents' importance. }

A bottom-up approach to quantify resilience has to start from the agents.
In each organization, agents have a different \emph{importance}, $r_{i}$, that reflects their hierarchical status, reputation, embedding in the organization, knowledge, etc.
To obtain values for $r_{i}$ is a challenge in itself and depends on the available data.
Because there is no general solution, we resort to some guiding examples.

In the simplest case, importance is defined in the hierarchical structure of a team \citep{Schweitzer2022IChing}.
There are also ways to determine hierarchies based on interaction patterns.
In the absence of such information, we may utilize topological information from the reconstructed network (see also Figure~\ref{fig:change}). 
In a directed social network we can use the \emph{eigenvector} centrality of agents as a measure of their reputation \citep{Bonacich1987,schweitzer2014}.  %
For undirected networks, 
 \emph{coreness} \citep{Seidman1983} or weighted $k$-core centralities \citep{Garas-Havlin-2012} can quantify an agent's embeddedness in a network \cite{Verginer-ACS}, assuming that more important agents are closer to the core (see also Figure~\ref{fig:netw}). 
 These measures also estimate the robustness of an agent's network position against failure cascades.

For temporal networks different centrality measures can be used \cite{Scholtes2016a} (see also Figure~\ref{fig:centr2}).
\emph{Betweenness preference} \citep{Pfitzner2013} quantifies an agent's importance in communication processes.
Functional roles can be only partially inferred from communication patterns or specific topological embeddings of agents.   
Existing algorithms for role detection \citep{henderson-gallagher-2012-rolx} do not detect organizational roles, but classify network positions.

\paragraph{Social impact. }

What matters in an organization is not just the importance, $r_{i}$, but also the support or opposition an agent receives from others. 
Their influences are combined in an individual social impact, $I_{i}$.
The total impact of an agent is then the sum of its own importance and the social impact exerted by others, $q_{i}=r_{i}+I_{i}$.

Here, we define the social impact as $I_{i}=\sum_{j}w_{ij}r_{j}=I_{i}^{p}-I_{i}^{n}$.
The $w_{ij}$ denote the weighted and signed relations between agents, which can be positive, negative or zero.
$I_{i}^{p}$ is the sum of all positive contributions, while $I_{i}^{n}$ sums up the negative contributions \citep{latane1981psychology,nowak1990private,holyst2000phase}.
$I_{i}$ can become negative, reflecting the fact that an agent may not have a high esteem in an organization because it receives little support, but strong opposition from others.

\paragraph{Infer signed relations. }

To calculate the social impact, $I_{i}$, we also have to determine 
the signed relations $w_{ij}$.
For this, we apply the method described above in Section~\ref{sec:prob-appr}.
It returns for every pair of agents a weight and a sign to characterize their relationship. 

To conclude, our measure of the total importance, $q_{i}=r_{i}+I_{i}$,  combines different, but rather complete information about each agent, namely information about its topological embedding and about its activities because its repeated interactions with other agents determine its relations.  
$q_{i}$ aggregates in one value 
the positive, negative or neutral influences from all counterparties, weighted by their individual importance.
Hence, with $q_{i}$ we have a \emph{non-local measure} about the true impact an agent can have in the organization.

\subsection{Quantifying social resilience}
\label{sec:quant-adapt}

In order to obtain a measure for resilience, we have to solve two problems: (i) 
defining different proxies to measure robustness and adaptivity based on the available information about agents and their relations, (ii) determining a functional form for resilience dependent on the two dimensions.
Again, there is not the \emph{one} way of combining available information into meaningful measures.
Therefore in the following we list a number of candidates to quantify robustness, which we can choose from.
For adaptivity instead we provide only one measure based on the assumption that we have data available to construct a multi-edge temporal network. 

\paragraph{Topological robustness. }
As noted above, robustness can only be defined with respect to a specific shock.
A software developer team can be robust against an external shock, e.g., stronger legal regulations, but not against an internal shock, e.g., the dropout of a leading developer.
Therefore, we must consider different ways for defining the robustness of social organizations. 

Topological measures are often easy to calculate and reflect specific aspects.
The robustness against agent removal can be linked to agents'
\emph{coreness}~\citep{morone2019kcore,casiraghi2019probing,Seidman1983,Garas-Havlin-2012,Verginer-ACS} (see also Figure~\ref{fig:netw}).
It helps understanding cascading effects from removing a specific agent.
\emph{Centralization} \citep{Wasserman1994} takes the concentration of interactions in a few agents into account, which increases the systemic risk if these agents fail \citep{Casiraghi2021} (see also Figure \ref{fig:change}).
\emph{Betweenness preference} \citep{Pfitzner2013} and \emph{Eigengap} \citep{Masuda2013} indicate communication bottlenecks and identify gate keepers.
These measures can be used separately, as demonstrated for centralization \citep{Zanetti2013,Schweitzer2022IPP}, or in combination to quantify robustness.

We can further utilize higher-order models of temporal networks to capture robustness. 
The Second-Order Algebraic Connectivity, for instance, can be interpreted as a \emph{temporal-topological} robustness measure \cite{Scholtes2020}.

\paragraph{Structural robustness. }

A different measure of robustness is proposed by the concept of \emph{structural balance} as explained in Section~\ref{sec:netw-repr}.
It decomposes the network into triads and determines their balance $S_{ijk}$ by multiplying the signs of the signed relations, $\sign(\omega_{ij})$.
This approach has several shortcomings.
First, triads are evaluated \emph{independently}, i.e., the fact that each agent is likely part of different triads at the same time is ignored.
Secondly, the different \emph{weights} of each signed relation, $\omega_{ij}$, are not taken into account.
Thirdly, the importance $r_{i}$ of the agents composing the triad is ignored.
That implies all triads have the same weight in estimating the robustness of the organization, which is not justifiable. 

Correcting for these shortcomings is an open discussion.
As a possible alternative 
we have proposed a new \emph{weighted balance} measure $T_{ijk}$  \citep{Schweighofer2020,Schweitzer2022IChing} that takes into account not only the \emph{signs} and the \emph{weights} of the signed relations, but also the impact of the \emph{agents} involved in the triad.
To determine the structural balance of the whole collective, we take the arithmetic mean, $\mean{T}=\sum T_{ijk}/N^{t}$, where $N^{t}=\lVert T_{ijk}\rVert$ is the total number of triads in the network.

\paragraph{Quantifying adaptivity. }

Ideally a maximally resilient system would have maximal robustness, i.e., it could withstand \emph{any} shock, and maximal adaptivity, i.e., \emph{if} a shock impacts the system it will always recover.
That means resilience $\mathcal{R}$ should increase both with robustness $R$ and adaptivity $A$, \(\mathcal R(R,A) \sim R\cdot A\).

\begin{figure}[htbp]\centering
	\includegraphics[width=0.45\textwidth]{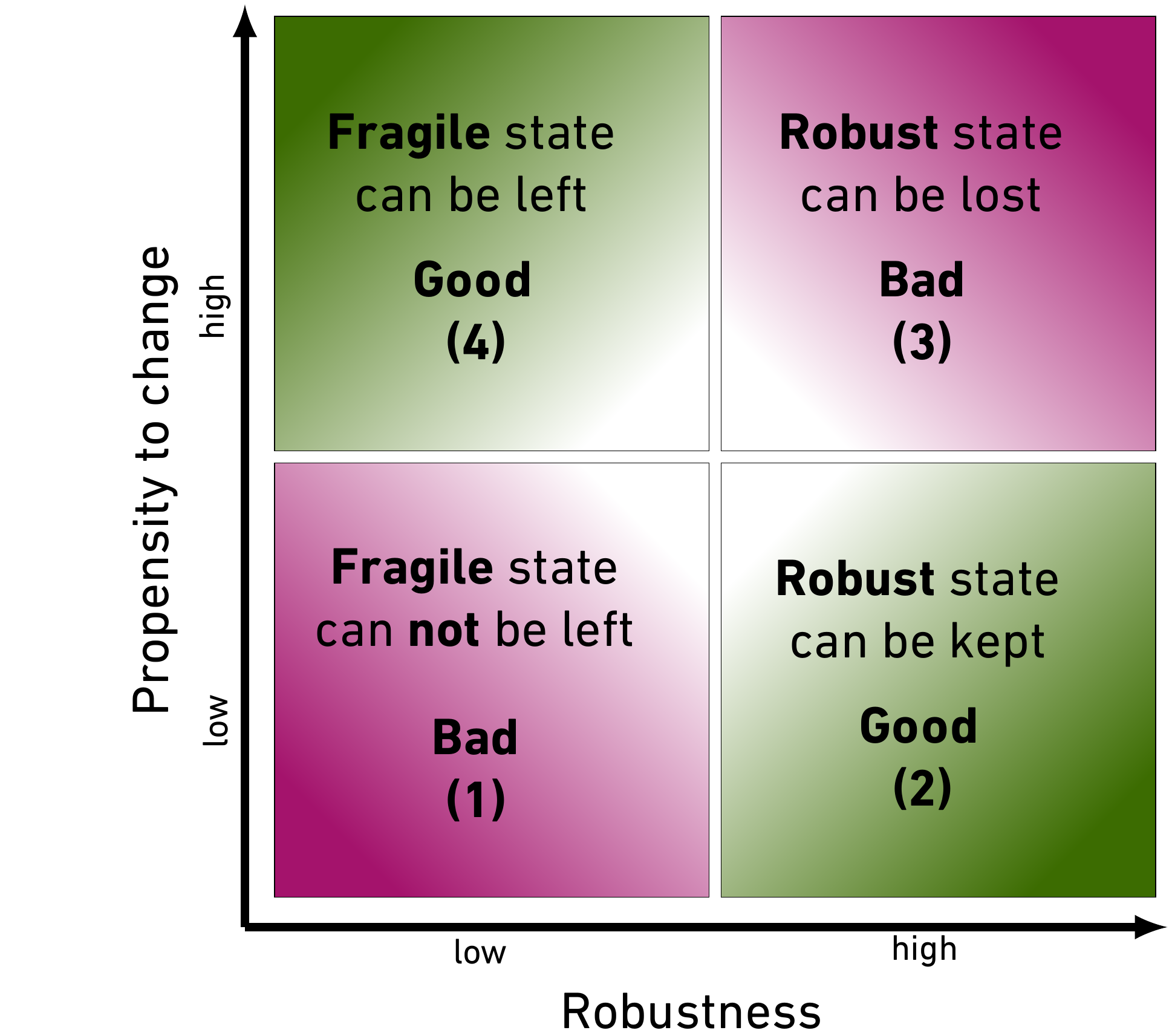}(a)
        \hfill
  		\includegraphics[width=0.45\textwidth]{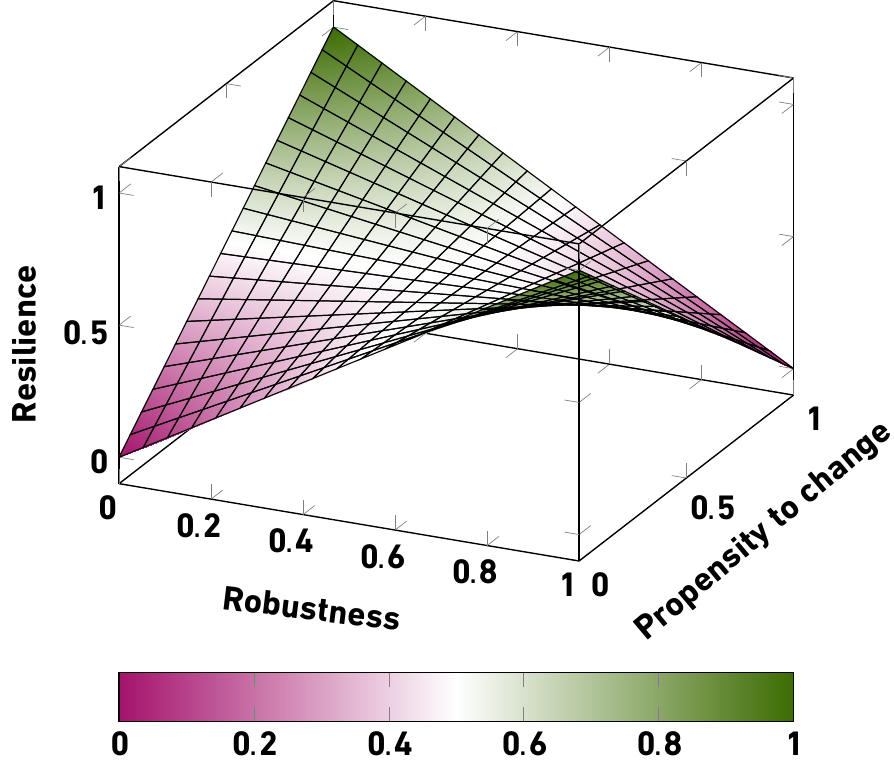}(b)
  \caption{Resilience $\mathcal{R}$ as a function of robustness $\hat R$ and propensity to change $\hat{P}$: (a) Qualitative assessment of different states. (b) Exemplary quantification of $\mathcal{R}(\hat R,\hat P)$. \cite{Schweitzer2022IPP}}
  \label{fig:square}
\end{figure}

Adaptivity does not simply mean ``dynamics''.
Instead, it refers to the \emph{ability} of the organization to attain different states, which we also call \emph{potential}.
But will the system actually attain these alternative states at random, without a response to a shock?
If so, we call this the \emph{propensity to change} to indicate that it is independent of the quality of the current state.
It turns out that the propensity to change is a two-edged sword. 
If a team is in a bad shape, it should be able to leave such bad state.
Then a high propensity to change allows the team to attain other, and likely better, configurations.
On the other hand, if the team has reached a good state, it should be interested in keeping it.
A high propensity to change would be counter productive because the good state could be easily lost.
This means that resilience should \emph{increase} with the propensity to change if the system is in a  \emph{bad} state, and \emph{decrease} in a good state.
Figure~\ref{fig:square}(a) illustrates the problem.

\paragraph{A functional form. }

A decomposition of resilience into robustness and adaptivity, \(\mathcal R(R,A) \sim R\cdot A\), rests on the fact that we can 
capture the {potential to change} of the system independently from its {propensity to change}, which is in fact not possible.
Therefore, we use the propensity to change $\hat P$ as an empirical proxy for adaptivity.

Our measure of potentiality~\citep{Zingg2019}, introduced in Sect.~\ref{sec:prob-appr}, allows us to proxy this propensity. 
It quantifies the probability distribution of states attainable by a system at a given point in time.
The larger the potentiality, the larger is the number of alternative states attainable by the system, and {the more likely is the system to change towards one of them}.
The smaller the potentiality, the smaller the number of states attainable and {the smaller the probability the system will move away from the current state}.

These considerations have determined us to
propose the following functional form for the resilience of social organizations: \mbox{$\mathcal{R}(\hat R, \hat P)=\hat R(1-\hat P)+\hat P(1-\hat R)$} \citep{Schweitzer2022IChing,Schweitzer2022IPP}.
The quantity $\hat P$ is a convenient transformation of potentiality $P$.
I.e., low values of $\hat P$ (below $0.5$) map to a state with low propensity to change, while large values of $\hat P$ (above $0.5$) map to a state with high propensity to change.
The lowest achievable potentiality is mapped to $\hat P=0$, while the highest to $\hat P=1$.  
Similarly, the value of robustness $R$ should be always positive and conveniently scaled between 0 and 1.
This can be achieved for most topology based measures.
For structural balance measures, however, $\mean{T}$ can become negative.
To use these measures, we have to map them as $\hat R=1/(1+e^{\beta \mean{T}})$, where $\beta=0.2$ gives a rather smooth mapping.
$\mean{T}=0$ would then be equivalent to $\hat R=0.5$.
Note that the function plotted in Figure~\ref{fig:square}(b) reflects the arguments summarized Figure~\ref{fig:square}(a).

\paragraph{Relation to ecological concepts. }

It is worth to get back from here to the discussion about ``potential'' and ``connectedness'' as constituents of ecological resilience in Section~\ref{sec:ecological-systems}.
Potential shall define the number of possible alternatives states.
But so far it was only a conceptual proposal because of the lack of operationalization.
This gap is closed by our concept of adaptivity which indeed can be calculated and also compared across different systems \citep{Zingg2019}.
Connectedness refers to the robustness of the system, capturing topological aspects.
Again, with our measure of robustness we are able to calculate and to compare the robustness of different systems.
Moreover, both adaptivity and robustness can be monitored over time, making our resilience measure an instantaneous early warning signal.

Most interesting is the relation between low connectedness and high resilience, on the one hand, and high connectedness and low resilience, on the other hand, discussed in Section~\ref{sec:ecological-systems} \citep{HollingGunderson2002}. 
This was presented together with the hypothesis about the ``adaptive cycle'', which emerges if low connectivity is met by high potentiality.
While this adaptive cycle was considered a ``metaphor'' \citep{Carpenter2001} or a ``thinking tool'', we are able to demonstrate its existence in data about real world organizations \citep{Schweitzer2022IPP}.

\paragraph{Resilience as a compromise. }

Our framework reflects that 
high resilience requires both, the maintenance of a valuable organizational structure to withstand shocks, and the ability to change this structure quickly if needed.
Reasons to change can result from internal or from external problems, for instance from an incapable management or from governmental restrictions.
The resilient organization has to achieve conditions under which it can respond even without prior knowledge about the shock.
Instead of rigidity, it needs fluidity.
But instead of fragility, it also needs stability, dependent on the situation.
Hence, the maximum resilience should be a compromise to balance these different requirements in an efficient manner.

  \parbox{\textwidth}{

    \begin{custombox}{Conclusions}{black!50}
      Turning concepts into measures is the hardest part of modeling.
      There are always different options to operationalize measures, dependent on available information.
      Topological measures alone are not enough to estimate the robustness and adaptivity of social organizations.
      Instead, we need to quantify the impact of agents, to correct structural balance.
      Optimal resilience is a compromise between robustness and adaptivity.
    \end{custombox}

  }

\section{Network construction and interventions}
\label{sec:how-do-we}

So far we have translated our concepts for the robustness and adaptivity of social organizations into measures.
As the last step we have to discuss possibilities of obtaining the data needed to calculate these measures.
If we achieve to have a calibrated generative model of the social organization, we can address the problem of \emph{system design} \citep{Schweitzer2019-bigger}.
That means, we can test how possible intervention strategies impact the organization's resilience. 

  \parbox{\textwidth}{

    \begin{custombox}{Questions}{black!50}
      \begin{itemize}
      \item How do we construct networks from data?
      \item What type of data is needed to calculate our resilience measure?
      \item How can we obtain such data from repositories of social organizations?
      \item Is there a way to validate our generative model?
            \item How can we control resilience using network interventions?  
      \end{itemize}
    \end{custombox}
  }

\subsection{Data acquisition and analysis}
\label{sec:what-data}

We want to emphasize that our methodology inverts the usual \emph{supply driven} approach found in computational social science.
This starts from the data given, often collected without a clear purpose and a research question in mind, to subsequently squeeze out interesting features.
In contrast, our \emph{demand driven} approach has first identified in four steps shown in Figure~\ref{fig:frame} what  data will be needed to inform our models.
Then, utilizing this data we can infer information about \emph{agents} and their properties, but also about their \emph{interactions} with others, as shown in Figure~\ref{fig:outline}.

Such data \emph{cannot} directly provide the input for our models and is \emph{not} sufficient to simply estimate social resilience.
Instead, it has to be pre-processed, before we can construct the networks that are essential for our framework.
These networks are never given, and their generation and subsequent statistical interpretation bears some of the most overlooked problems in modeling social organizations.

\paragraph{Extract interactions. }

\begin{figure}[htbp]
  \centering
\includegraphics[width=0.45\textwidth]{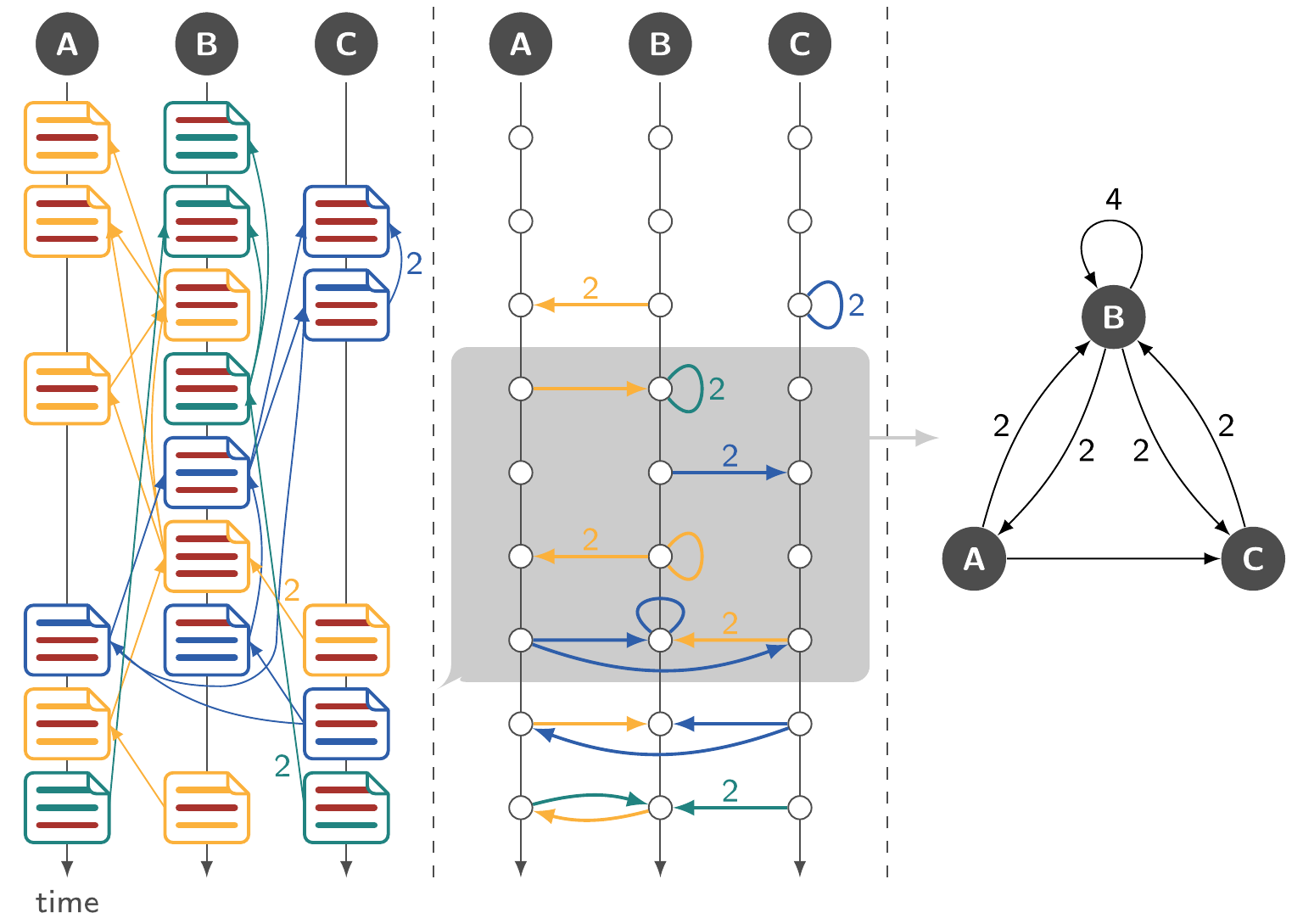}  
  \caption{Extracting the collaboration network of developers A, B, C using \texttt{git2net} \citep{gote2019git2net}. }
  \label{fig:git2net}
\end{figure}

One of our reasons to study software developer teams as prototypes of social organizations is the availability of vast \texttt{git} repositories.
These contain fine-grained records of all changes made to the software, together with information who changed it, what was changed and when.
We developed a software package, \texttt{git2net} \citep{gote2019git2net,gote2021analysing}, that is able to extract this information, to create bipartite networks and their projections into an interaction network between developers (see Figure~\ref{fig:git2net}).

We note that, in addition to the co-editing network, i.e., the collaboration network of developers, we can obtain additional information about the social organization.
For instance, analyzing the \emph{amount} of code changes can quantify the productivity of developers \citep{Scholtes2016b} and analyzing the \emph{sequence} of code changes gives insights into their hierarchical structure. 

\texttt{git2net} can be also used to mine other \texttt{git} repositories, e.g., for publications.
Additionally we have developed the rule-based disambiguation tool \texttt{gambit} to solve the persistent problem of \emph{name disambiguation} occurring in most real-world user data \citep{Gote2021}.

\paragraph{Time window detection. }

To obtain interaction networks, a sliding window approach aggregates interactions over a certain time interval.
Choosing the right window size is a problem in itself, because the window size impacts the network density and subsequently all topological analyses.

Often we have no data about interactions and need to infer them from time series of observed events.
For instance, from co-location data, i.e., observations about two individuals $i,j$ acting at times $t_{i}$ and $t_{j}$ at a given place, we need to detect the time interval $\Delta t= \abs{t_{i}-t_{j}}$. 
Only observations with a $\Delta t$ lower than a given threshold $\Delta t^{\mathrm{thr}}$ will count as interactions \citep{Mavrodiev2021}.
Such considerations are important to quantify, e.g., the transmission of information within a social organization.

\paragraph{Analyzing temporal data. }

The dynamics of \emph{temporal networks} crucially depend on $\Delta t$.
The problem, who can potentially influence whom, requires to reconstruct temporal \emph{paths} of various lengths \citep{Gote2020}, on which our networks can be generated.

We have developed different software packages to support the analysis of temporal networks.
They are combined in the toolbox 
\texttt{pathpy} \citep{pathpy}.
It implements, for instance, statistical techniques to find optimal graphical models for the causal topology.
These models balance model complexity with explanatory power for empirically observed paths in relational time series.
As part of \texttt{pathpy}, 
\texttt{MOGen} \citep{Gote2020} is a multi-order generative model to statistically evaluate paths of
various lengths.
It can be used to improve the computation of different temporal centrality measures in case of insufficient observations.

\paragraph{Infer signed relations. }

Signed relations $\omega_{ij}$ are instrumental to calculate the social impact and subsequently the robustness of the organization.
As described above, our framework uses \emph{interaction} data to infer signed relations.
This method can be enhanced by taking additional data sets into account that can provide information about the positive or negative relations between individuals.

If written or spoken text is available, we can use \emph{sentiment analysis} to obtain information about the emotional content \citep{Garcia2016,Schweitzer2013}, to infer social relations.
\emph{Natural language processing} (NLP) provides an extended tool box to further extract information about opinions, attitudes, or ideological positions \citep{Abercrombie2020,Russo2022}.
These can help quantifying the social impact that individuals exert on others.

\paragraph{Information from collaboration platforms. }

Another important source of information are online collaboration platforms, such as \texttt{slack}, \texttt{zoom}, or \texttt{GitHub}.
In addition to interaction data and text messages, they often provide  information about attention, e.g., via \texttt{likes}, about declared trust, recommendations, and activity patterns.

Based on reconstructed collaboration networks, one can analyze the presence of social mechanisms like \emph{reciprocity}, \emph{homophily}, \emph{triadic closure} \citep{Brandenberger2019,Rivera2010}, or of other \emph{motifs} \citep{Xuan2015}.
This information can be used to further characterize the importance of agents and their signed relations and to estimate their impact on the resilience of the organization.

\paragraph{Network regression. }

If the topological information is sufficient to reconstruct an additional network layer, it can be utilized for the network regression outlined above.
To facilitate the computation, we have developed an $R$ package \texttt{ghypernet} \citep{ghypernet2022}. 
It implements gHypEG, the network ensemble considering propensities.
In addition to network regressions, the package can be used to infer significant relations from observed interactions \citep{Casiraghi2017}.

Once gHypEG is calibrated, we can also compute our \emph{potentiality} measure even for large ensembles.
\texttt{SciPy} \citep{scipy2020} provides an efficient implementation for computing the entropy of a given multinomial distribution.

\paragraph{Calibration and validation. }
To find the optimal \emph{combination} for the different measures mentioned above is recognized as an open problem.
Symbolic regression \citep{gplearn2022,diveev-shmalko-2021-symbol-regres-method} and other 
machine learning (ML) techniques are increasingly used to find solutions.
In many cases, \emph{ground truth data} is not available.
Then, we have to rely on in-sample and out-of-sample predictions to aggregate different information in a meaningful manner.

This issue becomes relevant if we, for instance, want to improve the importance measures for agents.
If reliable aggregation methods are not available, we have to resort on determining the $r_{i}$ values from topological measures, combined with dynamic processes as, e.g., in feedback centralities, as we will do in the following example. 

\subsection{Resilience and control}
\label{sec:resilience-control}

\emph{Improving} the resilience of complex systems implies that, to some extent, we are able to \emph{influence} such systems in a way that their functionality and their stability is maintained, or even enhanced.
The formal model of social organizations described above allows testing such intervention strategies.

\paragraph{Top-down control. }

Generally we distinguish between bottom-up and top-down interventions \citep{schweitzer-2020-design}.
The latter mostly focus on the boundary conditions for organizations, either to prevent shocks or to enhance their business environment. 
These can be financial measures during the Corona crisis, but also legislative measures to ensure fair competition.

In general, to use the top-down approach, one needs to identify global \emph{control parameters} which is a challenge on its own.
Often they  can be derived from the known macroscopic or system dynamics.
As a major conceptual drawback, control parameters usually reflect limitations of stability, rather than of resilience. 

While the top-down approach is discussed in macro-economics and recently in macro-prudential regulations, we are interested in the bottom-up approach which is more in line with the complex systems philosophy.

\paragraph{Bottom-up interventions. }

Our \emph{bottom-up} approach to resilience uses interventions targeting specific agents and their interactions \citep{Schweitzer2020}.
\emph{Structural} interventions focus on the interaction structure, basically changing the adjacency matrix of the network.
\emph{Functional} interventions change, for instance, the interaction rules to affect timing of interactions \citep{Sanhedrai2022}.

\begin{figure}[htbp]
  \centering
  \includegraphics[width=0.5\textwidth]{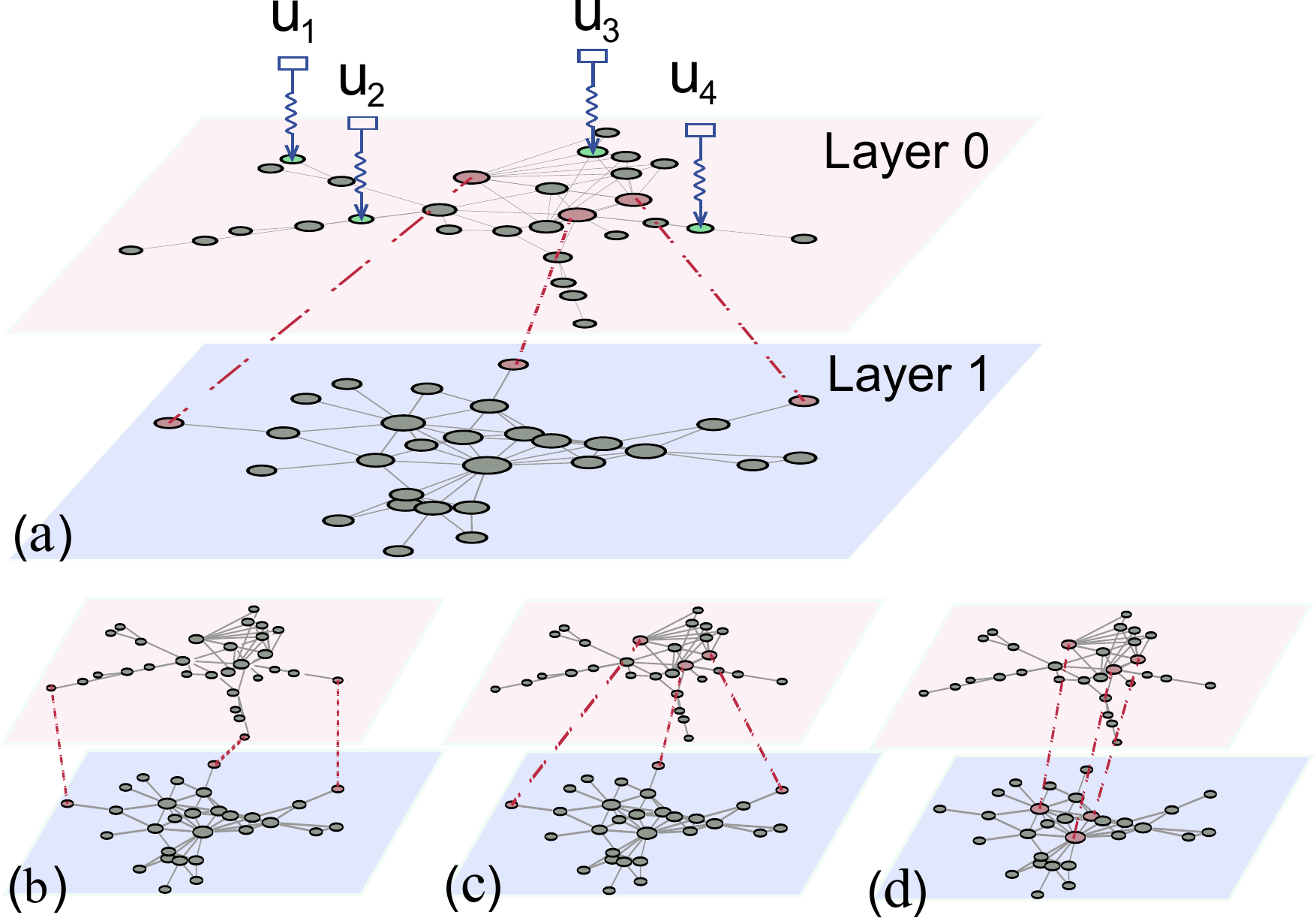}
  \caption{Network control in a two-layer network. (b-d) Different couplings between the two layers, dependent on the peripheral or central position of agents. \citep{Zhang2016b} }
  \label{fig:control}
\end{figure}

\emph{Dynamical} interventions instead influence the internal state of nodes, i.e., the agents. 
Such measures can include nudging or mechanism design \citep{schweitzer-2020-design}, but the most promising way for us are network interventions (see Figure~\ref{fig:control}).
They require to first identify the driver nodes, i.e., those agents that should be targeted, and secondly to decide about the type and the amount of interventions \citep{Liu2011,Zhang2016b}.
Often these interventions change the agents' utilities $u_{i}$ using control signals.

\paragraph{Indirect influence. }

To get access to agents, we can utilize the multi-layer structure of the organization. 
For instance, if one layer contains friendship relations and the second one task assignments, the friendship layer can be used to influence the work relations.

If the impacted agent responds appropriately, changes can propagate through the network, this way influencing agents that were not targeted directly.
As the example of Figure~\ref{fig:control} shows, we can target agents at the periphery of the network to impact agents in the core \citep{morone2015influence,Zhang2016b,Zhang2019}.

\paragraph{Influence on decisions. }

Network interventions only control a small number of agents at a comparably low cost, while utilizing the systemic feedback.
But the method requires a model of the organization to forecast the impact.  
Further, it assumes that the agent's utility is known.
For the latter we can have at least reasonable assumptions.

Rational agents want to keep or even increase their impact, $q_{i}$, by either increasing their importance, $r_{i}$, or by decreasing a negative social impact, $I_{i}$, they experience.
But changing signed relations or maintaining collaborations is costly.
Agents may decide to leave the organization if their costs exceed their benefits.
Conversely, they may decide to stay if their benefits have been increased. 

Changing agents' utility has the advantage of influencing these decisions.
If agents leave or reorganize their links, this changes the network topology and impacts the dynamics in each layer.
Consequently, both the robustness and the adaptivity of the organization are impacted.
This can lead to counter intuitive effects.
For instance, removing some agents may \emph{stabilize} the organization~\citep{casiraghi2020improving,Schweitzer2020}.
While this is known in human resource management, models are hardly able to reproduce such behavior.

\begin{figure}[htbp]
   \includegraphics[width=0.3\textwidth]{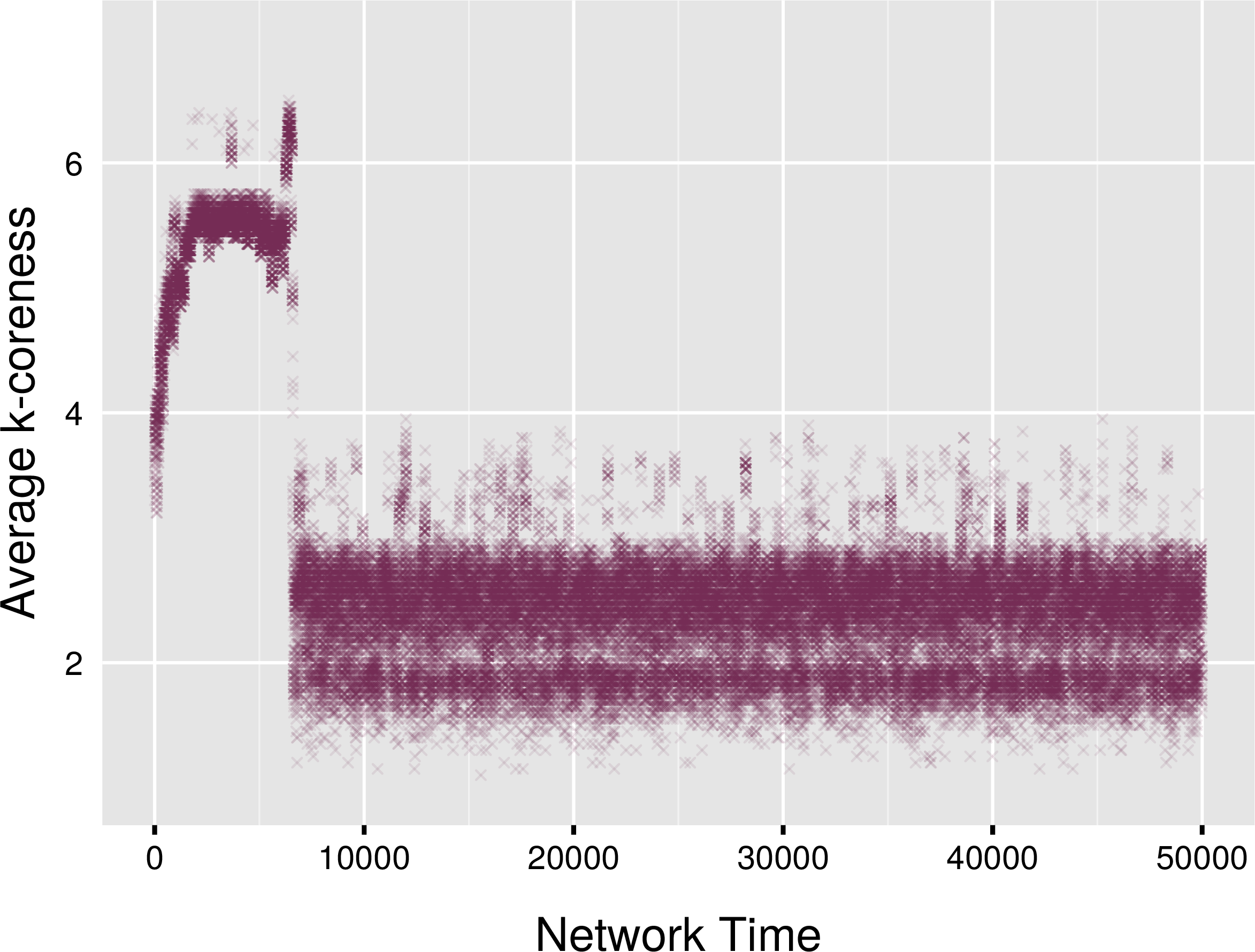}\hfill
   \includegraphics[width=0.3\textwidth]{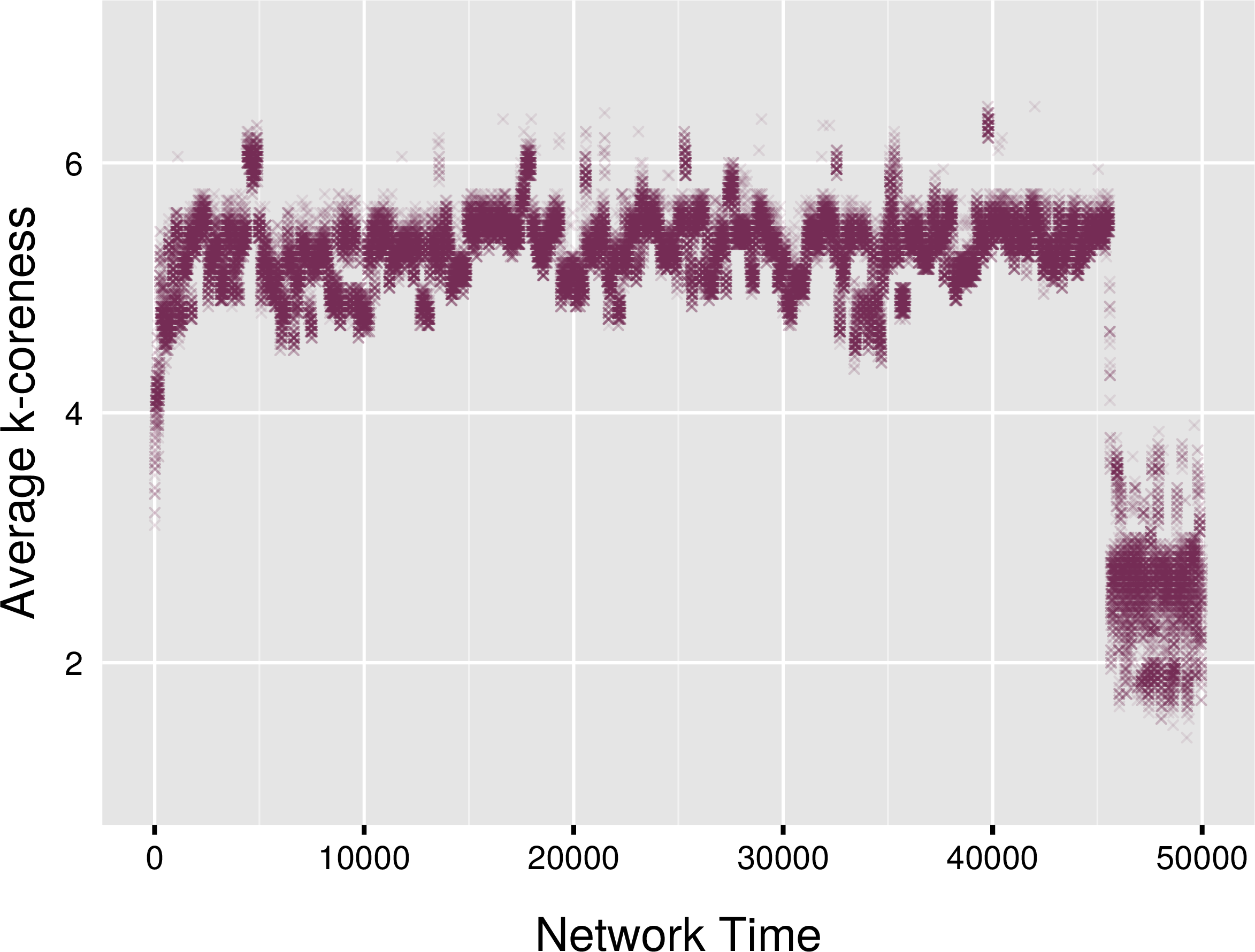}\hfill
\includegraphics[width=0.3\textwidth]{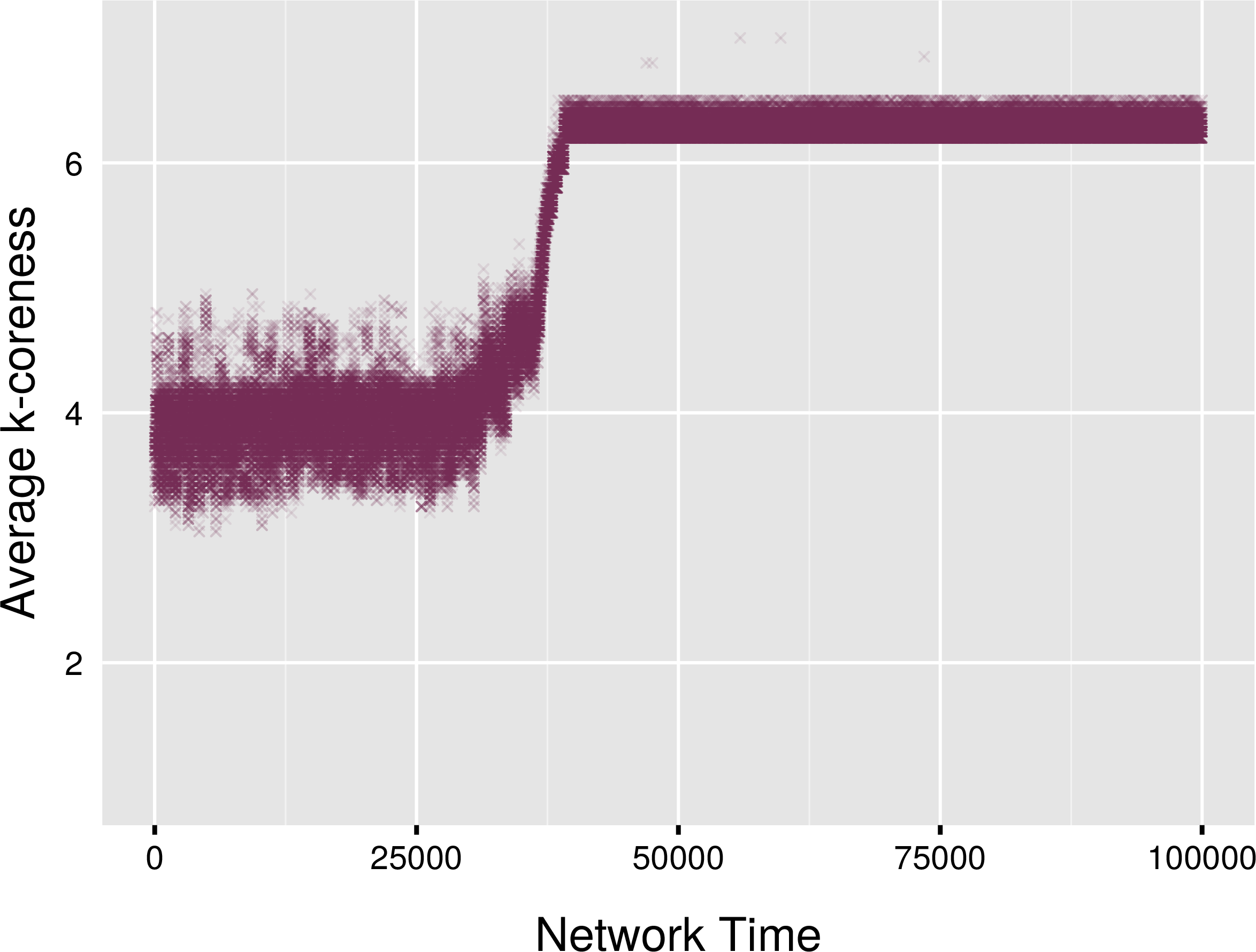}

\hfill \includegraphics[width=0.33\textwidth]{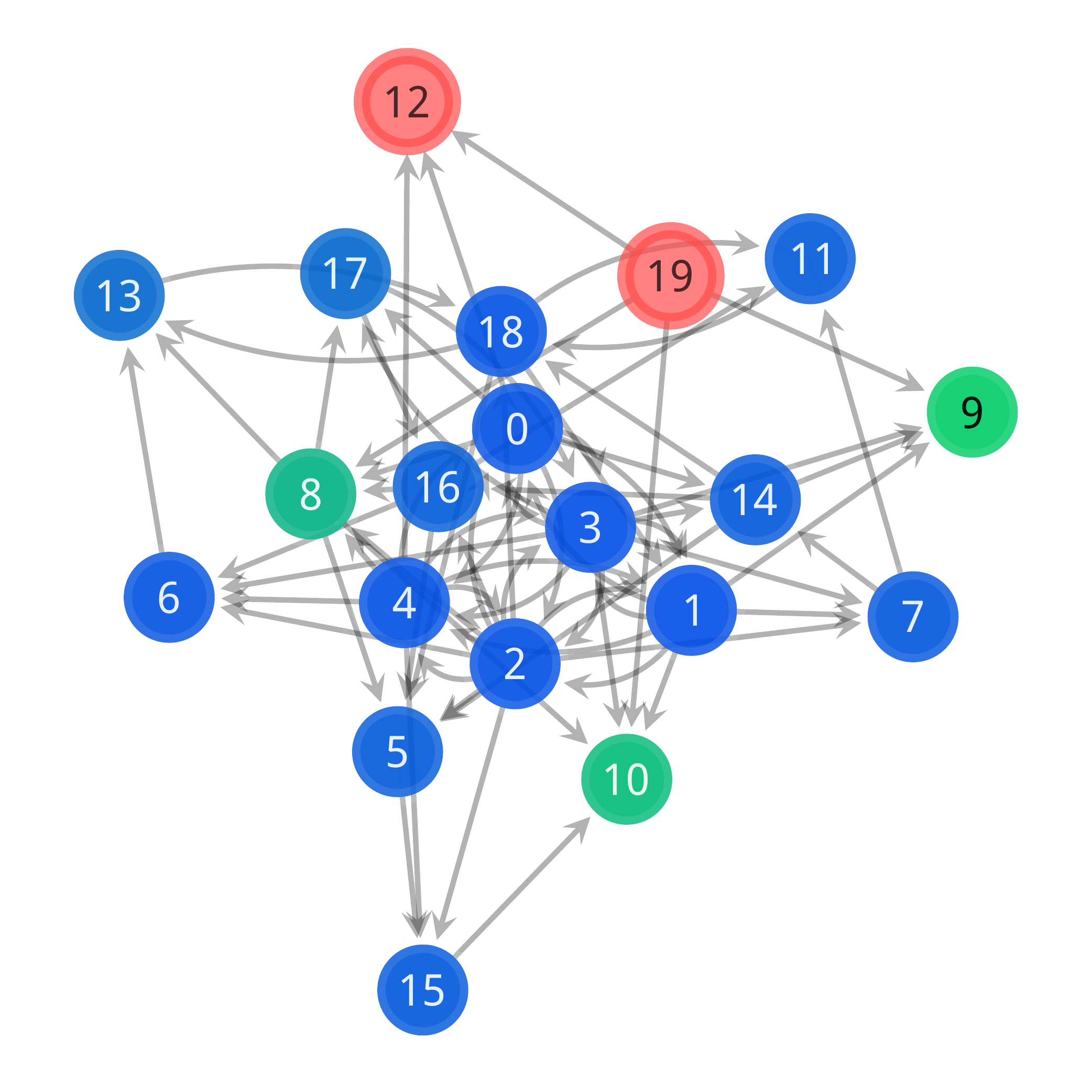}   
      \includegraphics[width=0.33\textwidth]{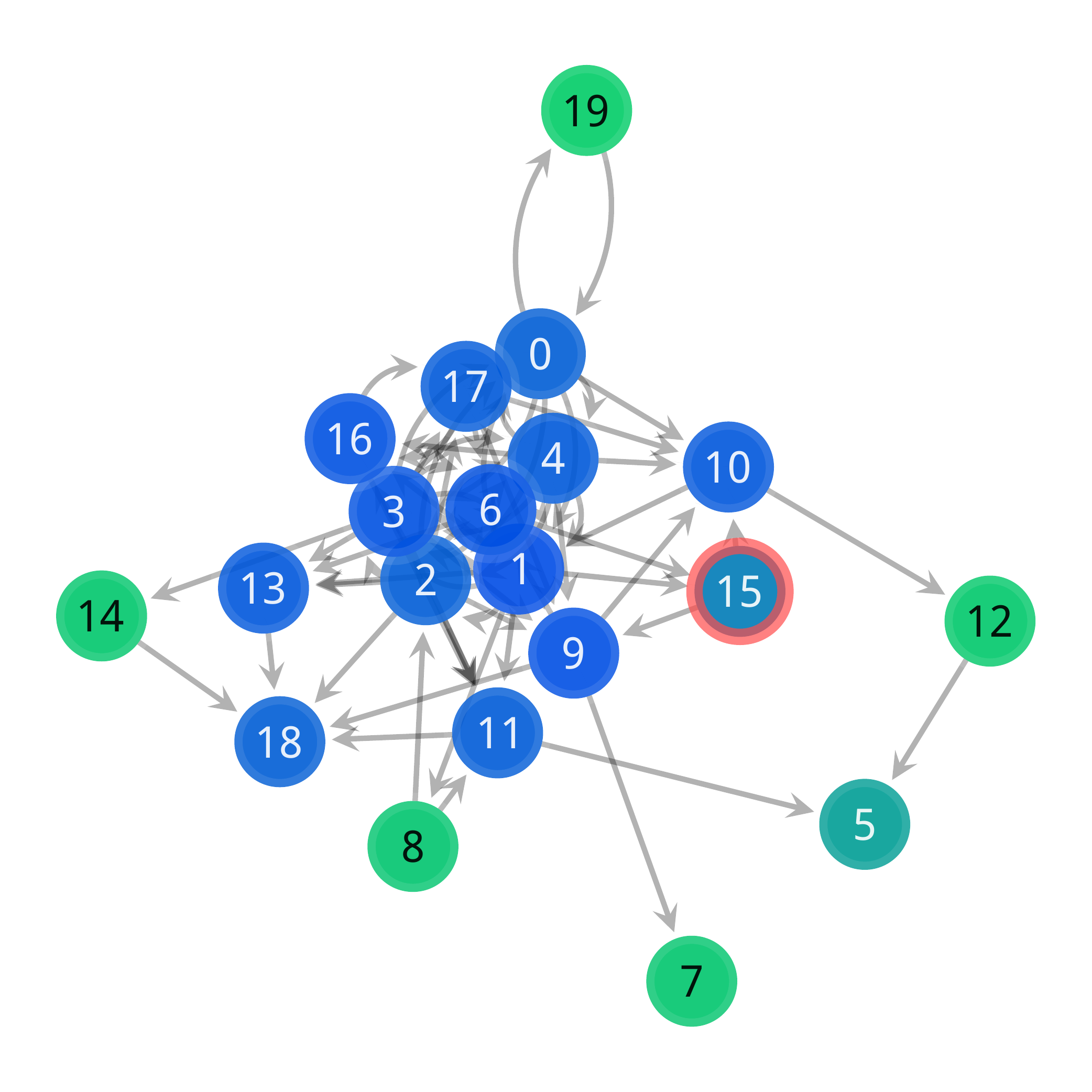}
    
    \caption{Network interventions to prevent the breakdown of a social network, indicated by the drop of $k$-coreness. (a) No intervention, (b) peripherial agents targeted, (c) agents close to the core targeted \cite{casiraghi2020improving,Schweitzer2020}. }
    \label{fig:3cases}
  \end{figure}

  \parbox{\textwidth}{

    \begin{custombox}{Conclusions}{black!50}
      We provide a whole tool box for mining and analyzing data of social organizations.
      In particular, interaction data can be obtained from repositories.
      Other tools allow to calculate temporal centralities to characterize communication,
      and to infer propensities for interacting individuals.
      Which of the different measures are calculated depends on the available data.
      There are various ways to proxy robustness, adaptivity  and resilience of social organizations.
            Network interventions allow to improve the resilience of organizations. 
    \end{custombox}
  }

\section{Conclusions} 
\label{sec:how}

\subsection{What is resilience?}
\label{sec:what-resilience}

\paragraph{Structural and dynamic dimensions. }

Summarizing our tour through the modeling of social organizations, some  important insights should be noted.
First, resilience is a concept that combines \emph{two dimensions}, robustness and adaptivity.
Robustness, as the structural dimension, captures the ability of a social system to withstand shocks.
Adaptivity, as the dynamic dimension, captures the ability of a social system to recover from shocks.
Neither maximal robustness, nor maximal adaptivity alone are sufficient to warrant  resilience for social organizations. 
Both dimensions create a tension, because increasing robustness may lower adaptivity and the other way round.  
Therefore, a resilient state is a \emph{compromise} balancing the influence of both dimensions.

This insight is important because, following arguments from engineering, resilience is too often just treated as a synonym for stability.
This leads to the conclusion that maximizing resilience means maximizing robustness.
Such a perspective may hold for designed infrastructure systems, but not for self-organizing systems such as social organizations.

\paragraph{Resilience measure. }

To turn a concept into a measure requires operationalization which points to a different problem domain \cite{Helfgott2018}.
Even if we agree about our resilience concept, there may be different proposals to operationalize it.
They have to solve two problems.
Firstly, the functional form of resilience dependent on robustness and adaptivity should be specified.
Secondly, measures for robustness and adaptivity have to be proposed and subsequently operationalized.

The latter is the real difficulty.
What should we measure to quantify robustness or adaptivity?
A system may be robust against some specific shocks but will fail for others. 
Therefore, the question cannot be answered without an appropriate \emph{formal model} of the social organization.
In this paper, we made an operationalization proposal based on networks which can be constructed from data.
In general, these are multi-edge, temporal, multiplex and dynamic networks. 
From these networks, topological information can be used to calculate robustness.
Using the ensemble approach, we are further able to calculate adaptivity.

\paragraph{Resilience as an emergent property. }

For our modeling framework of social organizations we have adopted the complex systems perspective, in general, and the \emph{complex networks} approach, in particular.
This implies to explain \emph{resilience} as an \emph{emerging property} of the social organization.
Following the bottom-up approach, we have to focus on the \emph{micro level} of interacting agents.
Measures for resilience need to be derived from this perspective.
It requires to characterize agents in some detail regarding their importance, their signed relations and their social impact on others.
Simple network measures that treat agents as dots to just calculate their network position are not sufficient to estimate robustness, even less to understand 
adaptivity as the dynamic component of resilience.

Resilience as a systemic property has to be constantly maintained, which requires the \emph{activity} and the \emph{cooperation} of the members of the social organization.
Conversely, big threats to social resilience are not coming only from external shocks, but from internal challenges as well.
As our model framework demonstrates, 
negative signed relations and negative social impact hamper robustness and adaptivity.
Biased interactions, the lacking integration of newcomers and low connectivity undermine the conditions under which resilience can be established.

\subsection{The policy dimension}
\label{sec:policy-dimension}

\paragraph{Science versus policy. }

Our framework for modeling the resilience of social organizations helps to understand 
under which conditions resilience is lost.
Our reasoning does not refer to the loss of ``robustness'' and ``adaptivity'' in an abstract manner.
It instead relates these losses to the underlying properties of agents and their dynamic interactions.
Only this way we are able to propose network interventions for improving resilience.

But do these models, while successful from a scientific perspective, benefit policy makers in any way?
Are we able to tell them what to do?
To put this challenging question into perspective, we remind on some preconditions and some findings.

Our models focus on a specific type of social organizations, namely teams of collaborating members sharing a common goal.
This means that we are not considering societies and, hence, our models neither aim at, nor are suited for, making suggestions on how to improve the resilience of societies against political, economic or environmental shocks.
Delimitation was the first step for developing our framework.
With these restrictions in mind, our models indeed support some general insights.

\paragraph{Awareness. }

The most important insight is probably about the role of awareness for what resilience really means.
This requires distinguishing it from concepts of robustness, stability, functionality, or optimality.
Resilient systems are not obtained by maximizing or optimizing specific functions or key figures.
A resilient organization has to withstand \emph{various kinds} of shocks and to recover from them.
That means it needs to be prepared for the \emph{unknown}, instead of being specialized to fit the known.

This addresses a policy issue:
Instead of improving resilience, organizations have strong incentives to rather improve performance as the most visible indicator of success.
This reminds on the classical conflict between short-term benefits and long-term deterioration and points to the 
misallocation of limited resources needed for maintaining resilience.
Ideally, a social organization should be able to \emph{anticipate} possible shocks to some degree, and to prepare in advance for this, also by securing resources. 
This requires \emph{collective awareness}, a state of consciousness that is based on continuously analyzing and recognizing the situation inside and outside the organization.

\paragraph{Flexibility. }

Next to robustness, our models highlight the role of \emph{adaptivity}.
It proxies \emph{the number of options} that an organization may have to respond to shocks.
Consequently, we measured adaptivity by potentiality.
It does not imply that these options are taken, but that they exist in a current situation.
Resilience depends on \emph{alternatives}.
That means, concepts like \emph{flexibility} or \emph{fluidity} become increasingly important.
We remind on the  concept of \emph{adaptive capacity}  which already refers to the ability of an organization to \emph{adapt} either in preparation, or in response to perturbations.

\paragraph{Quantifying resilience principles. }

Ten years ago, \citet{biggs2012} identified
seven principles for building resilient socio-ecological systems:
\begin{enumerate}[noitemsep,label=(\arabic*)]
\item maintain diversity and redundancy,
\item manage connectivity,
\item manage slow variables and feedbacks,
\item foster complex adaptive systems thinking,
\item encourage learning,
\item broaden participation,
\item promote poly-centric governance systems.
\end{enumerate}
These principles already highlight the importance of a \emph{complex systems perspective}, the role of adaptivity and decentralized control. 
But now we provide a \emph{modeling framework} for social resilience where 
formal models allow to quantify the value of \emph{redundancy} and \emph{connectivity} using 
multi-edge and multi-layer networks.
They show how agents' diversity, i.e., their \emph{heterogeneity}, their social impact and  their signed relations impact social resilience.

From a broader perspective, our paper wishes to contribute to a better concept of \emph{resilience management}.
This requires \emph{both} an understanding of the system that should be managed and an active involvement of those who are managing and those being managed.
Social organizations are a prime example for those systems.
We are the system elements, the agents, of our own social organization.
We are in the position to change our organization to improve resilience.
At the same time, we are also affected by these changes, as well as by internal and external shocks.
Our modeling framework helps to raise \emph{attention} for the role of diversity and feedback processes, the power of decentralized network interventions and collective learning.
In the end, however, it depends on us how much of these insights can be implemented in our social organizations.

\subsection*{Acknowledgements}

The authors thank  A. Garas, D. Garcia, P. Mavrodiev and S. Schweighofer for early discussions. %

\small \setlength{\bibsep}{1pt}

\end{document}